\documentclass{iopjournal}
\usepackage{bm,color,amsmath,txfonts}
\usepackage{graphicx}
\usepackage{siunitx}
\usepackage{subfigure}
\usepackage{verbatim}
\usepackage{dcolumn}
\usepackage{epsf}
\usepackage{xcolor}
\usepackage{hyperref}
\usepackage{hhline}
\usepackage{float}
\usepackage{enumerate}
\usepackage{bbm}
\usepackage{lipsum}
\usepackage{mathrsfs}
\usepackage{amssymb}
\usepackage{ulem}
\usepackage{amsmath}
\usepackage{subeqnarray}
\usepackage{cases}
\usepackage{amsfonts}
\usepackage{graphicx,braket}
\usepackage[dvipsnames]{xcolor}
\usepackage{calc}
\usepackage{epsfig}
\usepackage{color}
\usepackage{tikz}
\usepackage{graphicx,array}
\usepackage{dcolumn,soul}

\usepackage[figuresright]{rotating}%
\usepackage{algorithm, algorithmicx, algpseudocode}
\usepackage{listings}%
\usepackage{hyperref}

\begin{document}
	
	\articletype{Topical Review} 
	
	\title{Quantum Magnonics: Quantum States Generation and Applications}
	
	\author{Zi-Xu Lu$^{1*}$, Xuan Zuo$^{1*}$, Xin-Lei Hei$^{2*}$, Gang Liu$^{1\dag}$, Zeng-Xing Liu$^{3\dag}$, Qi Guo$^{4\dag}$, Peng-Bo Li$^{2\dag}$, and Jie Li$^{1\dag}$}
	
	\affil{$^1$School of Physics, Zhejiang University, Hangzhou 310027, China} \\
	\affil{$^2$Ministry of Education Key Laboratory for Nonequilibrium Synthesis and Modulation of Condensed Matter, Shaanxi Province Key Laboratory of Quantum Information and Quantum Optoelectronic Devices, School of Physics, Xi'an Jiaotong University, Xi'an 710049, China} \\
	\affil{$^3$School of Electrical Engineering $\&$ Intelligentization, Dongguan University of Technology, Dongguan 523808, China} \\
	\affil{$^4$State Key Laboratory of Quantum Optics Technologies and Devices, and College of Physics and Electronic Engineering,  Shanxi University, Taiyuan 030006, China} \\
	\affil{*These authors contributed equally to this review. $^{\dag}$Authors to whom any correspondence should be addressed.}
	
	\email{gangliuphys@gmail.com; liuzx@dgut.edu.cn; qguo@sxu.edu.cn; lipengbo@mail.xjtu.edu.cn; jieli007@zju.edu.cn}
	
	\keywords{quantum magnonics, cavity magnonics, optomagnonics, superconducting qubits, quantum science and technology}

	\begin{abstract}
		Hybrid systems based on magnons in ferromagnetic materials, such as yttrium iron garnet, have achieved remarkable development in the last decade. These include the coupling of magnons to microwave and optical photons, superconducting qubits, phonons, spins, the center-of-mass motion of a ferromagnet, etc. Here, we review both the experimental and theoretical progress in this field, focusing on the generation of magnonic quantum states and their applications in a broad range of fields. Since the strong coupling is a prerequisite for achieving coherent quantum control of magnons and preparing magnonic quantum states, we start by introducing representative strong-coupling experiments in cavity magnonics, then review a series of protocols for creating various magnonic quantum states, such as Fock, cat, squeezed, and entangled states, and discuss their potential applications in macroscopic quantum studies, quantum information science, quantum sensing, magnonic quantum devices, dark matter detection, and so on. Finally, we summarize the review and give an outlook for the future study of quantum magnonics. 
	\end{abstract}

	\section{Introduction}
	
	Hybrid systems based on magnons have achieved rapid and significant progress in the past decade or so~\cite{Naka19,Clerk20,Li20,Aws21,Yuan22,Bauer22,Zuo24}. This process has been accelerated ever since the strong coupling between magnons in an yttrium-iron-garnet (YIG) crystal and microwave cavity photons was experimentally achieved~\cite{Huebl13,Tabuchi14,Zhang14}, benefitting from many excellent properties of the YIG material, {such as low dissipation rate and high spin density (several orders of magnitude higher than that of atomic ensembles).} On the one hand, high spin density can greatly enhance the cavity--magnon coupling strength because it is proportional to the square root of the number of spins. On the other hand, the YIG, especially the YIG sphere, has so far the smallest magnon dissipation rate among all the ferromagnetic and antiferromagnetic materials~\cite{Chumak26}.  These two facts make YIG an ideal choice for realizing the cavity--magnon strong coupling, where reversible energy exchange between magnons and cavity photons is faster than their energy dissipation to the environment. The strong coupling is a prerequisite for realizing coherent control of magnons and for preparing magnonic quantum states, and thus is vital for activating quantum applications in magnonics.
	
	Another distinct advantage of the magnonic system is manifested as its exceptional ability to coherently couple with various quantum systems, including microwave and optical photons, superconducting qubits, phonons, spins, the external center-of-mass (COM) motion of a ferromagnet, etc (figure~\ref{hybrid}). {Such an excellent property makes the magnonic system a very promising platform to build hybrid quantum systems, combining individual strengths and complementary functionalities of different physical systems.} To date, the magnon mode in a YIG sphere has achieved strong coupling with the microwave cavity mode~\cite{Tabuchi14,Zhang14}, the superconducting qubit~\cite{tabuchi2015}, and the mechanical mode~\cite{Shen25}.  The strong coupling with an optical mode, manifested as the magnon-induced Brillouin light scattering (BLS), has not yet been realized primarily due to the weak single-photon optomagnonic coupling rate for a large-sized YIG sphere~\cite{Nakamura16,Zhang16,Haigh16}. Nevertheless, the optomagnonic strong coupling is promising to be reached in the near future by significantly reducing the mode volume and increasing the mode overlap of the optical and magnon modes, e.g., in a YIG micro disk~\cite{Disk26}.  
	
	Among all magnonic hybrid systems, the effective magnon--qubit strong coupling mediated by a microwave cavity~\cite{tabuchi2015} remains, so far, the only platform that has created and measured magnonic quantum states in a macroscopic ferromagnet, such as the detection of a single magnon~\cite{lachance-quirion2020}, the superposition state of a single magnon and vacuum~\cite{xu2023}, and magnon squeezing below vacuum fluctuations~\cite{weng2026}. Theories also indicate that other quantum states, e.g., entangled and cat states, and magnon  blockade effects can be generated using currently available parameters from the state-of-the-art experiments.  Therefore, given the importance and high feasibility of this platform in preparing various magnonic quantum states, the section of the magnon--qubit system is the core of this review. The optomagnonic system is another focal point, given its essential role in building magnonic quantum networks, where light, rather than microwave fields, should be used to transmit quantum information among remote quantum nodes, and in controlling and engineering magnonic quantum states using well-developed optical means.  Moreover, this system is an ideal arena to apply central concepts in quantum information science, such as quantum teleportation, quantum networks, and the Duan--Lukin--Cirac--Zoller (DLCZ) protocol, to the emerging field of quantum magnonics. Besides the above two systems, the magnon--phonon nonlinear interaction renders the magnomechanical system another platform to prepare macroscopic quantum states, e.g., the magnon--photon--phonon entanglement~\cite{Li18}, and microwave quantum states, e.g., entangled microwave fields~\cite{LJ20}.  Note that, since cavity magnomechanics has recently been reviewed in Ref.~\cite{Zuo24}, covering recent advances on both quantum states generation and applications, this part will no longer be included in this review. Reference~\cite{Zuo24} can serve as a supplement to this paper. Magnons can also couple to other degrees of freedom, such as the nitrogen-vacancy (NV)-center spins, and the COM motion of a levitated YIG sphere, via designed novel coupling mechanisms. Although relevant experiments have not yet been demonstrated, theories indicate that the spin--magnon--phonon tripartite coupling can be greatly enhanced~\cite{Hei2023} holding the potential for building hybrid quantum systems, and the ground-state cooling of the COM motion of a massive YIG sphere is reachable~\cite{Kani22} with important applications in macroscopic quantum studies.
	
	The review is organized as follows. Section~\ref{HQSs} introduces the framework of the hybrid quantum systems based on magnons, basically following the framework established in Ref.~\cite{Naka19} but with an extension of the coupling to spins and the COM motion. The following five sections introduce in turn the coupling of magnons to microwave cavity photons (section~\ref{electromag}), superconducting qubits (section~\ref{mag-qubit}), optical cavity photons (section~\ref{optomag}), NV-center spins (section~\ref{mag-spin}), and the COM motion of a YIG sphere (section~\ref{mag-COM}).  In each section, we first introduce the basic theory of the system and review representative experiments that are quantum-related, e.g., the strong-coupling experiment, which is the first step for preparing quantum states in the system. We then review a series of theoretical protocols for generating various magnonic quantum states, such as entangled, squeezed, cat, and Fock states, and discuss their broad applications in macroscopic quantum mechanics, quantum sensing, quantum networks, quantum teleportation, quantum illumination, magnonic quantum devices, dark matter detection, and so on. Finally, section~\ref{conc} summarizes this review and provides an outlook for future research on quantum magnonics.
	
	It is worth noting that although some other reviews~\cite{Naka19,Yuan22,Bauer22} also touched on magnonic quantum states and their applications, central concepts in quantum information and quantum sensing, e.g., quantum teleportation, quantum networks, quantum illumination, the DLCZ protocol, and the Bell test, and recent experimental achievements on magnonic quantum states, e.g., quantum superposition~\cite{xu2023} and squeezed~\cite{weng2026} states, were not involved. This review, together with Ref.~\cite{Zuo24}, aims to cover comprehensive and most recent advances, within the framework sketched in figure~\ref{hybrid}, on the generation of magnonic quantum states and their applications in related fields.

	\section{Hybrid quantum systems based on magnons}\label{HQSs}

	\subsection{Quantization of magnetization---magnons}
	
	In a ferromagnet comprising an ensemble of $N$ spins with quantum number $S$, the low-energy collective excitations are quantized as magnons. Based on the Heisenberg model on a lattice, the nearest-neighbor exchange interaction and an external magnetic field are incorporated. The Hamiltonian can be written as~\cite{Stancil09}
	\begin{align}
		H = -2 J \sum_{\langle i,j \rangle} \mathbf{S}_i \cdot \mathbf{S}_j - g_e\mu_B B \sum_i S_i^z,
		\label{Ht}
	\end{align}
	where $\mathbf{S}_i$ and $\mathbf{S}_j$ are the spin operators at lattice sites $i$ and $j$, respectively, $S_i^z$ denotes the $z$-component of $\mathbf{S}_i$, $J > 0$ is the exchange coefficient, $\langle i,j \rangle$ denotes pairs of nearest neighbors, $g_e$ is the Land\'{e} $g$-factor, $\mu_B$ is the Bohr magneton, and $B$ is the external magnetic field applied along the $z$-direction.

	The spin operators can be expressed in terms of bosonic creation and annihilation operators via the Holstein-Primakoff transformation
	\begin{align}
		S_i^+ &= \sqrt{2S} \sqrt{1 - \frac{a_i^\dagger a_i}{2S}}\, a_i,
		\qquad
		S_i^- = \sqrt{2S}\, a_i^\dagger \sqrt{1 - \frac{a_i^\dagger a_i}{2S}},
		\label{HP}
	\end{align}
	with $[a_i, a_j^\dagger] = \delta_{ij}$.
	For low magnon excitation numbers, i.e., $\sum\limits_{i} \langle a_i^\dagger a_i \rangle \ll 2 N S$, the square roots can be expanded to $S_i^+ \approx \sqrt{2S}\,a_i$, and $S_i^- \approx \sqrt{2S}\,a_i^\dagger$. Substituting these into the Hamiltonian~\eqref{Ht} and retaining up to the second-order terms yields
	\begin{equation}
		H = E_0 + 2JS \sum_{\langle i,j\rangle} \left(a_i^\dagger - a_j^\dagger \right) \left(a_i - a_j \right)
		+ g_e \mu_B B \sum_i a_i^\dagger a_i,
		\label{H_quad}
	\end{equation}
	where the constant ground-state energy
	$E_0 = -2J N Z S^2 - g_e \mu_B B N S$, with $Z$ being the coordination number.
	
	To diagonalize the Hamiltonian, lattice translation invariance is exploited by introducing Fourier transforms $
	a_i = \frac{1}{\sqrt{N}} \sum_{\mathbf{k}} e^{i\mathbf{k}\cdot\mathbf{r}_i} a_{\mathbf{k}}$, and $a_i^\dagger = \frac{1}{\sqrt{N}} \sum_{\mathbf{k}} e^{-i\mathbf{k}\cdot\mathbf{r}_i} a_{\mathbf{k}}^\dagger$, with \([a_{\mathbf{k}}, a_{\mathbf{k}'}^\dagger] = \delta_{\mathbf{k},\mathbf{k}'}\). Substituting them into equation~\eqref{H_quad} yields the final diagonal form
	\begin{equation}
		H = E_0 + \sum_{\mathbf{k}} \hbar\omega_{\mathbf{k}} \, a_{\mathbf{k}}^\dagger a_{\mathbf{k}},
	\end{equation}
	where the magnon dispersion relation is given by $\hbar\omega_{\mathbf{k}} = 4 J S Z \bigl(1 - \gamma_{\mathbf{k}}\bigr) + g_e\mu_B B$. The structure factor $\gamma_{\mathbf{k}} = \frac{1}{Z} \sum\limits_{\boldsymbol{\delta}} e^{i \mathbf{k}\cdot\boldsymbol{\delta}}$, with $\boldsymbol{\delta}$ being a vector to one of the $Z$ nearest-neighbor sites. For a simple cubic lattice ($Z=6$) and small $|\mathbf{k}|$, $\hbar\omega_{\mathbf{k}} \simeq 4JS a^2 |\mathbf{k}|^2 + g_e\mu_B B$, which exhibits the characteristic $|\mathbf{k}|^2$ dispersion and the uniform precession mode (i.e., the Kittel mode~\cite{Kittel48}) $\omega_{\rm FMR} = g_e\mu_B B/\hbar$ at $|\mathbf{k}| = 0$.  In most cavity magnonics experiments~\cite{Naka19,Yuan22,Bauer22,Zuo24}, the Kittel mode of a YIG sphere is studied, of which the Hamiltonian is denoted as $H /\hbar =\omega_{m} m^\dagger m$ throughout this paper.

	\begin{figure}[t]
		\includegraphics[width=0.7\linewidth]{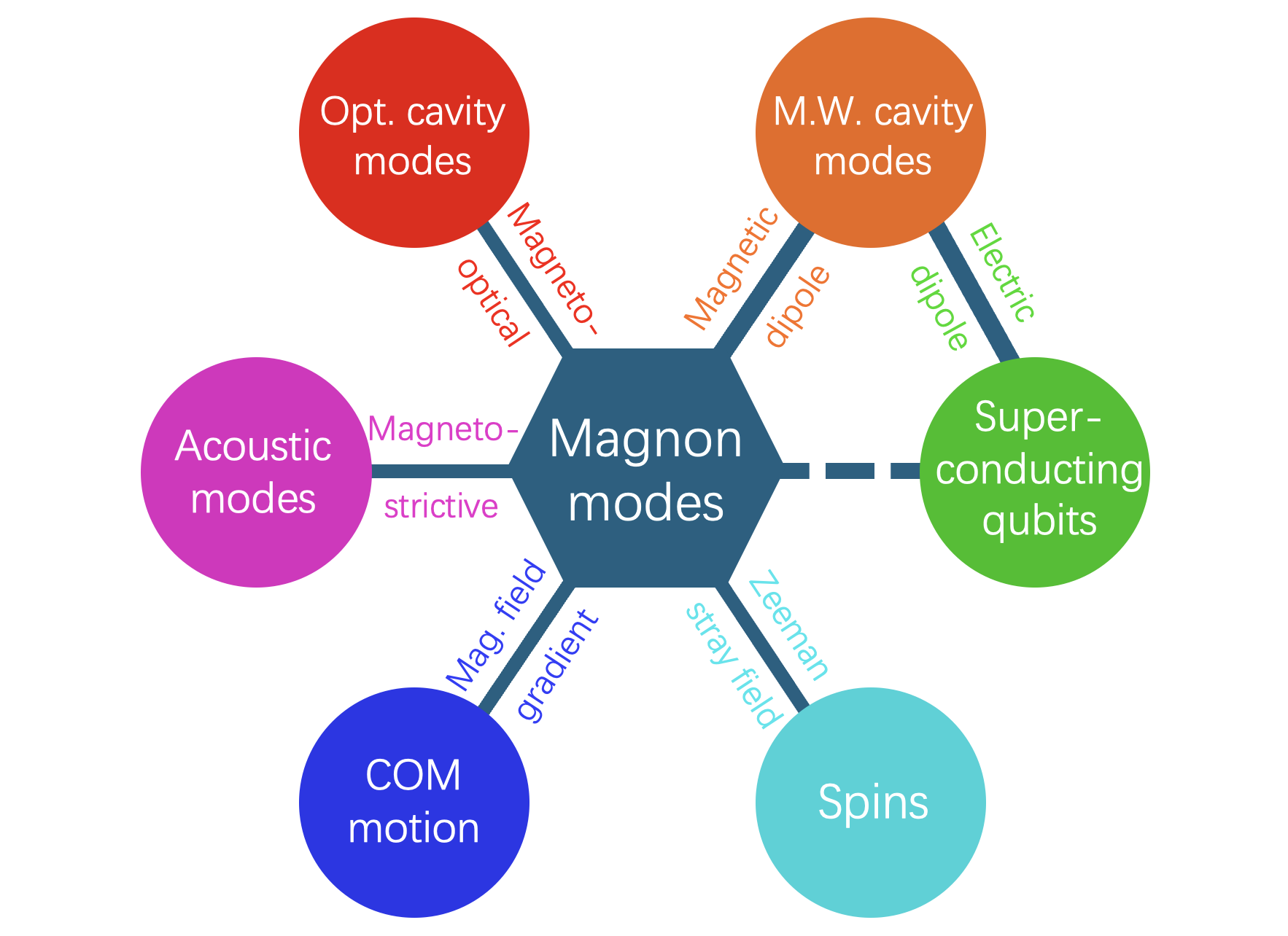}
		\centering
		\caption{Hybrid systems based on magnons, including the coupling of magnon modes to microwave (M.W.) cavity modes, optical (Opt.) cavity modes, acoustic modes, the COM motion of a YIG sphere, spins, superconducting qubits, etc. Thicker lines represent stronger coupling strength achieved in current experiments. Solid lines stand for direct coupling between two subsystems, while the dashed line denotes an effective (cavity-mediated) coupling. See section~\ref{hybridqs} for the coupling mechanisms. }
		\label{hybrid}
	\end{figure}

	\subsection{Hybrid quantum systems}\label{hybridqs}

	Magnons in ferromagnetic crystals, e.g., the YIG, show an extraordinary ability to coherently couple with different kinds of quantum excitations (figure~\ref{hybrid}). For example, they can couple to microwave cavity photons via the magnetic-dipole interaction, to optical photons through the magneto-optical effect, and to acoustic phonons by the magnetostrictive force. The microwave cavity can also couple to superconducting qubits through the electric-dipole interaction. Thus, an effective magnon--qubit coupling can be established via the mediation of the cavity.   In addition, magnons can also couple to solid-state spins, such as the NV center, through the Zeeman interaction between the stray field of the magnon mode and the spin, and to the COM motion of a levitated YIG sphere by means of an oscillating magnetic field gradient.  These couplings to diverse quantum systems form hybrid quantum systems based on magnons (figure~\ref{hybrid}), which integrate individual strengths of distinct physical systems, leading to enhanced performance and novel applications in quantum science and technology. 
	In particular, the microwave cavity--magnon system offers a new platform for studying strong interaction between light and matter, whereas the optomagnonic coupling is essential for building long-distance quantum networks with magnonic nodes connected by light. The magnomechanical coupling exploits the long lifetime of the acoustic modes for realizing quantum storage of short-lived magnonic quantum states. The effective magnon--qubit coupling introduces necessary nonlinearity from the superconducting qubit to the magnonic system, which plays a key role in preparing magnonic quantum states.  Furthermore, the coupling to both microwave and optical photons enables the microwave-to-optical transduction via magnons by exploiting the great tunability of the magnonic system, e.g., a tunable frequency dependent on an external magnetic field.
	
	\begin{figure}[t]
		\includegraphics[width=0.95\linewidth]{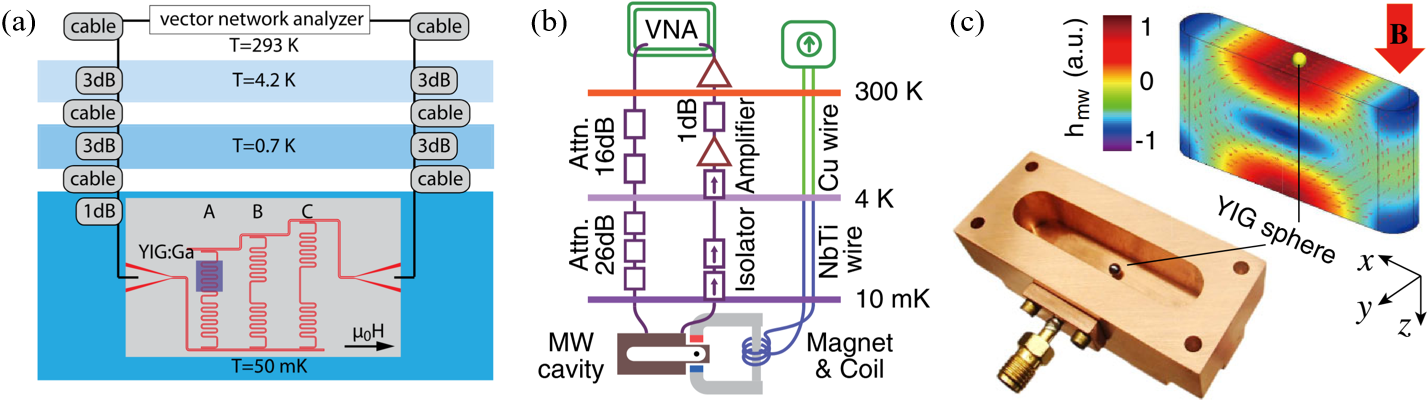}
		\centering
		\caption{(a) The (purple) gallium-doped YIG sample is cemented on top of one of the niobium microwave resonators. The experiment was performed at millikelvin temperatures. Reprinted figure (a) with permission from~\cite{Huebl13} Copyright 2013 by the American Physical Society. (b) A YIG sphere is mounted in a rectangular cavity made of oxygen-free copper. The cavity is placed inside a dilution refrigerator at 10 mK. Reprinted figure (b) with permission from~\cite{Tabuchi14} Copyright 2014 by the American Physical Society. (c) Top: Simulated microwave cavity TE$_{101}$ mode distribution. Red arrows and colors denote the magnetic field directions and amplitudes, respectively. Bottom: Half of the microwave cavity with a YIG sphere inside. Reprinted figure (c) with permission from~\cite{Zhang14} Copyright 2014 by the American Physical Society.}
		\label{SCExp}
	\end{figure}

	\section{Cavity electromagnonics}\label{electromag}

	\subsection{Cavity--magnon strong coupling}

	The coupling between the magnon mode and the microwave cavity originates from the Zeeman interaction between the spin ensemble and the cavity magnetic field. In a ferromagnet, $N$ spins precess uniformly, forming a collective magnon mode with an enhanced coupling strength $g_{\rm ma}=g_s\sqrt{N}$, where $g_s$ is the coupling strength of an individual spin to the cavity mode. The Hamiltonian of the two-mode cavity--magnon system is
	\begin{equation}
		\begin{aligned}
			H_{\rm ma}/\hbar =&\; \omega_{a} a^\dagger a + \omega_{m} m^\dagger m + g_{\rm ma}\left( a^\dagger m + a m^\dagger \right),
			\label{Hma}
		\end{aligned}
	\end{equation}
	where $a$ ($m$) is the annihilation operator of the cavity (magnon) mode, with the corresponding frequency $\omega_{a(m)}$, and $g_{\rm ma}$ is the cavity--magnon coupling rate. The condition for the strong coupling between the cavity and the magnon is that the coupling rate exceeds the dissipation rates of both subsystems. In this regime, coherent exchange between photons and magnons is enabled, which is a precondition required in many quantum protocols. The ferrimagnetic insulator YIG provides an ideal platform for reaching this regime, owing to its high spin density and low damping rate. Three landmark experiments~\cite{Huebl13,Tabuchi14,Zhang14} demonstrated strong coupling between magnons in the YIG and microwave cavity photons, thereby establishing the foundation of cavity magnonics~\cite{Bauer22}.
	
	The cavity--magnon strong coupling was first achieved in Ref.~\cite{Huebl13} using a planar microwave resonator, in which a gallium-doped YIG sample couples to a niobium superconducting microwave resonator at millikelvin temperatures (figure~\ref{SCExp}(a)). A coupling strength $g/2\pi = 450$ MHz was achieved, significantly exceeding both the photon decay rate $\kappa_a/2\pi=3$ MHz and the magnon decay rate  $\kappa_m/2\pi=50$ MHz. This work established that ferromagnetic spin ensembles are compatible with superconducting circuits---a crucial result for future on-chip cavity magnonic architectures. The single-excitation quantum regime was achieved in the following year~\cite{Tabuchi14}. An undoped single-crystal YIG sphere was cooled to 10 mK inside a three-dimensional (3D) copper cavity (figure~\ref{SCExp}(b)), and a clear normal-mode splitting was observed, corresponding to a cavity--magnon coupling strength $g/2\pi = 47$ MHz, much greater than the linewidth of the cavity (Kittel) mode of 2.7 MHz (1.1 MHz), yielding a cooperativity of $3\times10^3$~\cite{Tabuchi14}. Meanwhile, coherent coupling between microwave photons and magnons was realized at room temperature (figure~\ref{SCExp}(c)), and the ultrastrong coupling regime was accessed with a high cooperativity of 12600~\cite{Zhang14}.  

	\subsection{Magnon quantum states}

	The strong coupling offers the possibility to prepare magnonic quantum states in the system.  In this section, we review key protocols for preparing magnonic quantum states in cavity electromagnonics, including entangled, squeezed, and cat states. The cavity--magnon beam-splitter (BS) linear coupling (equation~\eqref{Hma}) can realize a coherent state swap, which however cannot, by itself, create quantum states. Nevertheless, nonlinearities from different mechanisms, e.g., magnetoelasticity and magnetocrystalline anisotropy, and external quantum driving fields can be introduced into the cavity--magnon system, which enable the generation of magnonic quantum states. Besides, quantum states can also be created or enhanced by exploiting the techniques of reservoir engineering and frequency modulation, as well as non-Hermitian and parity-time ($\cal{PT}$) symmetric physics. The magnonic quantum states in large-sized YIG spheres are macroscopic quantum states and contribute to fundamental studies of the quantum-to-classical transition and the test of unconventional decoherence theories at large scales~\cite{Leggett02,Bassi13,Weaver18}. Magnon squeezed states also enable quantum-enhanced detection of dark-matter axions~\cite{Flower19sq,Crescini20sq}.

	\begin{figure}[t]
		\includegraphics[width=0.92\linewidth]{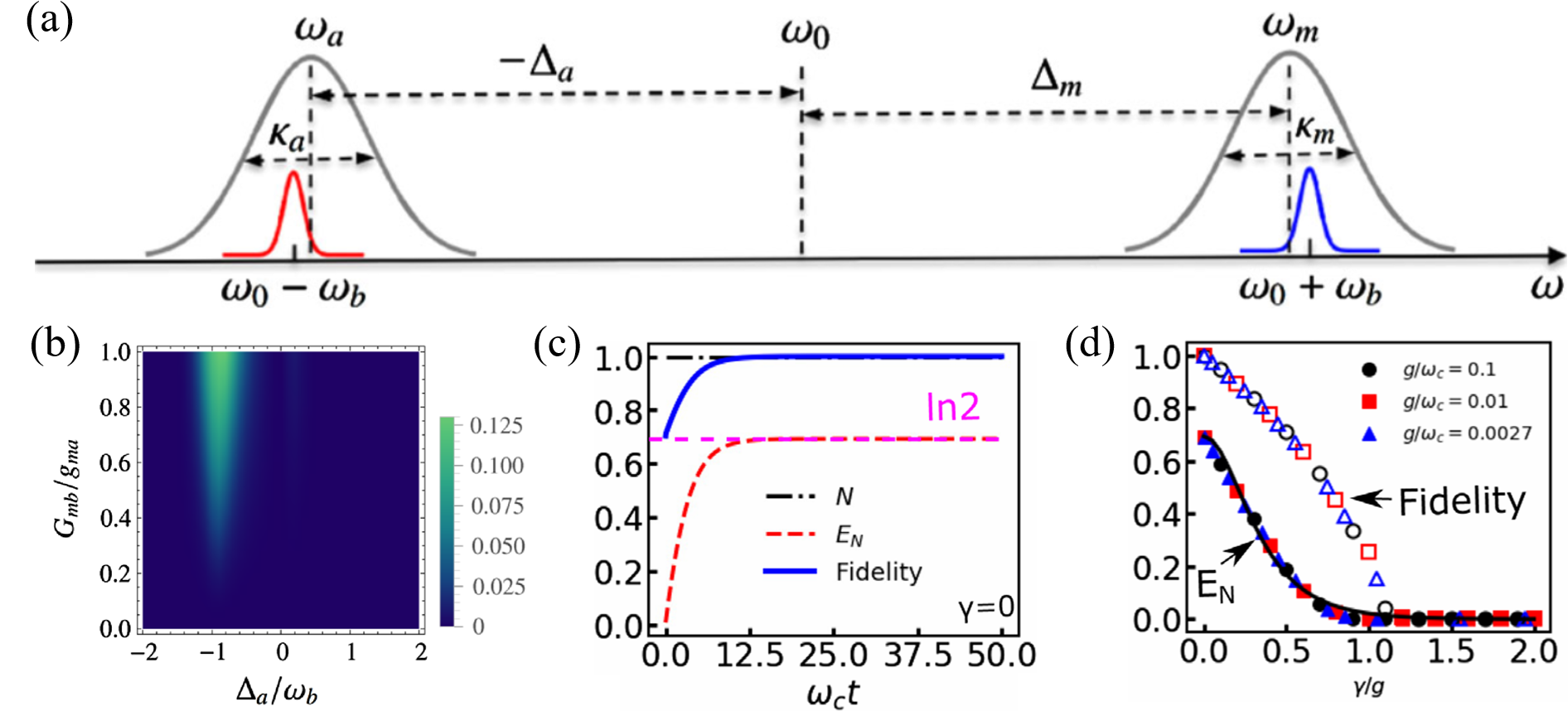}
		\centering
		\caption{{(a) Frequencies and linewidths of the system adopted to generate magnon--photon entanglement. A relatively strong microwave drive, red-detuned from the magnon resonance, is applied to simultaneously cool the mechanical motion and generate magnomechanical entanglement. The microwave cavity mode is tuned into resonance with the Stokes sideband to create magnon--photon entanglement.} (b) Cavity--magnon entanglement versus cavity-drive detuning $\Delta_a$ and effective magnomechanical coupling rate $G_{\rm mb}$ ($g_{\rm ma}$ is fixed). Reprinted figures (a)-(b) with permission from~\cite{Li18} Copyright 2018 by the American Physical Society. (c) Time evolution of the excitation number $N$ of the system, magnon--photon entanglement $E_N$, and the fidelity of the state with the Bell state in the $\cal{PT}$ symmetry-broken phase. (d) The entanglement and fidelity versus the magnon/photon damping rate. Reprinted figures (c)-(d) with permission from~\cite{Yuan20Bell} Copyright 2020 by the American Physical Society. }
		\label{En1}
	\end{figure}

	\subsubsection{Magnon--photon entanglement}\label{CMen}

	The first protocol of entangling cavity and magnon modes was given in Ref.~\cite{Li18} by exploiting a nonlinear magnetostrictive interaction.  Introducing the magnetostrictive coupling between magnons and phonons, which is a dispersive type for a large-sized YIG sphere~\cite{Zuo24}, the Hamiltonian of the system is given by
	\begin{equation}
		\begin{aligned}
			H/\hbar =\; H_{\rm ma}/\hbar + \frac{\omega_b}{2}\left(q^2 + p^2\right) + g_{\rm mb} m^\dagger m q + i\Omega\left(m^\dagger e^{-i\omega_0 t} - m e^{i\omega_0 t}\right),
		\end{aligned}
	\end{equation}
	where $q$ and $p$ are the dimensionless mechanical position and momentum, $\omega_{b}$ is the frequency of the mechanical mode, $g_{\rm mb}$ denotes the bare magnomechanical coupling rate, and $\Omega$ is the Rabi frequency of the drive directly applied to the magnon mode, e.g., via a loop antenna~\cite{Shen25}. The magnon mode is strongly driven, yielding a large steady-state amplitude $|\langle m\rangle|\gg1$ that allows for the linearization of the magnomechanical interaction~\cite{Vitali07}. Owing to the Gaussian nature of the input noises, the steady state is a three-mode Gaussian state, which is fully characterized by a $6\times6$ covariance matrix. The bipartite entanglement is then quantified by the logarithmic negativity~\cite{Vidal02,Plenio05}.
	
	Cooling the low-frequency mechanical mode to its quantum ground state is a prerequisite for generating entanglement in the system. This is accomplished by applying a red-detuned microwave drive with the detuning equal to the mechanical frequency, i.e., $\tilde{\Delta}_m \simeq \omega_b \gg \kappa_m$ (figure~\ref{En1}(a)), to activate the magnomechanical anti-Stokes scattering. Under moderate driving strength, the effective magnomechanical coupling $G_{\rm mb}$ is within the weak-coupling regime $G_{\rm mb} \ll \omega_b$, such that the rotating-wave approximation (RWA) applies, leading to an effective BS interaction $\propto \delta{m} \delta{b}^\dagger + \delta{m}^\dagger \delta{b}$, which is optimal for cooling the mechanical motion.  As the drive strength increases, the weak-coupling condition for the RWA no longer holds, such that the counter-rotating-wave terms $\propto \delta{m}^\dagger \delta{b}^\dagger + \delta{m} \delta{b}$ start to play a role, which create magnon--phonon entanglement. This entanglement subsequently transfers to the cavity--magnon subsystem when the cavity mode is tuned into resonance with the mechanical Stokes sideband, as shown in figures~\ref{En1}(a)-(b).

	Differing from the coherent coupling schemes, a steady magnon--photon Bell state can be generated in the cavity--magnon system via dissipative coupling~\cite{Yuan20Bell}. The underlying mechanism hinges on the spontaneous breaking of $\cal{PT}$ symmetry in the effective non-Hermitian Hamiltonian. Once the system enters the $\cal{PT}$ symmetry-broken phase, the eigenmode exhibiting gain acts as a dynamical attractor toward which the system evolves, irrespective of its initial state.  Within the single-excitation subspace, this attractor state corresponds precisely to a maximally entangled Bell state, yielding a logarithmic negativity of $E_{N} = \ln 2$ (figure~\ref{En1}(c)). Crucially, this entanglement is stationary and exhibits remarkable resilience to environmental dissipation (figure~\ref{En1}(d)). A high-fidelity Bell state, up to $\sim 97.85\%$ with feasible parameters, can be prepared and maintained when the dissipation rate is much smaller than the coupling rate. This work indicates that dissipative magnon--photon coupling can be harnessed to generate stationary noise-resilient entanglement.

	\subsubsection{Magnon--magnon entanglement}

\begin{figure}[t]
		\includegraphics[width=0.95\linewidth]{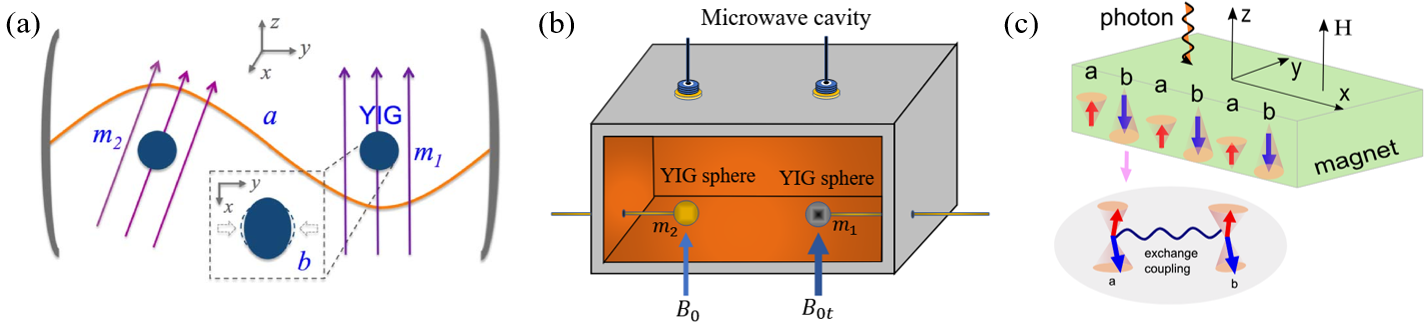}
		\centering
		\caption{(a) Two YIG spheres are placed inside a microwave cavity, each located near an antinode of the cavity field to achieve strong magnon--photon coupling. The bias magnetic fields are oriented such that the magnetostrictive interaction is activated only in one sphere, e.g., the right one. Reprinted figure (a) from~\cite{Li19MME} licensed under the \href{https://creativecommons.org/licenses/by/3.0/}{CC BY 3.0}. (b) Two YIG spheres in a microwave cavity are biased by a static magnetic field and a two-tone modulated field, respectively. Reprinted figure (b) with permission from~\cite{Hu24} Copyright 2024 by the Optical Society of America. (c) An antiferromagnet consists of two ferromagnetic sublattices with opposite magnetization directions. The two spins on the sublattices are coupled through the antiferromagnetic exchange interaction. Reprinted figure (c) with permission from~\cite{Yuan20} Copyright 2020 by the American Physical Society. }
		\label{MMEn}
	\end{figure}

	Apart from magnon--photon entanglement, the entanglement between two magnon modes residing in two macroscopic YIG spheres would be more appealing. This can be generated by exploiting the dispersive magnetostrictive interaction in a microwave cavity containing two YIG spheres~\cite{Li19MME} (figure~\ref{MMEn}(a)). The mechanism can be understood by using the results of Ref.~\cite{Li18} for a single YIG sphere supporting both a magnon mode ($m_1$) and a mechanical mode. As introduced in section~\ref{CMen}, the cavity--magnon ($m_1$) entanglement is generated under the resonant condition shown in figure~\ref{En1}(a). By further introducing the second YIG sphere supporting only a  magnon mode ($m_2$), which couples to the same cavity via a BS (state-swap) interaction, the two magnon modes therefore get entangled due to the mediation of the cavity. 
	
	Another approach for magnon--magnon entanglement is based on magnon Kerr nonlinearity~\cite{Zhang19}, which originates from magnetocrystalline anisotropy~\cite{Wang16}. The self-Kerr nonlinearity of one magnon mode can result in two-mode squeezing (TMS) between two magnon modes via the mediation of the cavity. In practice, however, achieving such a strong self-Kerr effect typically requires a drive power of tens of milliwatts~\cite{Shen22}, which inevitably induces heating (thermal noise). According to the experiment~\cite{Wu22}, a drive power on the order of 100 mW yields a temperature rise on the order of 10 K. The temperature rise caused by a power of tens of milliwatts would destroy almost all quantum states. This is relevant to all quantum-state protocols exploiting the magnon self-Kerr nonlinearity, where the accompanying strong heating effect must be taken seriously.

	Magnon--magnon entanglement can also be generated without requiring nonlinearity. Using an external squeezed microwave field and linear cavity--magnon state-swap interaction,  the squeezing of the driving field can be transferred to the two magnon modes, no matter the two magnets are placed in a single cavity~\cite{Nair20} or two separate cavities~\cite{Yu20}. Two-tone frequency modulation offers another energy-saving approach to achieving magnon--magnon entanglement~\cite{Xie23,Hu24}. Under the RWA in the rotating frame, frequency modulation induces specific sideband conditions that enable the desired coupling forms. Specifically, applying time-modulated external magnetic fields at two distinct frequencies to YIG spheres inside a microwave cavity (figure~\ref{MMEn}(b))---with the modulation frequencies tuned to the sum and difference of the magnon and cavity frequencies---simultaneously activates effective cavity--magnon TMS and BS interactions. The two magnon modes consequently get entangled via the cavity. The magnon--magnon entanglement was also studied in antiferromagnetic systems~\cite{Kamra19sq,Yuan20,Mousolou21}, where the exchange interaction directly induces parametric coupling (i.e., TMS) between the two magnon modes (figure~\ref{MMEn}(c)), thereby generating magnon--magnon entanglement.

	\begin{figure}[t]
		\includegraphics[width=0.95\linewidth]{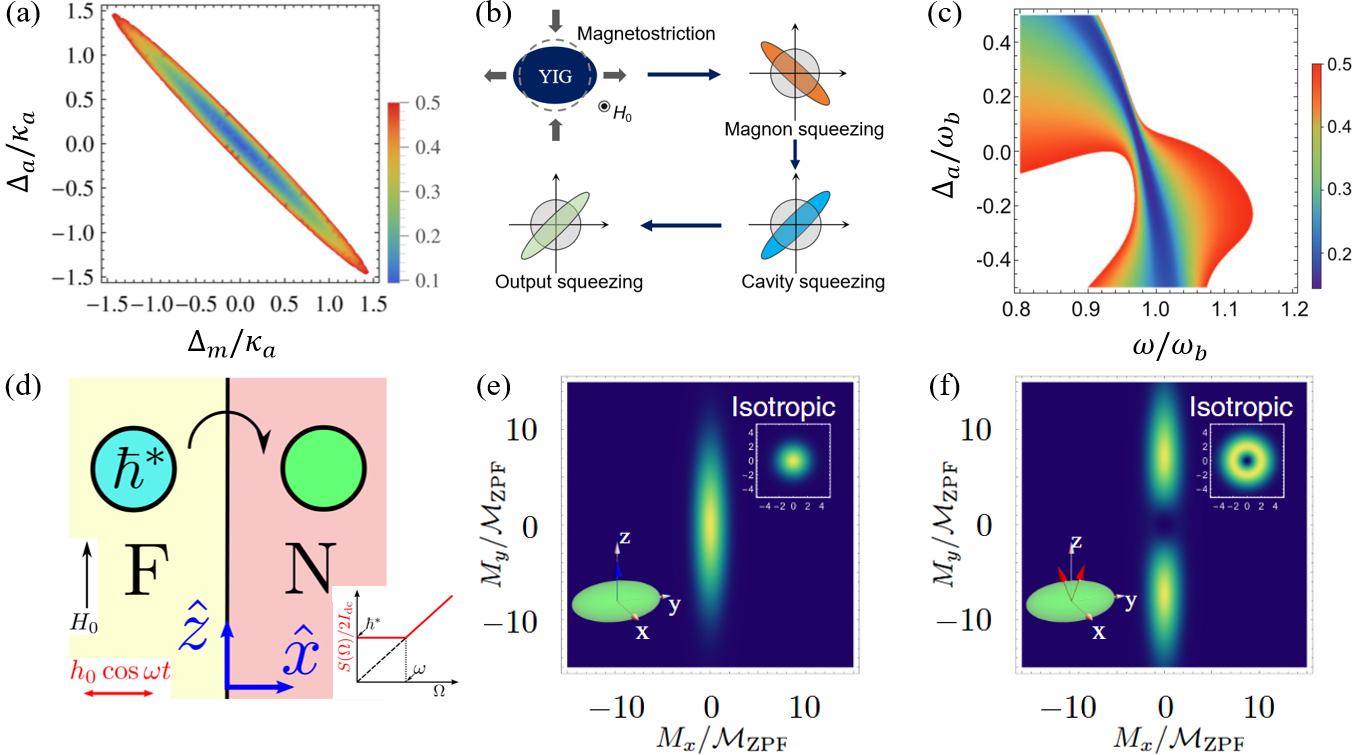}
		\centering
		\caption{(a) Variance of the magnon amplitude quadrature $\langle \delta x(t)^2 \rangle$ versus two detunings $\Delta_m$ and $\Delta_a$. Reprinted figure (a) with permission from~\cite{Li19b} Copyright 2019 by the American Physical Society. (b) Schematic diagram of the ponderomotive-like magnon squeezing, leading to a squeezed output field of the microwave cavity. (c) Noise spectral density of the output field  $S_W^{\rm out}(\omega)$ versus frequency $\omega$ and detuning $\Delta_a$. The blank areas in (a) and (c) denote the noise above the vacuum fluctuations. Reprinted figures (b)-(c) from~\cite{Li23sq} licensed under the \href{https://creativecommons.org/licenses/by/4.0/}{CC BY 4.0}. (d) Schematic of the ferromagnet and nonmagnetic conductor bilayer. A static magnetic field $H_0$ saturates magnetization while a coherent microwave field $h_0 \cos \omega t$ creates magnonic excitations. The latter annihilates at the interface creating excitations and injects a spin current into the nonmagnetic conductor. The inset shows the noise spectral density. Reprinted figure (d) with permission from~\cite{Kamra16sq} Copyright 2016 by the American Physical Society. (e)-(f) Husimi $Q$ function of the magnon state in a ferromagnet with shape anisotropy for (e) the ground state and (f) the state after subtracting one magnon. The inset corresponds to the isotropic case. Reprinted figures (e)-(f) from~\cite{Sharma21Cat} licensed under the \href{https://creativecommons.org/licenses/by/4.0/}{CC BY 4.0}. }
		\label{SqCat}
	\end{figure}

	\subsubsection{Magnon squeezed states}\label{sqzs}

	Squeezed states, with fluctuations in one quadrature reduced below the vacuum level, are typical representatives of nonclassical states. Magnon squeezed states can be generated by transferring squeezing from an external driving field~\cite{Li19b}. When a broadband squeezed vacuum field is injected into a microwave cavity, it modifies the noise characteristics of the cavity-field quantum fluctuations, resulting in a squeezed cavity mode. This squeezing is subsequently transferred to the magnon mode via the cavity--magnon BS interaction. Efficient transfer requires the resonance condition $\Delta_a=\Delta_m =0$ (in the frame rotating at the squeezed drive frequency $\omega_s$, figure~\ref{SqCat}(a)), strong coupling, and a much smaller magnon decay rate compared with the cavity decay rate~\cite{Li19b}. {Unlike transient squeezed states, steady-state magnon squeezing can be sustained under continuous driving~\cite{Kamra20sq}.}
	
	The dispersive magnomechanical interaction provides another approach to generating magnon squeezing~\cite{Li23sq}. Specifically, magnetostriction induces a deformation displacement of the YIG sample, proportional to the magnon excitation number, which in turn modulates the phase of the magnon mode. Therefore, through the displacement a unique correlation between the amplitude and the phase of the magnon mode is established. This amplitude-phase correlation results in magnon quadrature squeezing~\cite{Li23sq}. The underlying mechanism closely parallels ponderomotive squeezing of light induced by radiation pressure in optomechanics~\cite{Fabre94sq, Mancini94sq}.  The resulting squeezing is subsequently transferred to the cavity mode via the magnon--cavity BS coupling, and ultimately to the cavity output field (figure~\ref{SqCat}(b)), where it can be directly accessed and used as a quantum resource. Optimal squeezing is achieved under a strong red-detuned driving field with the detuning $0< \Delta_m \,{\equiv}\, \omega_m \,{-}\, \omega_d <\omega_b$, and the most pronounced squeezing appears around the mechanical frequency $\omega \simeq \omega_b$ (in the frame rotating at the drive frequency $\omega_d$, figure~\ref{SqCat}(c)). The squeezing grows with the magnomechanical cooperativity $C_{\rm m,b}=|G_{\rm mb}|^2/(\kappa_m \gamma_b)$, with $\gamma_b$ being the mechanical damping rate, and can potentially be strong.
	
	A distinct mechanism for magnon squeezing originates from the intrinsic magnetic anisotropy. The demagnetizing field, or equivalently the long-range dipolar interaction, induces shape-dependent anisotropic restoring torques on the transverse magnetization components~\cite{Kittel48,CostaFilho00sq}. In the quantum description, this corresponds to squeezed magnons with unequal fluctuations in their amplitude and phase quadratures~\cite{Kamra16sq,Kamra17sq}. Figure~\ref{SqCat}(d) illustrates a prototypical spin-pumping setup, in which a coherently driven ferromagnet ($F$) injects a spin current into an adjacent nonmagnetic conductor ($N$)~\cite{Kamra16sq}. Within the ferromagnetic thin film, the long-range dipolar interaction imposes unequal quantum fluctuations in the two magnon quadratures, thereby generating magnon squeezing. These squeezed magnons no longer carry the conventional spin angular momentum quantum $\hbar$; instead, they behave as new bosonic excitations with an effective spin $\hbar^* = \hbar(1+\delta)$, where $\delta$ quantifies the amplitude of the dipolar-interaction-mediated squeezing. Evaluation of the shot noise of this magnon-mediated spin current across the $F|N$ interface reveals a striking super-Poissonian spin transfer at low frequencies.
	
	Another approach to magnon squeezing is via parametric driving, as demonstrated in a recent experiment using YIG films~\cite{Hioki26sq}, where single- and two-mode magnon thermal squeezing were achieved. The squeezing is dynamically generated by applying an AC magnetic field at twice the magnon frequency, which parametrically deamplifies one quadrature of the magnetization fluctuations while amplifying the conjugate quadrature. Wigner function reconstruction reveals single-mode thermal squeezing. In the non-degenerate regime, two-mode thermal squeezing was observed, manifested by correlated fluctuations of magnons localized near the top and bottom surfaces of the film. {Parametric driving induced squeezing requires microwave pumping under appropriate parametric resonance conditions and is constrained by the parametric-oscillation threshold. In contrast, anisotropy-induced equilibrium squeezing does not require external driving fields but relies on sufficiently strong magnetic anisotropy.}

	\subsubsection{Magnon cat states}\label{catty}

	Schr\"odinger cat state~\cite{Schrodinger} was first proposed as a gedanken experiment discussing the boundary between the quantum and classical worlds, where a macroscopic object can be in a superposition of two macroscopically distinguishable states. Cat states find broad applications in, e.g., continuous-variable quantum information processing, bosonic quantum error correction, and quantum-enhanced sensing~\cite{Vlastakis13,Tatsuta19,Zheng26}. Correspondingly, the magnonic cat state represents a macroscopic quantum superposition of collective magnetic excitations involving a large number of spins. Magnon cat states can be generated by exploiting the shape anisotropy of the ferromagnet in a system consisting of a microwave cavity coupled to a ferromagnetic insulator~\cite{Sharma21Cat}. The anisotropy renders the ground state of the magnon mode a squeezed vacuum state, owing to the asymmetric zero-point fluctuations of the magnetization (figure~\ref{SqCat}(e)). Through parity measurements, realized by coupling the cavity to a superconducting qubit, or single-photon detection---specifically, using the cavity--magnon BS coupling, where detecting a single microwave photon corresponds to subtracting one magnon from the squeezed magnon state---one obtains a magnon cat-like state. Similar methods were used to prepare mechanical cat-like states in optomechanical systems~\cite{Shomroni20,Zhan20,Li26}. Such a cat-like state exhibits two separated wave packets in phase space, representing a quantum superposition of two distinct magnetization directions (figure~\ref{SqCat}(f)). The interference fringes around the origin become more evident when more magnons are subtracted from the squeezed vacuum state~\cite{Shomroni20,Zhan20,Li26}.

	
	\section{Magnon--superconducting--qubit system}
	\label{mag-qubit}

	In the preceding section, we have introduced coherent coupling between the magnon mode and the microwave cavity mode.~{This interaction, describing a linear coupling between two harmonic oscillators (equation~\eqref{Hma}), enables coherent exchange of excitations, but cannot, by itself, generate nonclassical states from initial classical states. Accessing the quantum regime in magnonics therefore necessitates the introduction of a controllable source of nonlinearity or anharmonicity, such as the magnetostrictive interaction discussed in section~\ref{CMen}.~Superconducting qubits offer another powerful and mature platform} by leveraging the nonlinearity of Josephson junctions, and the superconducting qubit can couple to microwave cavity photons via the electric-dipole interaction, thus forming a hybrid cavity--magnon--qubit (CMQ) system. The cavity-mediated effective magnon--qubit coupling brings necessary nonlinearity from the qubit to the magnonic system for creating magnonic quantum states.~This paradigm was pioneered in the experiment \cite{tabuchi2015}, where coherent coupling between a transmon qubit and a magnon mode in a YIG sphere was demonstrated, establishing the experimental foundation for quantum magnonics.~In this section, we first introduce some representative experiments demonstrated in the CMQ system, and then review related theoretical proposals for preparing various magnonic quantum states.
	
	\begin{figure}[t]
		\centering
		\includegraphics[width=0.99\linewidth]{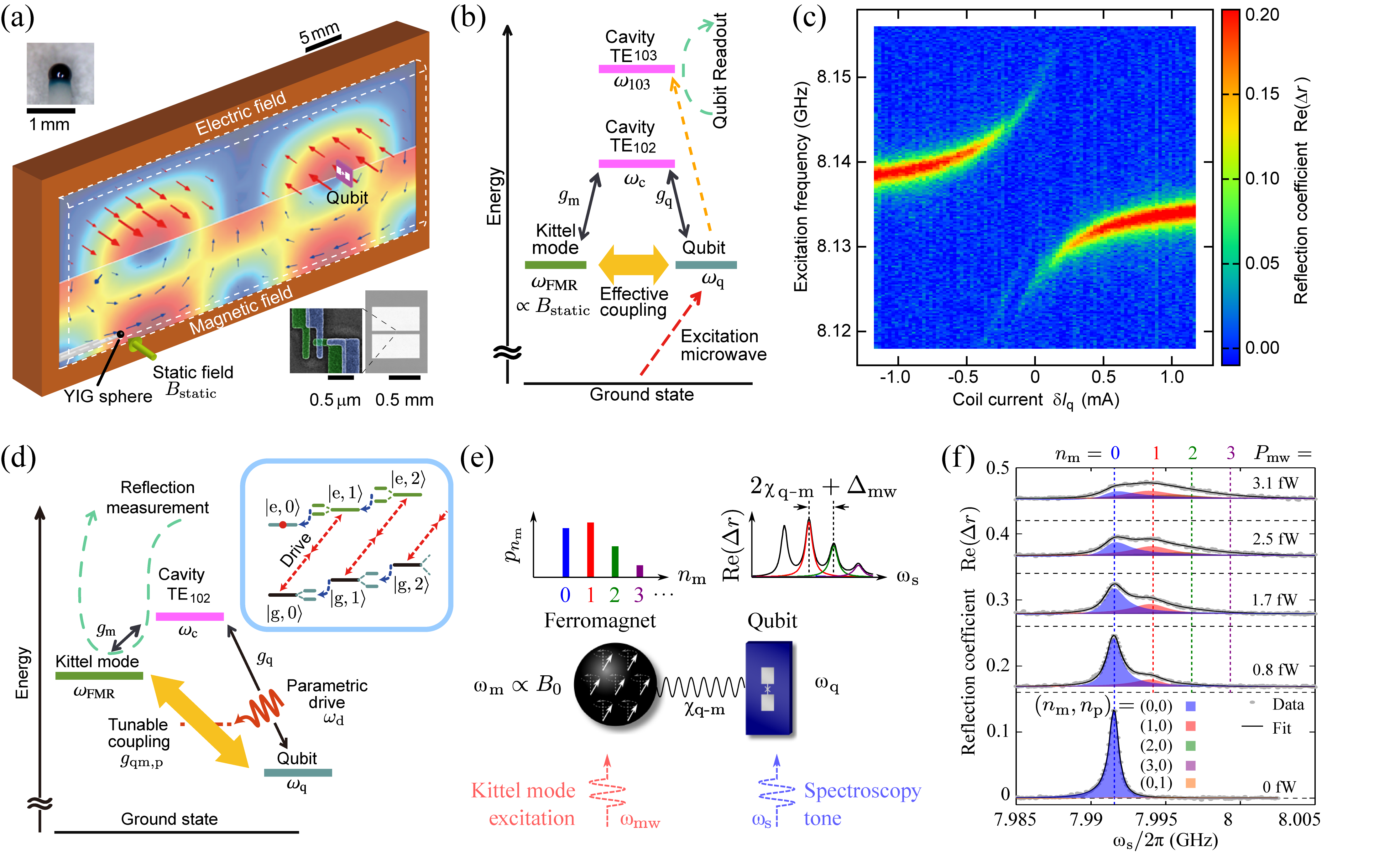}
		\caption{
			(a) The hybrid CMQ system.
			(b) Energy-level diagram of the cavity-mediated magnon--qubit interaction.
			(c) Avoided level crossing showing a vacuum Rabi splitting of $2g_{\rm qm}/2\pi = 20.0~\text{MHz}$ as the frequency of the Kittel mode is tuned through the qubit resonance.
			(d) Energy-level structure in the parametric drive scheme for dynamic control of the effective magnon--qubit coupling. Reprinted figures (a)-(d) with permission from~\cite{tabuchi2015} Copyright 2015 by the American Association for
			the Advancement of Science (AAAS).
			(e) Magnon-number-resolved qubit spectroscopy in the strong dispersive regime.	
			(f) Qubit absorption spectra measured with an increasing drive power $P_\mathrm{mw}$. Reprinted figures (e)-(f) with permission from~\cite{lachance-quirion2017} Copyright 2017 by the American Association for the Advancement of Science (AAAS).
		}
		\label{fig:qmfig1}
	\end{figure}

	\subsection{Theoretical background}

	The hybrid CMQ system typically comprises a 3D microwave cavity, a transmon qubit, and a single-crystal YIG sphere, as illustrated in figure~\ref{fig:qmfig1}(a). The qubit is placed near an antinode of the cavity electric field to maximize the electric-dipole coupling, whereas the YIG sphere is located at an antinode of the cavity magnetic field to maximize the magnetic-dipole interaction. The qubit can be described as an anharmonic oscillator 
	\begin{equation}
		H_{q}/\hbar =  \omega_{q} q^\dagger q + \frac{\alpha}{2}  q^\dagger q^\dagger q q,
	\end{equation}
	where $\alpha$ is the anharmonicity and $q^\dagger$ ($q$) are bosonic ladder operators. For a transmon qubit, sufficiently large anharmonicity can suppress leakage to higher levels, justifying truncation to the ground and first-excited states $\{|g\rangle, |e\rangle\}$, which leads the qubit Hamiltonian to $H_{q}/\hbar = \frac{1}{2}\omega_{q}\sigma_z$, with $\omega_{q}$ being the transition frequency between the states $|g\rangle$ and  $|e\rangle$.

	Within the cavity QED framework, the Hamiltonian of the tripartite CMQ system, under the RWA, is given by 
	\begin{equation}
		H/\hbar = \omega_{a} a^{\dagger} a + \omega_{m} m^{\dagger} m + \frac{\omega_{q}}{2}\sigma_{z}
		+ g_\mathrm{am}\!\left(a^{\dagger} m + a m^{\dagger}\right)
		+ g_\mathrm{aq}\!\left(a^{\dagger} \sigma_{-} + a \sigma_{+}\right),
		\label{eq:hybrid_hamiltonian}
	\end{equation}
	where $\sigma_{\pm}$ denote the raising and lowering operators, and $g_\mathrm{aq}$ represents the cavity-qubit coupling strength. The direct magnon--qubit coupling is negligibly small, but a strong effective magnon--qubit coupling can be established via the mediation of virtual cavity photons.
	When both the magnon and qubit systems are far detuned from the cavity, i.e., $g_\mathrm{aq}, g_\mathrm{am} \ll |\Delta_\mathrm{aq}|, |\Delta_\mathrm{am}|$, with $\Delta_\mathrm{aq} \equiv \omega_a - \omega_q$ and $\Delta_\mathrm{am} \equiv \omega_a - \omega_m$, the photon population remains negligible and the cavity can be adiabatically eliminated via the Fr\"ohlich-Nakajima transformation~\cite{blais2004,blais2021}. This yields an effective Jaynes--Cummings (JC) Hamiltonian
	\begin{align}
		H_\mathrm{JC}/\hbar = \omega_{\rm M} m^\dagger m + \frac{\omega_{\rm Q}}{2}\sigma_z + g_{\rm qm}\left(m^\dagger\sigma_- + m \sigma_+ \right),
		\label{eq:jc_hamiltonian}
	\end{align}
	where $g_{\rm qm} = \frac{g_\mathrm{aq}g_\mathrm{am}}{2}( \Delta_\mathrm{aq}^{-1} + \Delta_\mathrm{am}^{-1} )$ is the effective qubit--magnon coupling strength, and $\omega_{\rm M} = \omega_m - g_\mathrm{am}^2/\Delta_\mathrm{am}$ and $\omega_{\rm Q} = \omega_q - g_\mathrm{aq}^2/\Delta_\mathrm{aq}$ are the magnon and qubit effective frequencies, respectively.
	This interaction enables coherent exchange of the magnon and qubit excitations at a rate of $g_{\rm qm}$.

	Equation~\eqref{eq:jc_hamiltonian} governs two physically distinct operating regimes. In the (nearly) resonant regime, where $|\omega_{\rm Q} - \omega_{\rm M}| \lesssim g_{\rm qm}$, with strong coupling $g_{\rm qm} > \kappa_{m}, \gamma_{q}$ ($\gamma_{q}$ is the dissipation rate of the qubit), the eigenstates of the system are manifested as the magnon--qubit dressed states, with the accompanied anticrossing in the spectrum manifested as the magnon-vacuum-induced Rabi splitting of $2g_{\rm qm}$. In the dispersive regime, where $|\omega_{\rm Q} - \omega_{\rm M} | \gg g_{\rm qm}$, resonant exchange is suppressed and a perturbative treatment of the exchange interaction yields a dispersive-type interaction
	\begin{equation}
		H_\mathrm{disp}/\hbar = \omega_{\rm M} m^\dagger m + \frac{1}{2}\!\left[\omega_{\rm Q} + 2\chi_\mathrm{qm}\!\left(m^\dagger m + \frac{1}{2}\right)\right]\sigma_z,
		\label{eq:dispersive}
	\end{equation}
	with the dispersive shift $\chi_\mathrm{qm} = g_{\rm qm}^2/\Delta_\mathrm{qm}$, and $\Delta_\mathrm{qm} = \omega_{\rm Q} - \omega_{\rm M}$ is the effective qubit--magnon detuning. The coupling $\chi_\mathrm{qm}\, m^\dagger m\,\sigma_z$ enables a mutual dispersive interaction, where the qubit transition frequency is shifted by $2\chi_\mathrm{qm}$ per magnon, while the magnon frequency acquires a qubit-state-dependent shift of $\pm\chi_\mathrm{qm}$. This dispersive shift underpins the magnon-number-resolved spectroscopy.
	When the dispersive shift satisfies $2|\chi_{\rm qm}| > \gamma_{q}, \kappa_{m}$, the system enters the strong dispersive regime, in which individual magnon Fock states become spectrally resolvable, enabling magnon-number-resolved qubit spectroscopy~\cite{lachance-quirion2017} and single-shot detection of a single magnon~\cite{lachance-quirion2020}.

	\subsection{Magnon--qubit strong coupling}

	The coherent coupling between a magnon mode (the Kittel mode) and a superconducting transmon qubit was first demonstrated in a 3D copper cavity~\cite{tabuchi2015} (figure~\ref{fig:qmfig1}(a)). 
	A YIG sphere,  supporting the Kittel mode involving $\sim$$1.4\times10^{18}$ spins, and a transmon qubit are placed inside the cavity and both are coupled to the TE$_{102}$ mode of the cavity, while the TE$_{103}$ mode serves as an independent dispersive readout channel (figure~\ref{fig:qmfig1}(b)). 
	Both the magnon and the qubit are far detuned from the cavity, leading to an effective JC interaction of equation~\eqref{eq:jc_hamiltonian}.
	A pronounced avoided crossing was observed by sweeping the frequency of the Kittel mode through the qubit resonance (figure~\ref{fig:qmfig1}(c)), with a vacuum Rabi splitting of $2g_{\rm qm}/2\pi = 20.0$~MHz, substantially exceeding both the qubit and magnon dissipation rates $\gamma_{q}/2\pi = 1.2$~MHz and $\kappa_{m}/2\pi = 1.3$~MHz, placing the system well within the strong-coupling regime. 	
	At the degeneracy point, the bare states $|e,0\rangle$ and $|g,1\rangle$ hybridize into symmetric and antisymmetric superpositions of $|e,0\rangle$ and $|g,1\rangle$, confirming coherent hybridization at the single-quantum level. 
	Beyond this static interface, an off-resonant parametric microwave drive at $\omega_{d}=(\omega_{m}+\omega_{q})/2$ activates an effective two-photon transition $|g, n_m\rangle \leftrightarrow |e, n_m+1\rangle$ across the magnon ladder, with the corresponding Rabi splittings scaling as $2g_\mathrm{qm,p}\sqrt{n_m+1}$ (figure~\ref{fig:qmfig1}(d)), where $g_{\rm qm,p}$ is the effective coupling strength and scales linearly with drive power.
	
	\begin{figure}[t]
		\centering
		\includegraphics[width=\linewidth]{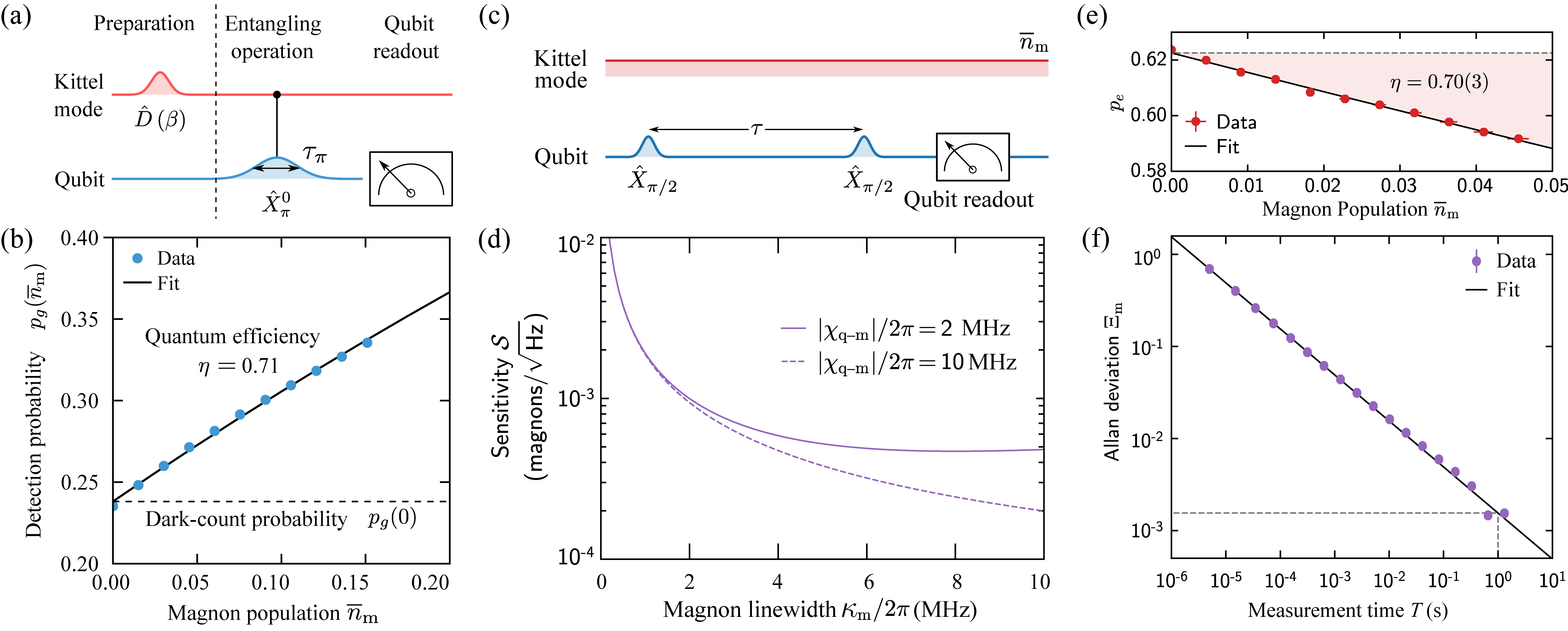}
		\caption{
			(a)~Protocol for entanglement-based detection of a single magnon. 	
			(b)~Detection probability $p_g(\bar{n}_m)$ as a function of the mean magnon number $\bar{n}_m$.	Reprinted figures (a)-(b)  with permission from~\cite{lachance-quirion2020} Copyright 2020 by the American Association for the Advancement of Science (AAAS).
			(c)~Pulse sequence used for the dissipation-based quantum sensing of magnons.
			(d) Magnon detection sensitivity $S$ calculated as a function of the magnon linewidth $\kappa_{m}/2\pi$, for amplitudes of the dispersive shift  $|\chi_\mathrm{qm}|/2\pi = 2$~MHz (solid line) and $10$~MHz (dashed line).
			(e) Probability of the qubit in the excited state $p_e$ as a function of the steady-state magnon population $\bar{n}_m$.		
			(f) Magnon population Allan deviation $\Xi_m(T)$ versus measurement time $T$. Reprinted figures (c)-(f) from~\cite{wolski2020} licensed under the \href{https://creativecommons.org/licenses/by/4.0/}{CC BY 4.0}.
		}
		\label{fig:qmfig2}
	\end{figure}

	\subsection{Magnon-number-resolved spectroscopy}

	With the strong magnon--qubit coupling achieved, a central challenge was then the characterization of the magnon quantum state. To this end, the hybrid system was operated in the strong dispersive regime ($2|\chi_\mathrm{qm}| > \gamma_{q}, \kappa_{m}$)~\cite{lachance-quirion2017}, where the dispersive interaction (equation~\eqref{eq:dispersive}) maps each Fock eigenstate $|n_m\rangle$ of the magnon mode onto a spectrally resolved qubit transition at $\omega_{\rm Q} + 2n_m\chi_{\rm qm}$, splitting the spectrum into a comb of peaks whose weights can be used to infer the magnon-number probability distribution $p_{n_m}$ (figure~\ref{fig:qmfig1}(e)). 
	Driving the Kittel mode into a coherent state, the qubit spectrum reveals a comb of peaks with Poissonian weights (figure~\ref{fig:qmfig1}(f)), from which the dispersive shift was extracted as $\chi_\mathrm{qm}/2\pi = 1.5 \pm 0.1$~MHz, satisfying $2|\chi_\mathrm{qm}| > \gamma_{q}, \kappa_{m}$. 
	When the drive power is turned off, the magnon mode is essentially in the vacuum state with the thermal occupation $\bar{n}_\mathrm{th} < 0.01$ at a temperature of 10 mK.
	At the lowest drive power of 79 aW, the resolved spectral weights correspond to a mean magnon number of $\bar{n}_m \simeq 0.026$, equivalent to a single spin flip among $5\times 10^{19}$ spins~\cite{lachance-quirion2017}.

	\subsection{Single-shot detection of a single magnon}

	The magnon-number-resolved spectroscopy requires averaging over many experimental repetitions and is therefore incapable of determining the magnon occupation within a single measurement window. 
	To overcome this limitation, Ref.~\cite{lachance-quirion2020} applied the photon-counting protocol of circuit QED~\cite{schuster2007} to the field of magnonics, and realized the first single-shot detection of a single magnon.
	The protocol encodes the magnon occupation into the qubit state via a three-step sequence (figure~\ref{fig:qmfig2}(a)). 		
	After initializing the qubit in $|g\rangle$ and displacing the Kittel mode into a coherent state $|\beta_m\rangle = \sum_{n_m} c_{n_m}|n_m\rangle$, a selective $\pi$-pulse $\hat{X}_\pi^0$, of which the frequency corresponds to the qubit frequency with the Kittel mode in the vacuum state, is applied to the qubit, yielding the following qubit--magnon entangled state 
	\begin{equation}
		\hat{X}_\pi^0\,|g,\beta_m\rangle = c_0\,|e,0\rangle + \sum_{n_m \geq 1} c_{n_m}|g,n_m\rangle.
		\label{eq:entangled_state}
	\end{equation}
	A subsequent qubit measurement projects the above entangled state into one of the two eigenstates: an outcome $|g\rangle$ heralds the presence of at least one magnon ($n_m \ge 1$), while $|e\rangle$ certifies the magnon vacuum. 
	Fitting the detection probability $p_g(\bar{n}_m) = \bar{\eta}(1-e^{-\bar{n}_m})+p_g(0)$ to the experimental data yields a quantum efficiency $\bar{\eta} = 0.71$ and a dark-count probability $p_g(0) = 0.24$, where the dark-count contribution $p_g(0)$ sets a floor on the detection probability (figure~\ref{fig:qmfig2}(b)). Based on these values, if the Kittel mode is in the Fock state $\ket{1}$, the detector clicks with a probability of 0.95 (ideally unity), demonstrating the single-shot detection of a single magnon. The above protocol essentially detects the presence of at least one magnon. The presence of exactly one magnon can be detected by using the conditional operation $\hat{X}_\pi^1$ that excites the qubit only when there is exactly a single magnon in the Kittel mode. For the magnon state, in which the probability of having more than one magnon is negligible, both protocols detect the presence of a single magnon.
	
	\begin{figure}[t]
		\centering
		\includegraphics[width=\linewidth]{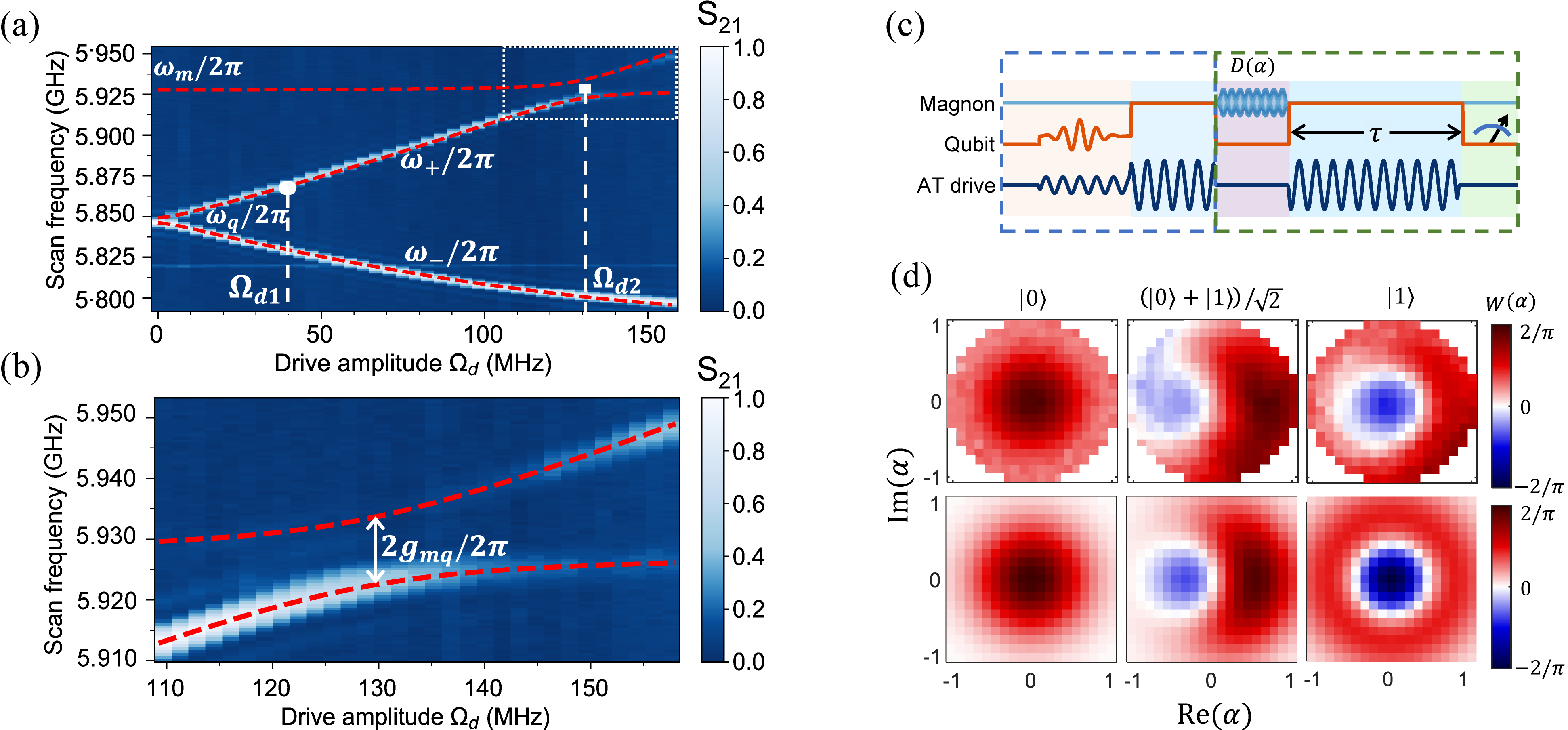}
		\caption{
			(a) Measured $S_{21}$ transmission spectrum as a function of the AT drive amplitude $\Omega_d$.
			(b) Avoided crossing between the magnon and the qubit.
			(c) Pulse sequence for magnon state preparation and Wigner tomography.
			(d) Wigner function tomography of the vacuum state $|0\rangle$, the superposition state $(|0\rangle+|1\rangle)/\sqrt{2}$, and the single-magnon state $|1\rangle$. Top row: measured data; Bottom row: analytic results of the corresponding ideal states. Reprinted figures (a)-(d) with permission from~\cite{xu2023} Copyright 2023 by the American Physical Society.
		}
		\label{fig:qmfig31}
	\end{figure}

	\subsection{Dissipation-based quantum sensing of magnons}

	Another magnon quantum sensing protocol exploiting magnon dissipation as the primary information channel was realized in Ref.~\cite{wolski2020}.
	Operating in the strong dispersive regime, stochastic fluctuations of the magnon number are transduced via the dispersive coupling $\chi_\mathrm{qm}\,m^\dagger m\,\sigma_z$  into a random qubit frequency shift, leading to a pure-dephasing rate proportional to the magnon population $\bar{n}_m$. This allows a standard Ramsey sequence to directly extract the steady-state magnon population $\bar{n}_m$ without requiring entanglement or single-shot detection (figure~\ref{fig:qmfig2}(c)). 
	In the weak-excitation limit ($\bar{n}_m \ll 1$), the probability of the qubit being in the excited state scales linearly as $p_e(\bar{n}_m) \approx p_e(0) \pm \eta\,\bar{n}_m$, yielding a detection efficiency of $\eta = 0.70(3)$ by fitting the data of figure~\ref{fig:qmfig2}(e). 
	Repeated single-shot readout of the qubit verified by an Allan-deviation analysis confirms shot-noise-limited performance, achieving an optimal magnon detection sensitivity of $S = 1.55(3)\times10^{-3}\,\mathrm{magnons}/\sqrt{\mathrm{Hz}}$ (figure~\ref{fig:qmfig2}(f)). 
	Importantly, the sensitivity analysis reveals that for values up to $\kappa_{m} \lesssim 4|\chi_\mathrm{qm}|$, the sensitivity scales as $S \sim 1/\kappa_{m}$, indicating that increasing the magnon dissipation improves the sensitivity and the sensing is governed primarily by the dissipation (figure~\ref{fig:qmfig2}(d)).

	\subsection{Superposition of a single magnon and vacuum}

	The previous magnon quantum sensing protocols exploit the dispersive coupling, which precludes fast resonant energy exchange and makes the generation of magnonic quantum states challenging within the coherence time of the system. Operating in the resonant-coupling regime, Ref.~\cite{xu2023} deterministically prepared the single-magnon state and the superposition of a single magnon and vacuum. The experiment innovatively utilized the second excited state $\ket{f}$ of the transmon and exploited the Autler--Townes (AT) effect~\cite{autler1955} to adjust the qubit frequency {\it in situ}, namely, applying a strong control field to drive the $\ket{e}\to\ket{f}$ transition of the transmon qutrit to hybridize these levels into a pair of dressed states $\ket{\pm}$, thereby rendering the effective $|g\rangle\leftrightarrow|+\rangle$ transition frequency continuously tunable with the drive amplitude (figure~\ref{fig:qmfig31}(a)). By tuning the qubit frequency to be resonant with the magnon mode, i.e., to the swap point, the strong magnon--qubit coupling $g_{\rm qm} /2\pi = 5.55$ MHz (figure~\ref{fig:qmfig31}(b)) enables a full magnon--qubit excitation swapping in only 45 ns, well within the magnon lifetime $T_{1,m} = 128$~ns.	
	When the system is initialized in $|0\rangle \otimes |g\rangle$, i.e., the magnon in the vacuum state and the qubit in the ground state, applying a $\pi$ pulse ($\pi/2$ pulse) to the transmon followed by a state swap deterministically prepares the single-magnon  state $|1\rangle$ (the superposition state $(\ket{0} + \ket{1})/\sqrt{2}$). The magnon quantum states can be characterized by performing Wigner tomography implemented via a magnon displacement operation $D(\alpha)$ followed by a qubit-assisted magnon parity measurement $P= e^{i \pi m^\dagger m}$, i.e., $W(\alpha) = (2/\pi)\mathrm{Tr}[D(-\alpha)\rho D(\alpha)P]$~(figure~\ref{fig:qmfig31}(c)).
	The Wigner tomography for three generated magnon states is shown in figure~\ref{fig:qmfig31}(d), where the single-magnon state $\ket{1}$ and the superposition state $(\ket{0} + \ket{1})/\sqrt{2}$ exhibit pronounced negative values of the Wigner function---an unambiguous signature of nonclassicality. The fidelities of the reconstructed density matrices for the single-magnon and superposition states are $0.815\pm 0.008$ and $0.942\pm 0.009$, respectively.

	\begin{figure}[t]
		\centering
		\includegraphics[width=\linewidth]{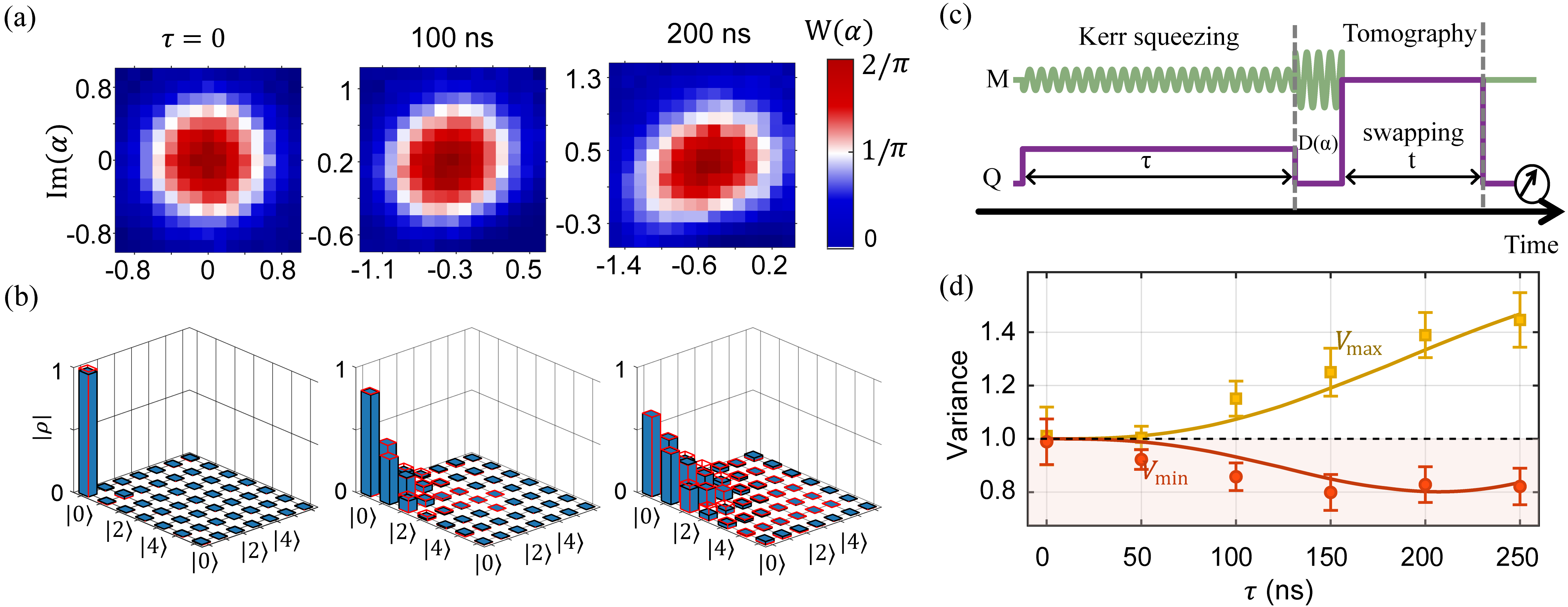}
		\caption{		
			(a) Experimentally measured Wigner functions for the vacuum state ($\tau = 0$) and squeezed states at $\tau = 100$ and 200~ns. 
			(b) Reconstructed density matrices (blue bars) for the vacuum state ($\tau = 0$) and squeezed states at $\tau = 100$ and 200~ns, compared with numerical simulations (red bars). 
			(c) Pulse sequence for Kerr squeezing and Wigner tomography. 
			(d) Minimum and maximum quadrature variances versus evolution time $\tau$. The shaded region indicates the squeezing below the vacuum level, with a maximum squeezing of $\sim$1.0~dB around $\tau \approx 150$~ns. Reprinted figures (a)-(d) with permission from~\cite{weng2026} Copyright 2026 by the Springer Nature.
		}
		\label{fig:qmfig32}
	\end{figure}

	\subsection{Magnon quantum squeezing (experiment)}\label{sqzex}

	Apart from the aforementioned single-magnon state and its superposition with the vacuum state, the magnon mode, as a continuous-variable system, can also exhibit other quantum characteristics, such as quantum squeezing, where the noise in one quadrature of the magnon mode is less than the vacuum fluctuations. The magnon squeezing at the quantum level has recently been achieved in the experiment~\cite{weng2026}. The squeezing mechanism resembles that of the Kerr-induced squeezed light in the field of quantum optics~\cite{kerr_squeezing,kerr_state,kerrwigner}. Specifically, the magnon--qubit dispersive coupling results in an effective self-Kerr nonlinearity of the magnon mode, characterized by the Hamiltonian 
	\begin{align}
		H_\mathrm{eff}/\hbar = \omega_m' m^\dagger m - \delta\,\left(m^\dagger m \right)^2
	\end{align}
	where $\delta/2\pi = 0.36$ MHz is the self-Kerr coefficient measured in the experiment and the effective magnon frequency $\omega'_m= \omega_m + \delta + \chi$, where $\omega_m$ is the original magnon frequency and $\chi$ is the qubit-induced dispersive shift. The generated magnon squeezed state is characterized by both Wigner tomography (figure~\ref{fig:qmfig32}(a)) and density-matrix reconstruction (figure~\ref{fig:qmfig32}(b)). The Wigner tomography essentially amounts to measuring the expectation value of the magnon displaced parity operator of the state. The magnon parity measurement typically requires the magnon--qubit state swapping, and the associated pulse sequence for Wigner tomography is shown in figure~\ref{fig:qmfig32}(c). One feature distinguishing Kerr-induced squeezing from parametric squeezing is manifested as the shearing and distortion of the Wigner function in phase space (figure~\ref{fig:qmfig32}(a)). The experiment observed a minimum variance of the magnon quadrature $V_{\rm min} = 0.799 \pm 0.068$ at $\tau =150$ ns (figure~\ref{fig:qmfig32}(d)), corresponding to about 1 dB squeezing below the vacuum level. Stronger squeezing could be achieved by enhancing the self-Kerr interaction and essentially the magnon--qubit coupling strength. Moreover, stronger squeezing corresponds to a larger mean magnon number.
	This work represents the first experimental observation of magnon quantum squeezing in a macroscopic ferrimagnet, which finds applications in, e.g., quantum-enhanced detection of dark-matter axions \cite{Crescini20sq}.

	\subsection{Magnon cat states}

	Beyond the quantum-related experiments introduced above, the strong nonlinearity in the CMQ system has motivated a growing number of proposals for engineering different kinds of magnonic quantum states, including magnon cat, squeezed, and entangled states, and blockade effects. Some of them are based on currently available parameters from the experiments and thus very promising to be realized in the near future.
	Besides the protocol introduced in section~\ref{catty} by exploiting the anisotropy of the ferromagnet and photon detection, the qubit--magnon hybrid system shows a greater capability of generating magnonic cat states. The related proposals fall into mainly two distinct categories.
	The first is the conditional-displacement (CD) framework~\cite{he2023cat,he2024scat,he20264cat,li2026arxiv}, which utilizes a qubit-state-dependent force to displace the Kittel mode along opposite directions in phase space, and a subsequent qubit measurement projects the magnon mode into a cat state. The second is the two-magnon driven-dissipative mechanism~\cite{hou2024cat,liu2025cat}, which exploits an engineered nonlinear coupling that confines energy exchange to magnon pairs, and allows the competition between coherent driving and qubit dissipation to autonomously stabilize a magnonic cat state in the long-time limit.

	\begin{figure}[t]
		\centering
		\includegraphics[width=\linewidth]{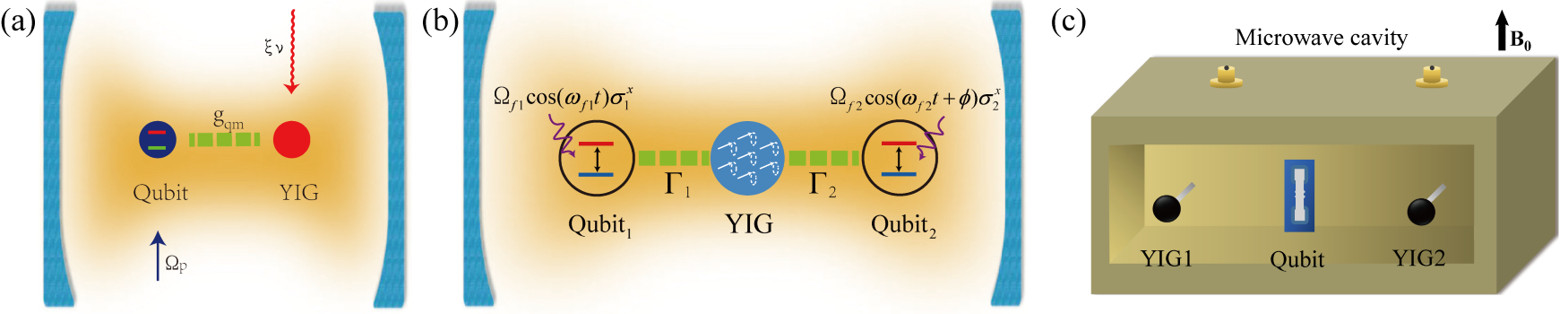}
		\caption{
			(a) A qubit and a magnon mode in a YIG sphere are coupled via a microwave cavity, where the magnon and the qubit are driven by a Floquet drive and a classical microwave field, respectively. Reprinted figure (a) with permission from~\cite{he2023cat} Copyright 2023 by the American Physical Society.
			(b) Two qubits and a magnon mode are coupled via a cavity, where Floquet drives $\Omega_{f_1}\cos(\omega_{f_1}t)\sigma_1^x$ and $\Omega_{f_2}\cos(\omega_{f_2}t+\phi)\sigma_2^x$ are applied to the two qubits, with a tunable relative phase $\phi$ to control the rotational symmetry of the multi-component cat state. Reprinted figure (b) with permission from~\cite{he20264cat} Copyright 2026 by the American Physical Society.
			(c) A qubit and two magnon modes in two YIG spheres are coupled via a cavity, where the qubit is resonantly driven by a microwave field. Reprinted figure (c) with permission from~\cite{li2026arxiv} by the authors.
		}
		\label{fig:qmfig4}
	\end{figure}

	Following the CD mechanism, a Floquet-engineering scheme~\cite{he2023cat} employs a low-frequency periodic drive to modulate the magnon frequency (figure~\ref{fig:qmfig4}(a)), while a strong microwave field is used to resonantly  drive the qubit. Adiabatic elimination of the cavity mode yields an effective CD interaction, and a subsequent projective measurement of the qubit projects the magnon onto an even or odd cat state whose coherent amplitude and fidelity are both tunable through the Floquet drive strength. This mechanism was subsequently extended to a two-qubit system to prepare a multi-component magnonic cat state~\cite{he20264cat} (figure~\ref{fig:qmfig4}(b)), where each superconducting qubit is driven by a Floquet field and virtual cavity-photon exchange induces an effective CD-type interaction between the magnon mode and the qubit by neglecting far off-resonant contributions under the RWA. Notably, the relative phase difference between the two Floquet drives fully controls the rotational symmetry of the state in phase space, enabling the deterministic generation of four-component magnonic cat states. 
	Extending the CD mechanism to multi-mode magnonic systems, a protocol for generating two-mode magnonic cat states has recently been offered~\cite{li2026arxiv} based on a system involving two YIG spheres and a transmon qubit coupled to a common microwave cavity (figure~\ref{fig:qmfig4}(c)). By adiabatically eliminating the cavity and applying a resonant qubit drive, an effective CD interaction is engineered between the qubit and each magnon mode. Working in the magnon--magnon strong-coupling regime and considering two identical magnon frequencies and coupling strengths to the cavity, the two magnon modes hybridize into a bright mode and a dark mode. A projective qubit measurement then collapses the bright mode into a cat state while the dark mode remains in its initial vacuum state. This corresponds to a two-mode cat state of two original magnon modes residing in two YIG spheres, which share strong non-Gaussian entanglement. This multi-mode CD architecture is in principle scalable to prepare three- or more-mode magnonic cat states.
	Another distinct approach is based on the two-magnon driven-dissipative mechanism. Along this line, Ref.~\cite{liu2025cat} applies two transverse drives to the qubit and an additional drive to the magnon mode, such that the interplay of the resulting longitudinal and transverse magnon--qubit couplings in the dressed-state basis generates an effective two-magnon nonlinear interaction. The spontaneous emission of the qubit then autonomously steers the magnon mode into a steady even or odd cat state depending on the parity of the initial magnon state. This scheme is inherently robust against qubit dephasing since the qubit relaxes to its ground state in the steady state. 
	Beyond the cavity-mediated magnon--qubit platform, proposals based on a magnon--qubit directly coupled configuration have also been offered to prepare magnonic cat states via either CD or driven-dissipative mechanisms~\cite{he2024scat,hou2024cat,kounalakis2022cat}.

	\subsection{Magnon blockade}
	\label{blockade}

	Magnon blockade, a phenomenon that the excitation of one magnon suppresses the subsequent excitation of a second one, is one of the main manifestations of the quantum properties of the magnonic system. The blockade effect was originally explored in photonic systems \cite{first_Exp_blockade}. As an analogue, magnon blockade not only provides a route for probing strong correlations and quantum nonlinearities of collective spin excitations, but also offers promising opportunities for designing single-magnon emitters~\cite{liu2026quantum}. Magnon blockade is typically achieved in hybrid systems in which magnons interact with a strongly nonlinear quantum system, e.g., a superconducting qubit~\cite{liu2019magnon,xie2020quantum}, enabling the engineering of effective interactions and quantum interference effects that suppress multi-magnon excitations.	
	
	In the following, we review a series of magnon blockade protocols based on the qubit--magnon hybrid system. In the most general situation, both the qubit and magnon are driven by classical microwave fields, with the driving Hamiltonians given by $H_{\text{dri}} = \Omega(\sigma_+ e^{-i\omega_d t}+\sigma_- e^{i\omega_d t}) + \xi(m^\dagger e^{-i\omega_p t}+m e^{i\omega_p t})$, where $\omega_d$ and $\omega_p$ are the corresponding drive frequencies and $\Omega$ and $\xi$ denote the related drive strengths. The full Hamiltonian of the system is therefore $H = H_{\rm JC} + H_{\text{dri}}$ (cf. equation \eqref{eq:jc_hamiltonian}). The dissipation and dephasing effects of the system can be studied by adopting the master-equation approach. Solving the master equation in the steady-state limit provides access to the statistical properties of the magnon excitations, which can be quantified via the equal-time second-order correlation function
	\begin{equation}
		g^{(2)}(0)=\frac{\langle m^\dagger m^\dagger m m \rangle}{\langle m^\dagger m \rangle^2}.
	\end{equation}
	The quantity $g^{(2)}(0)<1$ signifies the antibunching of the magnonic field, which is a nonclassical feature. In the strong blockade regime, the probability of the simultaneous presence of two magnons is negligibly small, corresponding to $g^{(2)}(0)\ll 1$. Therefore, evaluation of this correlation function from the steady-state magnonic state provides a quantitative criterion for identifying magnon blockade.

	\begin{figure}[t]
		\centering
		\includegraphics [width=0.8\linewidth]{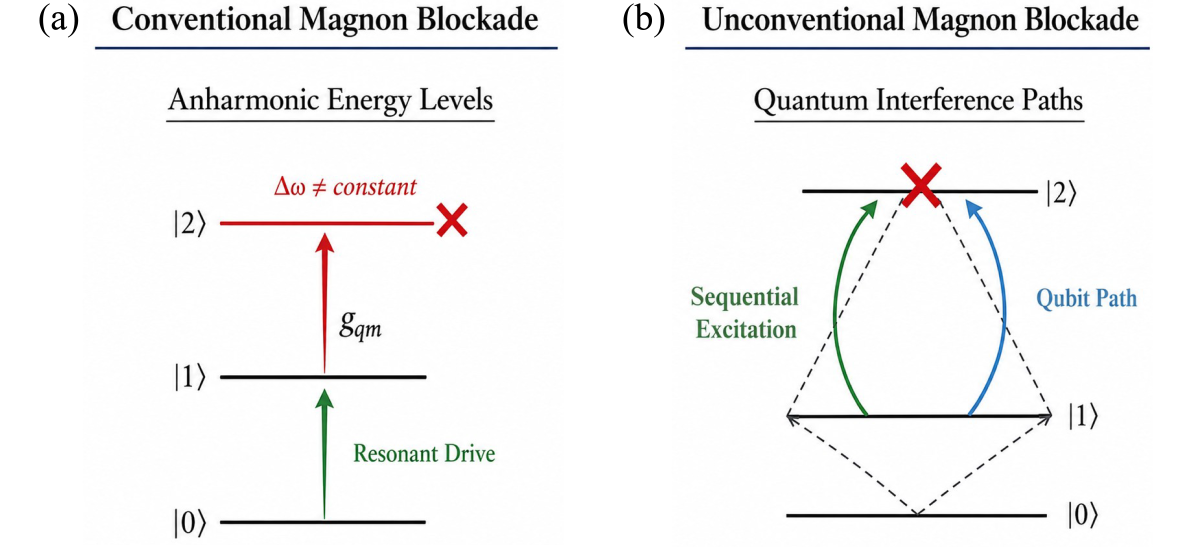}
		\caption{Two physical mechanisms for achieving magnon blockade: (a) the CMB induced by anharmonic energy levels, and (b) the UMB arising from destructive quantum interference between different excitation pathways.}
		\label{fig:qmfig5}
	\end{figure}

	Magnon blockade arises from two distinct physical mechanisms, namely, conventional and unconventional blockade (figure ~\ref{fig:qmfig5}). Conventional magnon blockade (CMB) is governed by the anharmonicity of the energy spectrum induced by the strong qubit--magnon coupling (figure~\ref{fig:qmfig5}(a)). The hybridization between the qubit and the magnon leads to dressed states with non-uniform energy level spacing, such that the transition from the vacuum state $\ket{0}$ to the single-magnon state $\ket{1}$ can be resonantly driven, while the subsequent transition to the two-magnon state $\ket{2}$ becomes off-resonant. Consequently, once a single magnon is excited, further excitations are energetically suppressed, producing a blockade effect that relies on strong coupling and large spectral nonlinearity.
	In contrast, unconventional magnon blockade (UMB) originates from destructive quantum interference between multiple excitation pathways, such as sequential excitation and qubit-mediated processes, to the two-magnon state (figure~\ref{fig:qmfig5}(b)), which effectively cancels the probability amplitude of the two-magnon state $\ket{2}$ without requiring strong anharmonicity. The former is an energy-selective suppression mechanism, while the latter is an interference-based suppression, offering a more flexible route to realizing magnon blockade in hybrid quantum systems.

	\begin{figure}[t]
		\centering
		\includegraphics [width=0.75\linewidth]{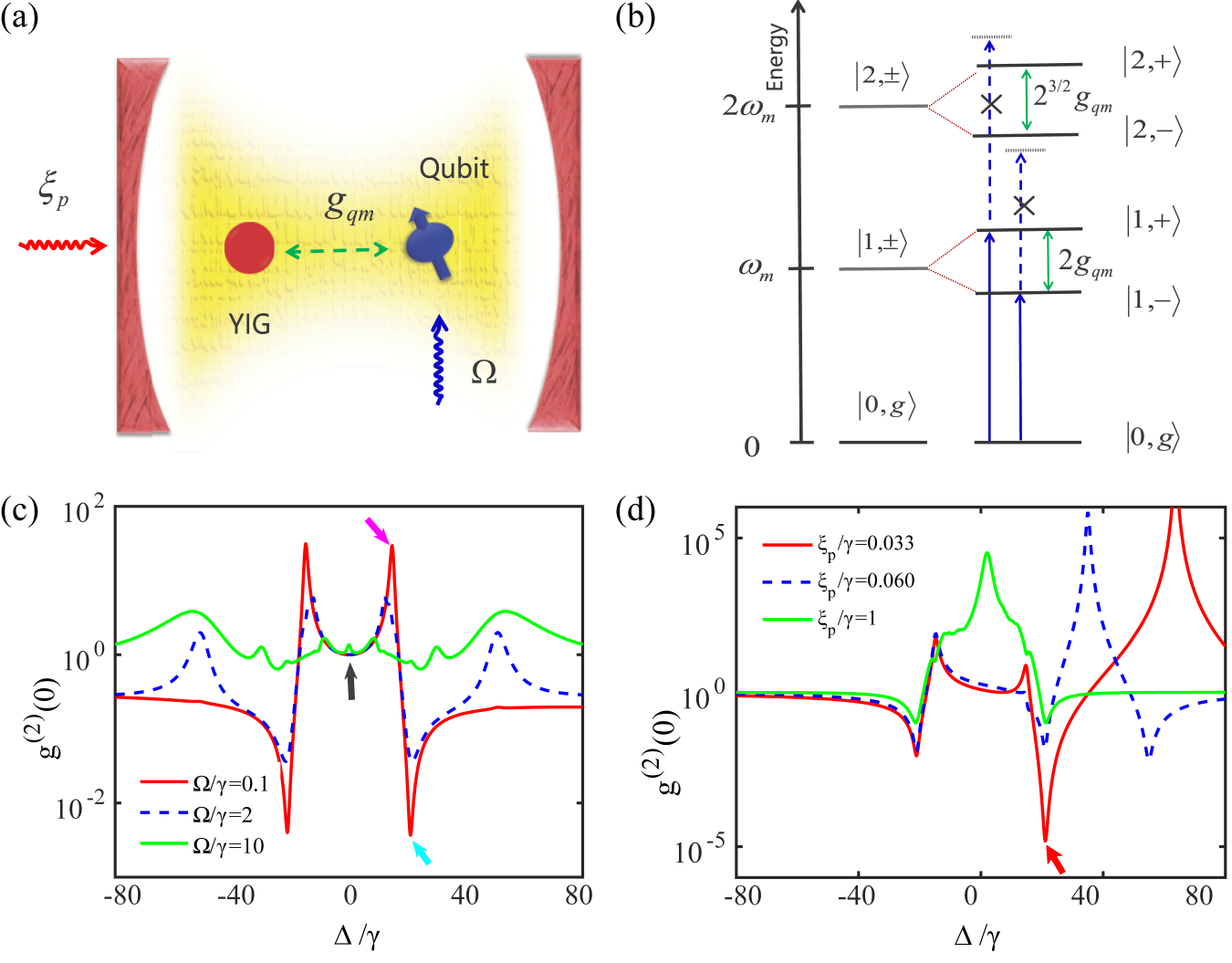}
		\caption{
			(a) The hybrid magnon--qubit system used for achieving magnon blockade, where both the qubit and the magnon are driven.  (b) Energy-level diagram in the dressed-state picture, showing the anharmonic spectrum induced by the magnon--qubit coupling. (c)-(d) Steady-state second-order correlation function $g^{(2)}(0)$ of the magnon mode versus magnon/qubit-drive detuning for different drive strengths. Reprinted figures (a)-(d) with permission from~\cite{liu2019magnon} Copyright 2019 by the American Physical Society.		
		}
		\label{fig:qmfig6}
	\end{figure}

	\subsubsection{Conventional magnon blockade}
	
	The first CMB protocol was given in Ref.~\cite{liu2019magnon} based on the CMQ system where both the magnon and qubit are driven by microwave fields (figure~\ref{fig:qmfig6}(a)). The qubit--magnon hybridization transfers the strong intrinsic nonlinearity from the qubit to the otherwise harmonic magnon mode, generating an anharmonic dressed-state spectrum (figure~\ref{fig:qmfig6}(b)) that selectively excites the single-magnon state while suppressing higher-order transitions. 
	Figures~\ref{fig:qmfig6}(c)-(d) present the results of the correlation function $g^{(2)}(0)$, which reveal a strong dependence of magnon statistics on the driving detuning and strengths.
	Pronounced antibunching with $g^{(2)}(0)\ll1$ is found at specific detunings, with a minimum value approaching $g^{(2)}(0)\sim10^{-5}$, indicating near-ideal magnon blockade, whereas bunching appears when higher-order transitions become resonant. The blockade effect, however, degrades rapidly with an increasing thermal magnon occupation, indicating the requirement of cryogenic temperatures for observing magnon blockade.
	This work thus establishes a foundational framework for exploring magnon blockade and a new route towards single-magnon sources in hybrid magnonic systems.		
	The same mechanism was subsequently extended to the strong dispersive regime~\cite{liu2025dispersive}, where the magnon-number-dependent ac-Stark shift (equation~\eqref{eq:dispersive}) creates an effective anharmonicity and suppresses two-magnon transitions, yielding $g^{(2)}(0)\sim0.04$ under optimal conditions. The advantage of this approach in the dispersive regime lies in the fact that it is naturally compatible with qubit readout and quantum non-demolition measurement of magnon numbers~\cite{lachance-quirion2020,wolski2020}, thus providing a more flexible platform for single-magnon manipulations.
	
	Magnon blockade has also been studied based on the direct magnon--qubit coupling without a cavity mediator. 	A scheme based on the direct transversal coupling between the magnon and a transmon qubit achieves $g^{(2)}(0)\sim10^{-7}$~\cite{jin2023magnon}. In their scheme, a weak driving (probe) field is applied to the magnon mode (the qubit).  It is found that the optimal blockade regime occurs when the transversal coupling strength equals the average detuning and the strength of the probe field is about three times that of the driving field. 	In addition, magnon blockade was studied in a magnon--spin hybrid system where a magnon mode is directly coupled to an NV-center spin~\cite{zhao2025magnon}. By driving both the magnon and the spin and introducing frequency detuning between them, this approach has the potential to reach $g^{(2)}(0)\sim10^{-8}$. Moreover, it relaxes the stringent requirement on the coupling strength, making the blockade accessible in the weak-coupling regime.

	\begin{figure}[t]
		\centering
		\includegraphics [width=\linewidth]{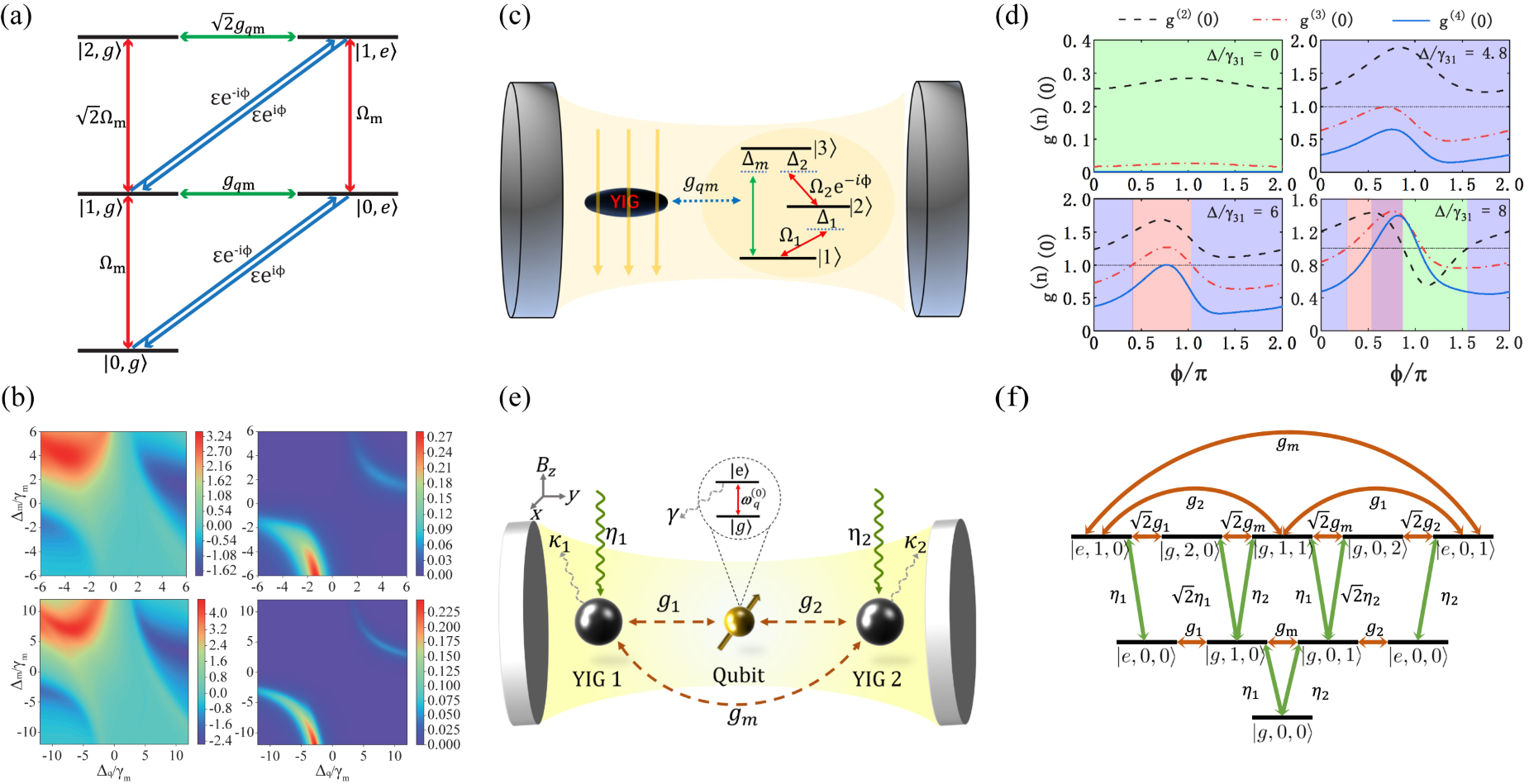}
		\caption{(a) Destructive interference of the transition paths prevents the two-magnon excitation.
			(b) Steady-state logarithmic second-order correlation function $\log _{10}g^{(2)}(0)$ (mean magnon number) versus two detunings in the left (right) column. Reprinted figures (a)-(b) with permission from~\cite{xie2020quantum} Copyright 2020 by the American Physical Society.
			(c) A magnon mode is coupled to a three-level fluxonium qubit via virtual cavity photons.
			(d) The second-, third-, and fourth-order correlation functions $g^{(2)}(0)$, $g^{(3)}(0)$, and $g^{(4)}(0)$ as a function of the phase $\phi$. Reprinted figures (c)-(d) with permission from~\cite{wu2021phase} Copyright 2021 by the American Physical Society.
			(e) Two YIG spheres coupled to a superconducting qubit used to achieve two-mode magnon blockade. 
			(f) Energy-level diagram illustrating multiple excitation pathways within the truncated few-magnon subspace. Reprinted figures (e)-(f) with permission from~\cite{fan2023nonclassical} Copyright 2023 by the American Physical Society.
		}
		\label{fig:qmfig7}
	\end{figure}

	\subsubsection{Unconventional magnon blockade}

	Another fundamentally different approach based on interference, i.e., the UMB, was first studied in Ref.~\cite{xie2020quantum}. Working in the weak-driving limit and truncating the Hilbert space to the lowest excitation states, the steady-state wavefunction is expanded in terms of few-excitation states, yielding $g^{(2)}(0) \propto |C_{2,g}|^2/|C_{1,g}|^4$, where $|C_{n,g}|^2$ represents the probability of the magnon--qubit system being in the state $|n,g\rangle$. This directly links magnon antibunching to the suppression of the two-magnon probability amplitude. Perfect blockade is achieved by enforcing $C_{2,g}=0$, which imposes constraints on system parameters and driving fields that guarantee destructive interference among multiple transition pathways to the two-magnon state (figure~\ref{fig:qmfig7}(a)). Numerical results (figure~\ref{fig:qmfig7}(b)) confirm strong antibunching over a broad parameter regime with significantly reduced requirements on the coupling strength. The results highlight that quantum interference provides a powerful and flexible route to engineer nonclassical magnon states, and enhances the feasibility of experimental realization of single-magnon sources.
	
	Such an interference-based mechanism was adopted in several other studies. In a protocol based on a hybrid system where two qubits couple to one magnon mode, both CMB and UMB can be simultaneously achieved by tuning the coupling strength ratio and detuning ratio~\cite{li2021tunable}. An alternative origin of the UMB in an anisotropic ferromagnet was studied in Ref.~\cite{xie2025unconventional}, where the ground state corresponds to an equilibrium squeezed vacuum. Applying a magnetic field to introduce displacement, destructive interference between squeezing and displacement pathways leads the two-magnon probability amplitude to zero. Remarkably, this UMB is robust against magnon decay due to its equilibrium nature. Furthermore, Ref.~\cite{xu2021conventional} systematically studies different mechanisms of magnon blockade in a magnon--qubit hybrid system under two driving fields, i.e., the strong-coupling CMB, the weak-coupling UMB, and the moderate-coupling UMB. The results indicate that quantum interference can significantly relax the requirement on the coupling strength.

	\subsubsection{Other magnon blockade effects}

	Beyond single-magnon statistics, Ref.~\cite{wu2021phase} extended the magnon blockade to the multi-magnon regime in a hybrid system where the magnon mode is coupled to a three-level $\Delta$-type fluxonium qubit via virtual cavity photons (figure~\ref{fig:qmfig7}(c)). The overall phase $\phi$ of the $\Delta$-type closed-loop configuration serves as a control knob. For appropriate detunings and driving strengths, varying $\phi$ enables a systematic transition among single-magnon blockade ($g^{(2)}(0)<1$), two-magnon blockade ($g^{(2)}(0)\geq1, g^{(3)}(0)<1$), and three-magnon blockade ($g^{(3)}(0)\geq1, g^{(4)}(0)<1$) (figure~\ref{fig:qmfig7}(d)), providing a versatile platform for generating tailored magnonic quantum sources. The underlying mechanism is the interference between direct magnon excitation and qubit-mediated indirect excitation pathways, which suppresses specific $n$-magnon states. 
	
	Unconventional two-mode magnon blockade and quantum-correlated magnon pairs were systematically studied in a system comprising two YIG spheres and a superconducting qubit coupled to a microwave cavity (figure~\ref{fig:qmfig7}(e))~\cite{fan2023nonclassical}. Adiabatic elimination of the cavity field yields two effective qubit--magnon couplings $g_1$, $g_2$ and an effective magnon--magnon coupling $g_m$. In the weakly driving regime, destructive interference among multiple excitation pathways (figure~\ref{fig:qmfig7}(f)) suppresses specific two-magnon occupations. By tuning the ratio of the coupling rates and of the driving strengths, parameter regimes are identified where both magnon modes exhibit strong antibunching and simultaneously the correlation between the two modes violates the Cauchy--Schwarz inequality $g_{12}^{(2)}(0)\leq\sqrt{g_{11}^{(2)}(0)g_{22}^{(2)}(0)}$, indicating a genuine nonclassical correlation.
	
	Magnon blockade can also show a nonreciprocal behavior, where magnon blockade becomes direction-dependent~\cite{wang2022dissipation,huang2024nonreciprocal,zhang2025nonreciprocal,zhang2024nonreciprocal,feng2026nonreciprocal}. For instance, Ref.~\cite{wang2022dissipation} proposes a scheme combining both coherent and dissipative qubit--magnon couplings mediated by a common waveguide. The dissipative coupling introduces a phase that depends on the drive direction, leading to a stark contrast in $g^{(2)}(0)$ for opposite propagation directions, thereby realizing nonreciprocal magnon blockade. Reference~\cite{huang2024nonreciprocal} proposes nonreciprocal UMB exploiting the Barnett effect in a spinning YIG sphere. By tuning the Barnett shift via controlling magnetic field direction, magnon antibunching occurs for one magnetic field direction while being suppressed for the opposite. By further tuning parameters, nonreciprocal CMB and UMB can be simultaneously achieved and a switch between them is also possible. These approaches provide the possibility of realizing chiral single-magnon devices.

	\begin{figure}[t]
		\centering
		\includegraphics[width=\linewidth]{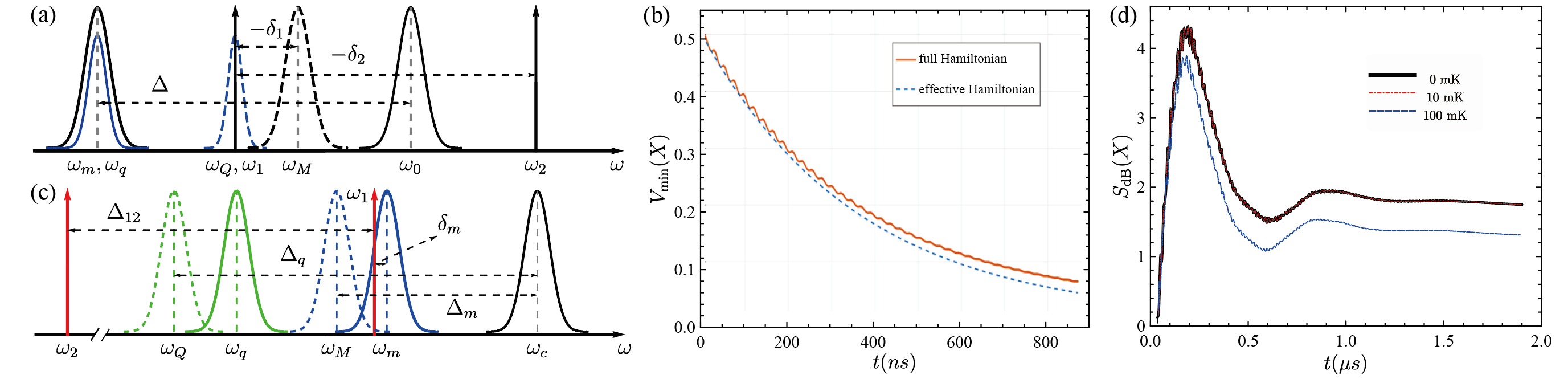}
		\caption{(a) Frequencies of the CMQ system and two driving fields adopted in the protocol~\cite{guo2023pra}, {to realize an effective parametric-amplification Hamiltonian for the magnon mode.} ~(b) Minimum variance $V_{\mathrm{min}}(X )$ of the magnon quadrature as a function of time $t$. Reprinted figures (a)-(b) with permission from~\cite{guo2023pra} Copyright 2023 by the American Physical Society.~(c) Frequencies of the CMQ system and two driving fields used in the protocol \cite{liu2026pra}, {to simulate the quantum Rabi model near the critical point, which yields magnon squeezing}. ~(d) The degree of magnon squeezing $S$ (dB) as a function of time $t$. Reprinted figures (c)-(d) with permission from~\cite{liu2026pra} Copyright 2026 by the American Physical Society. }
		\label{fig:qmfig8}
	\end{figure}

	\subsection{Magnon squeezing (protocols)}

	Apart from various mechanisms introduced in {section~\ref{sqzs}} leading to magnon squeezing, the CMQ system offers another practical and promising platform to prepare magnon squeezed states by appropriately manipulating the superconducting qubit. The basic idea is to engineer an effective two-magnon process, which induces the squeezing of the magnon mode.
	
	Based on the CMQ system, Ref.~\cite{guo2023pra} adopts a two-tone driving protocol, where the qubit is driven by two microwave fields with frequencies $\omega_{1}$ and $\omega_{2}$ and driving strengths $\Omega_{1}$ and $\Omega_{2}$ (figure~\ref{fig:qmfig8}(a)). Considering the situation where the qubit and the magnon are resonant, $\omega_m = \omega_q \equiv \omega$, and far detuned from the cavity, $\Delta = \omega_0 - \omega \gg g_{\rm aq}, g_{\rm am}$, and by appropriately choosing the two drive frequencies and strengths, an effective parametric amplification Hamiltonian of the magnon mode can be constructed, no matter the qubit is initially in the ground or excited state, given by
	\begin{eqnarray} 
		H_{\mathrm{eff}}/\hbar = \chi\left(m^{2}+m^{\dagger 2} \right),
	\end{eqnarray}
	where $\chi = g_{\rm qm}^2/(4\Omega_{2})$, and $g_{\rm qm} = g_{\rm aq} g_{\rm am}/\Delta$ is the effective qubit--magnon coupling. This Hamiltonian describes a two-magnon process that leads to magnon quadrature squeezing (figure~\ref{fig:qmfig8}(b)). Numerical results indicate that the squeezing is robust against the dissipation rates of the magnon and qubit, and against the bath temperature, remaining below vacuum fluctuations for temperatures up to 0.3~K.

	The magnon squeezing based on two-tone driving can be further enhanced by engineering an effective Rabi-type interaction and operating the system near a quantum critical point \cite{liu2026pra}. Differing from Ref.~\cite{guo2023pra}, the driving scheme is designed to simulate the quantum Rabi model. By applying two microwave drives to the qubit and designing an appropriate frequency spectrum of the system~(figure~\ref{fig:qmfig8}(c)), an effective Rabi-type Hamiltonian  can be derived
	\begin{eqnarray} 
		H_{\mathrm{Rabi}}/\hbar = \delta_{m}m^\dag m + \frac{\mathcal{E}_2}{2}\sigma_z + g\left(m + m^\dag \right)\sigma_x,
	\end{eqnarray}
	where $\delta_{m}$ is the effective magnon detuning with respect to the first drive, $\mathcal{E}_2$ is the Rabi frequency of the second drive, and $g = g_{\rm qm}/2$ is the coupling strength. This Hamiltonian exhibits a quantum phase transition from the normal phase to the superradiant phase at the critical point $g_c\equiv 2g/\sqrt{\delta_{m}\mathcal{E}_2}=1$. By operating the system in the normal phase but close to the critical point and projecting onto the ground state of the qubit, an effective Hamiltonian of the magnon mode can be obtained 
	\begin{eqnarray} 
		H_{\mathrm{eff}}/\hbar=\delta_{m}m^\dag m-\frac{\delta_{m} g_c^2}{4}\left(m^\dag + m \right)^2.
	\end{eqnarray}
	This Hamiltonian describes a parametric amplification-like process that generates magnon squeezing. Magnon squeezing above 4 dB can be achieved using currently available parameters (figure~\ref{fig:qmfig8}(d)).
	
	The above two protocols based on two-tone driving offer different mechanisms for generating magnon squeezing with respect to the effective-Kerr approach used in the experiment~\cite{weng2026} (section~\ref{sqzex}) exploiting the magnon--qubit dispersive interaction.

	\subsection{Magnon entanglement}

\begin{figure}[t]
		\centering
		\includegraphics[width=\linewidth]{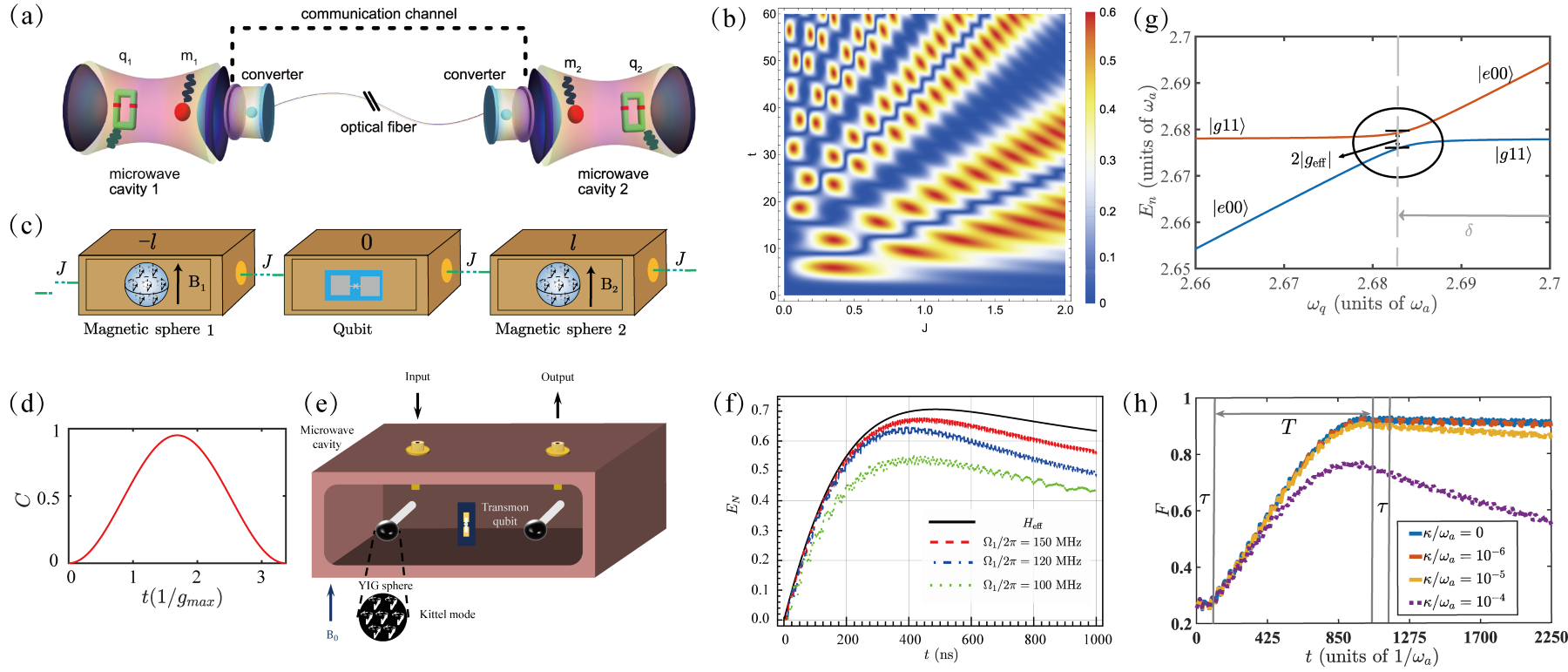}
		\caption{(a) Two remote CMQ systems coupled via microwave-optical converters and an optical fiber. (b) Magnon entanglement as measured by the concurrence versus fiber coupling strength $J$ and time $t$. Reprinted figures (a)-(b) with permission from~\cite{luo2021nonlocal} Copyright 2021 by the Optical Society of America. (c) Schematic of the entanglement generation using a cavity array as a quantum channel. (d) Magnon entanglement as measured by the concurrence $C$ versus time $t$. Reprinted figures (c)-(d) with permission from~\cite{ren2022prb} Copyright 2022 by the American Physical Society. (e) Two YIG spheres coupled to a transmon qubit in a single cavity used to generate magnon entanglement. (f) Entanglement between two magnon modes as a function of time. Reprinted figures (e)-(f) with permission from~\cite{yang2026arxiv} Copyright 2026 by the American Physical Society. (g) Energy levels and the avoided level crossing of the states $\ket{e00}$ and $\ket{g11}$. (h) Time evolution of the fidelity of the GHZ state under various dissipation rates. Reprinted figures (g)-(h) with permission from~\cite{qi2022pra} Copyright 2022 by the American Physical Society. }
		\label{fig:qmfig9}
	\end{figure}

	The magnon--qubit hybrid system also shows a powerful capability for the generation of magnonic entangled states. Reference~\cite{luo2021nonlocal} proposes a scheme for generating nonlocal entanglement between two magnon modes residing in separate microwave cavities (figure~\ref{fig:qmfig9}(a)). It is relevant for distributed quantum networks where quantum nodes are spatially separated. Each cavity contains a YIG sphere and a superconducting qubit. The communication channel is established via microwave-optical photon converters connected by an optical fiber.  The cavity modes are adiabatically eliminated to create an effective coupling between the qubits and magnons. The system operates in the single-excitation manifold, and the core mechanism relies on the qubit as a mediator to facilitate the entanglement between the two remote magnons. Magnon entanglement is nonlocally established and its dynamics  as a function of the fiber coupling strength are shown in figure~\ref{fig:qmfig9}(b).
	
	Reference \cite{ren2022prb} proposes to establish nonlocal magnon entanglement using a cavity array as a quantum channel. The system comprises a one-dimensional array of coupled microwave cavities, a superconducting transmon qubit, and two YIG spheres (figure~\ref{fig:qmfig9}(c)). The qubit is located at site $n=0$, while the two YIG spheres are placed symmetrically at sites $n=-l$ and $n=l$. The magnon and qubit interact with their respective cavities. When both the qubit and magnon frequencies lie outside the cavity array's frequency band, it seeds a localized photonic mode, forming a bound state with binding energy $\delta$ and localization length $\xi$. This localized photonic mode acts as a virtual cavity, mediating an effective interaction between the qubit and magnon, and between the two magnons, if they lie within the localization length. By adiabatically eliminating the photonic component, an effective qubit--magnon Hamiltonian is obtained as
	\begin{eqnarray} 
		H_{\text{int}}/\hbar = \sum_{j=1,2}g_{01}\left(m_j^\dagger q + m_j q^\dagger \right) + g_{12}\left(m_1^\dagger m_2 + m_1 m_2^\dagger \right),
	\end{eqnarray}
	where $g_{01}$ ($g_{12}$) is the effective coupling strength. For an initial state of the system $|1\rangle_{q}|0\rangle_{m1}|0\rangle_{m2}$, the two magnon modes evolve into a Bell state $(|01\rangle+|10\rangle)/\sqrt{2}$. The concurrence reaches a maximum value close to unity at an optimal time with realistic parameters (figure~\ref{fig:qmfig9}(d)).
	
	Differing from Refs.~\cite{luo2021nonlocal,ren2022prb}, where long-range magnon entanglement is established, Ref.~\cite{yang2026arxiv} provides a more feasible protocol to entangle two magnon modes in two YIG spheres that, together with a transmon qubit, are coupled to a single microwave cavity (figure~\ref{fig:qmfig9}(e)). Similarly, as in the protocols \cite{guo2023pra,liu2026pra}, the qubit is driven by a two-tone field. Adiabatic elimination of the cavity leads to an effective JC interaction between the qubit and each magnon mode and an effective BS coupling between the two magnon modes. By appropriately choosing the frequencies and strengths of the two driving fields, an effective interaction Hamiltonian of the magnon--qubit system can be derived
	\begin{eqnarray} 
		H_{\rm {int}}/\hbar  \approx  g_ {\rm{eff}} \left(m_1^\dag m_2^\dag + m_1 m_2 \right) \sigma_{z}, 
	\end{eqnarray}
	where $g _ {\mathrm {eff}}$ is the effective coupling between the two magnon modes. Clearly, this is a TMS interaction that generates entanglement between two magnon modes, no matter the qubit is initially in the ground or excited state. The scheme is powerful in the sense that strong entanglement (with logarithmic negativity reaching 0.68, figure~\ref{fig:qmfig9}(f)) can be achieved using fully feasible parameters. An entanglement detection scheme was also provided by performing a joint Wigner-function tomography of the two magnon modes exploiting their dispersive coupling to the qubit.
	
	The generation of entanglement can also be extended from magnonic systems to multipartite entanglement in the CMQ  system~\cite{qi2022pra}. Unlike standard approaches that rely strictly on the JC model, this protocol exploits the counter-rotating terms inherent in the ultrastrong coupling regime to access new transition channels. The protocol exploits avoided level crossings at specific resonance points derived using the high-order Fermi's golden rule to generate entangled states. For instance, at the resonance condition $\omega_q = \omega_a + \omega_m + \delta$, with $\delta$ being the frequency shift from the exact double-resonant point to the point of the avoided level crossing for $\ket{g11}$ and $\ket{e00}$ (figure~\ref{fig:qmfig9}(g); $\ket{\cdot, \cdot, \cdot}$ denotes the state of the qubit, photon, and magnon in sequence), an effective Hamiltonian $\hbar g_{\rm eff} (\ket{e00}\bra{g11} + \ket{g11}\bra{e00})$ can be derived, with  $g_{\rm eff}$ being the effective coupling strength. This enables the generation of the Bell state $(\ket{00} + \ket{11})/\sqrt{2}$ of the photon--magnon subsystem. 	
	For the Greenberger--Horne--Zeilinger (GHZ) states, the system is operated at a point where $\omega_q = \omega_a - \omega_m + \Delta$, with a corresponding frequency shift $\Delta$, leading to an effective Hamiltonian $\hbar g'_{\rm eff} (\ket{e11}\bra{g20} + \ket{g20}\bra{e11})$ with the effective coupling strength $g'_{\text{eff}}$. 	
	Subsequent qubit rotations combined with adiabatic frequency tuning drive the system from the initial state $\ket{g00}$ to the GHZ state $(\ket{g00} + \ket{e11})/\sqrt{2}$.  
	The results show that the fidelity of the GHZ state can be high when the dissipation rates of the system are sufficiently small (figure \ref{fig:qmfig9}(h)).
	
	Several other protocols have been offered for entanglement generation in the magnon--qubit hybrid system.~For example, a protocol for preparing magnonic NOON states was provided exploiting Floquet engineering \cite{qi2023pra}. Extending to a multi-level superconducting system, the complex magnon-superconducting hybrid system could also be used to generate magnon entanglement \cite{kong2021prb,kong2022prr,yan2024generating}.

	\section{Cavity optomagnonics}\label{optomag}
	
	\subsection{The system and theoretical model}
	
	\begin{figure}[t]
		\centering
		\includegraphics[width=0.95\textwidth]{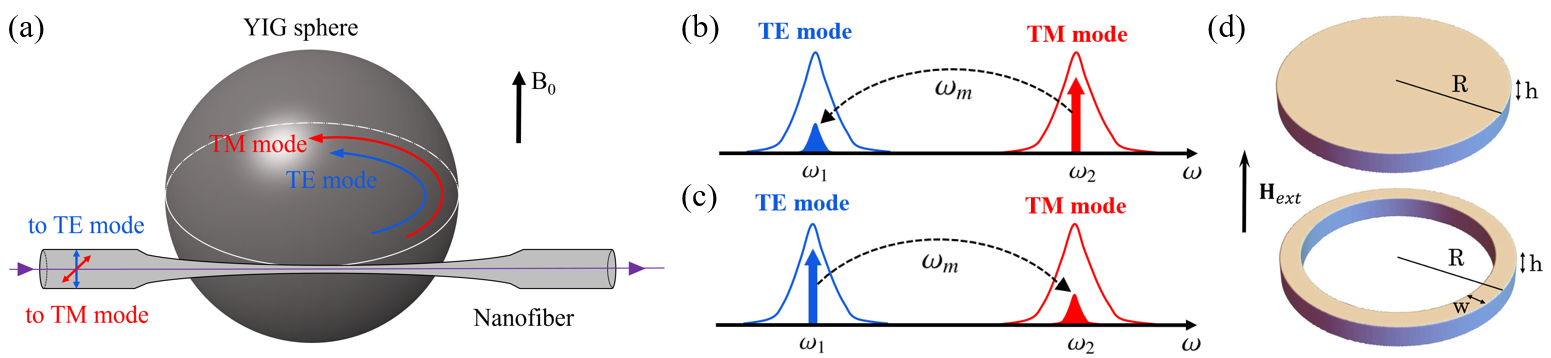}
		\caption{(a) Schematic diagram of cavity optomagnonic architecture where a YIG sphere supports a magnon mode and two WGMs with different polarizations, i.e., the TM- and TE-polarized modes. (b) Optomagnonic Stokes scattering, where a TM-polarized photon converts into a TE-polarized photon by creating a magnon. (c) Optomagnonic anti-Stokes scattering, where a TE-polarized photon converts into a TM-polarized photon by annihilating a magnon. (d) A YIG micro disk (upper) and ring (lower), which support both the magnon mode and WGMs. Reprinted figure (d) from~\cite{Disk26} licensed under the \href{https://creativecommons.org/licenses/by/4.0/}{CC BY 4.0}.}
		\label{fig1}
	\end{figure}

	Cavity optomagnonics studies the magneto-optical interaction between magnons and optical cavity photons. The typical optomagnonic system is a YIG sphere which supports both a magnetostatic mode, e.g., the Kittel mode, and two optical whispering gallery modes (WGMs) with different polarizations (figure \ref{fig1}(a)). Due to the magnon-induced {Brillouin light scattering (BLS)}~\cite{Nakamura16,Zhang16,Haigh16}, the photons in a WGM are scattered by lower-frequency magnons (typically in gigahertz), yielding two optical sidebands of which the frequencies with respect to the WGM equal to the magnon frequency. The scattering probability is maximized when the triple-resonance condition is fulfilled~\cite{Nakamura16,Zhang16,Haigh16,Nakamura18,Haigh21}, i.e., the scattered photons enter another WGM of the YIG sphere. Due to the selection rule~\cite{Sharma17,PAP17,Nakamuranjp,Haigh18} imposed by the conservation of the angular momenta of WGM photons and magnons, the BLS exhibits a pronounced asymmetry in the Stokes and anti-Stokes scattering strength.  In addition, the selection rule causes different optical polarizations of the two WGMs, e.g., the transverse-magnetic (TM)- and transverse-electric (TE)-polarized WGMs (figure \ref{fig1}(a)).  
	
	The Hamiltonian of the optomagnonic system with an optical drive can be written as
	\begin{equation}\label{Inter}
		H_{\rm om}/\hbar=\omega_{m}m^{\dagger} m+\omega_{1}a^{\dagger}_{1}a_{1}+\omega_{2}a^{\dagger}_{2}a_{2}+ g_{\rm om} \left(a_{1}^{\dagger} a_{2} m^{\dagger} +a_{1} a_{2}^{\dagger} m \right) +i E_{j}\left(a_{j}^{\dagger}e^{-i\omega_{p_{j}}t}- {\rm H.c.} \right)
	\end{equation}
	where $m$ ($m^{\dagger}$) is the annihilation (creation) operator of the magnon mode with frequency $\omega_{m}$ which is tunable by changing the bias magnetic field $B_0$, and $a_{j}$ ($a_{j}^{\dagger}$), $j=1,2$, are the annihilation (creation) operators of the WGMs with resonance frequencies $\omega_{j}$ which satisfy the relation $\omega_{j}\gg\omega_{m}$ and the triple-resonance condition $|\omega_{2}-\omega_{1}|=\omega_{m}$. 
	Here, without loss of generality, the $a_{1}$ ($a_{2}$) mode is assumed to be the TE (TM) mode of a certain WGM orbit.  It is also assumed that the magnon-induced BLS occurs only between the TM and TE modes with the same WGM index, i.e., the orbital angular momentum of the WGM photons is conserved~\cite{Nakamura16,Nakamuranjp} where the frequency of the TM mode is higher than that of the TE mode, $\omega_{\rm TM} > \omega_{\rm TE}$, due to the geometrical birefringence~\cite{Zhang16,Nakamura18}. Note that the optomagnonic interaction is a three-wave process and the single-photon optomagnonic coupling rate $g_{\rm om}$ is typically weak, on the order of $\sim$Hz, for a submillimeter YIG sphere~\cite{Nakamura16,Zhang16,Haigh16,Nakamura18,Haigh21}. However, the effective optomagnonic coupling strength can be significantly improved by strongly driving either the TM- or TE-polarized WGM, and $E_{j}=\sqrt{2P_{j}\kappa_{j}^{e}/\hbar\omega_{p_{j}}}$ corresponds to the coupling strength between the $j$th WGM  (with an external decay rate $\kappa_{j}^{e}$) and the drive field with frequency $\omega_{p_{j}}$ and power $P_{j}$.

	The optomagnonic Hamiltonian $H_{\rm om}$, in the frame rotating at the drive frequency $\omega_{p_{j}}$, is given by
	\begin{equation}
		H_{\rm om}/\hbar=\Delta_{1}a^{\dagger}_{1}a_{1}+\Delta_{2}a^{\dagger}_{2}a_{2}+\omega_{m}m^{\dagger}m+g_{\rm om}\left(a_{1}^{\dagger} a_{2} m^{\dagger}+a_{1} a_{2}^{\dagger} m \right)+i E_{j}\left(a_{j}^{\dagger}-a_{j} \right),
	\end{equation}
	where $\Delta_{j}=\omega_{j}-\omega_{p_{j}}$ ($j=1,2$) is the WGM-drive detuning. Consider that the WGM $a_{2}$ is resonantly driven at frequency $\omega_{2}=\omega_{1}+\omega_{m}$, and in this case, the operator $a_{2}$ can be treated classically as a number $\alpha_{2}\equiv\langle a_{2}\rangle=E_{2}/\kappa_{2}$, where $\kappa_{j}$ is the total decay rate of the $j$th WGM. This leads to the following linearized Hamiltonian in the interaction picture
	\begin{equation}\label{tms}
		H_{\rm TMS}/\hbar=G_{1}\left(a_{1}^{\dagger}m^{\dagger} +a_{1} m\right)
	\end{equation}
	where $G_{1}=g_{\rm om}\alpha_{2}$ is the effective optomagnonic coupling. The above Hamiltonian enables a TMS interaction creating the entanglement between the WGM $a_1$ and the magnon mode. This corresponds to the Stokes scattering process,
	where pump (TM-polarized) photons convert into lower-frequency sideband (TE-polarized) photons by emitting magnons (figure \ref{fig1}(b)). 
	Similarly, when the WGM $a_1$ is resonantly driven at frequency $\omega_{1}=\omega_{2}-\omega_{m}$, the effective optomagnonic Hamiltonian is given by
	\begin{equation}\label{bs}
		H_{\rm BS}/\hbar=G_{2}\left(a_{2}^{\dagger}m+a_{2} m^{\dagger}\right)
	\end{equation}
	where $G_{2}=g_{\rm om}\alpha_{1}$ is the effective optomagnonic coupling, with $\alpha_{1}\equiv\langle a_{1}\rangle=E_{1}/\kappa_{1}$. The above Hamiltonian is of the BS form, corresponding to the anti-Stokes scattering where pump (TE-polarized) photons convert into higher-frequency anti-Stokes (TM-polarized) photons by annihilating magnons (figure \ref{fig1}(c)).

	\subsection{Coupling enhancement}\label{CEnh}

	Equations (\ref{tms})-(\ref{bs}) indicate that a strong pump field can significantly enhance the optomagnonic coupling strength $G$. {However, there are restrictions on such enhancement---a too strong drive may cause significant heating and instability, leaving the optomagnonic strong coupling currently unrealized in experiments.} To circumvent these limitations, strategies~\cite{Disk26,Alm18,SV18,SV21,SV22} have been proposed to increase the single-photon optomagnonic coupling strength $g_{\rm om}$. The core is to appropriately design the optomagnonic configuration, or exploit material properties, to simultaneously reduce the mode volume and increase the mode overlap of the optical and magnon modes. Along this line, Ref.~\cite{SV18} indicates that a nonhomogeneous magnetic ground state, namely, a magnetic vortex can couple to WGMs much more strongly because of its rich spatial chiral and polar texture. Reference~\cite{SV21} designs an optomagnonic crystal, similar to the optomechanical crystal~\cite{Vahala}, which is a periodically patterned structure at the microscale and can colocalize both the magnon and optical modes. This design greatly increases the mode overlap and the bare optomagnonic coupling is predicted to reach the kHz level. Reference~\cite{SV22} indicates that a magnetized epsilon-near-zero (ENZ) medium can support an ultra-strong bare optomagnonic coupling at the ENZ frequency due to a drastic enhancement of the magneto-optical response. Recently, Ref.~\cite{Disk26} suggests that an optomagnonic micro disk or ring (figure \ref{fig1}(d)) can notably increase the mode overlap and the bare coupling can be up to $\sim4.5$ kHz for a micro disk with radius $R=5\:\mu$m and thickness $h=1\:\mu$m. 
	
	Conditioned on the significant experimental improvement on the optomagnonic coupling strength, one of the promising applications of cavity optomagnonics is the microwave-to-optics conversion~\cite{NakamuraPRB,Op20,GS22,LiF23,LPR24,Libb}, where the magnon mode serves as an intermediary that coherently converts microwave photons into optical photons. The advantage of the approach using magnons is mainly manifested as its great tunability---the magnon frequency can be continuously tuned via an external magnetic field. Note that the current low conversion efficiency is principally due to the small bare optomagnetic coupling in the experiment using a large-sized YIG crystal. Future improvements should focus on miniaturization to enhance the bare optomagnetic coupling, e.g., employing a YIG micro disk, as suggested in Ref.~\cite{Disk26}. 
	
	In the following, we review a series of proposals based on cavity optomagnonic systems focusing on quantum states generation and their potential applications in quantum science and technologies.

	\begin{figure}[t]
		\centering
		\includegraphics[width=0.95\textwidth]{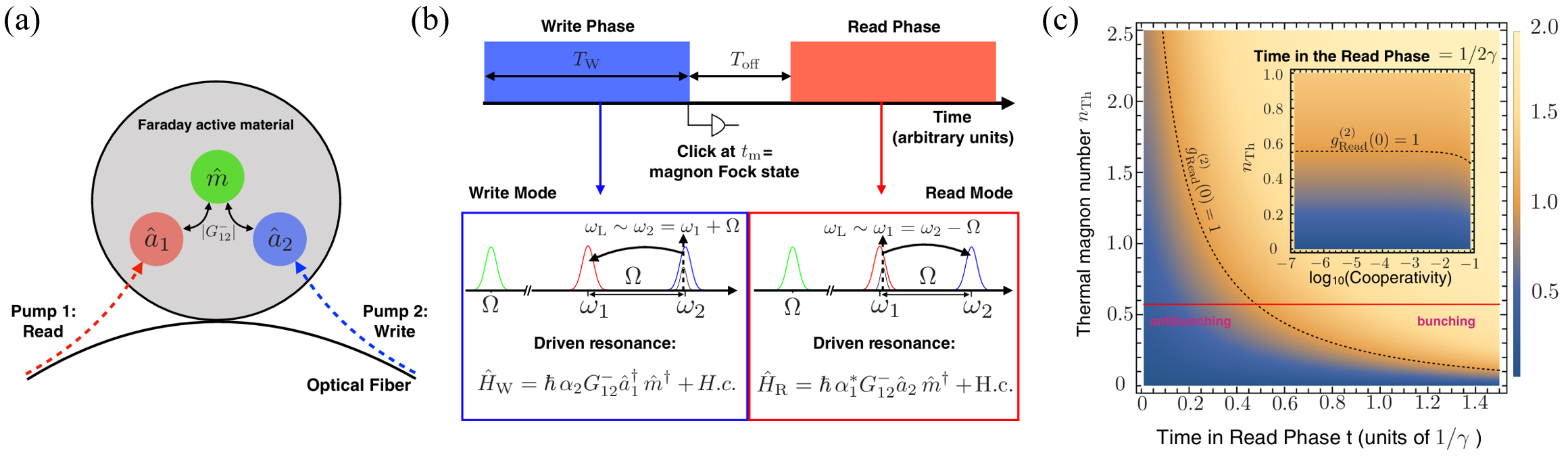}
		\caption{(a) The optomagnonic system used to prepare and read out magnonic Fock states by sending two optical pulses in sequence. (b) Pulse sequence used in the heralding protocol. The write pulse is sent to weakly generate the TMS interaction, creating a single magnon conditioned on the detection of a single Stokes photon. After a free evolution time $T_{\rm off}$, the read pulse is sent to generate BS (state-swap) interaction to optically read out the single-magnon state. (c) Second-order correlation function $g^{(2)}_{\rm Read}(0)$ of the anti-Stokes field conditioned on the measurement of a Stokes photon versus the duration time in the reading process and thermal magnon number. Inset: $g^{(2)}_{\rm Read}(0)$ as a function of cooperativity associated with the read pulse and thermal magnon number. Reprinted figures (a)-(c) with permission from~\cite{Fock} Copyright 2019 by the American Physical Society.}
		\label{fig2}
	\end{figure}

	\subsection{Magnonic non-Gaussian states}

	Non-Gaussian states~\cite{NG}, which can not be expressed as any convex mixture of Gaussian states, manifest many novel and unique features that break the laws of classical physics and can thus be exploited to explore many fundamental issues of quantum mechanics, such as the quantum-to-classical transition. Moreover, these states are essential components for achieving the universality in continuous-variable (CV) quantum computation~\cite{NG}. Below we introduce several protocols of preparing magnonic non-Gaussian states, namely magnonic Fock states, cat states, single-magnon-added coherent states (single-MACS) and -magnon-added thermal states (-MATS).

	\begin{figure}[t]
		\centering
		\includegraphics[width=0.97\textwidth]{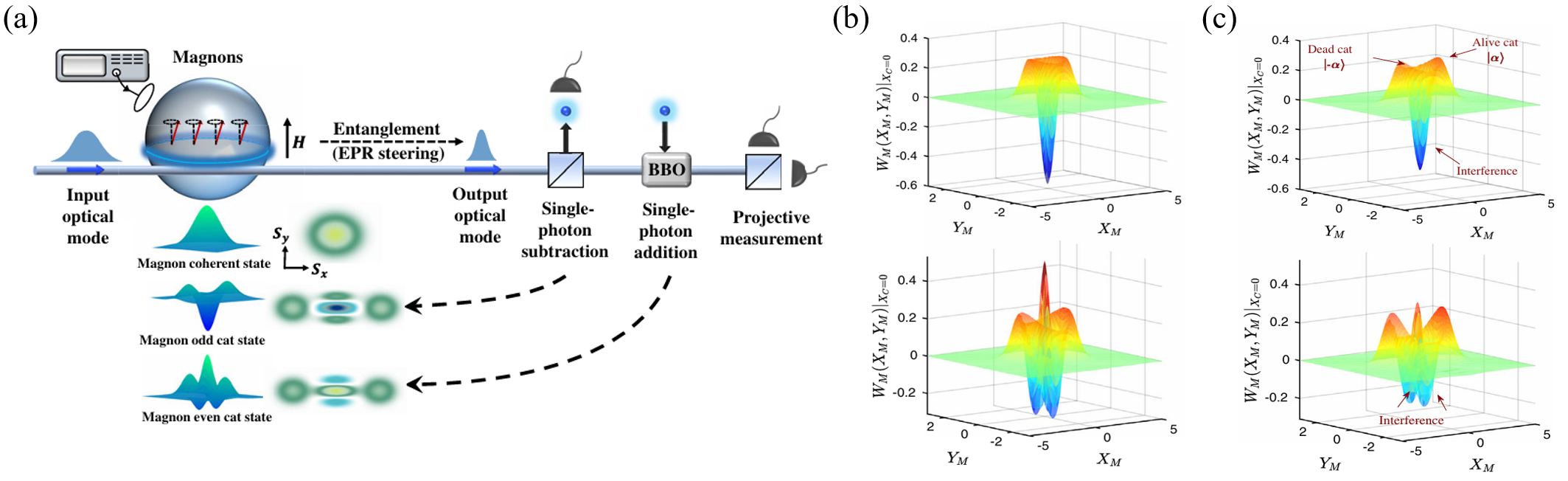}
		\caption{(a) Schematic diagram of remote generation of a magnonic cat state. The activation of the optomagnonic Stokes scattering, together with single-photon subtraction/addition operations and projective measurements, prepares the magnon mode in a cat state. (b)-(c) Wigner functions of the projected magnon mode with an outcome $X_{C}=0$ for the TMS parameter $r=0.1$ in (b) and $r=0.2$ in (c). Upper row corresponds to the case of subtracting a single photon from the optical output mode, yielding an odd cat state, while lower row corresponds to the case of subtracting a single photon and then adding a single photon on the optical output mode, giving an even cat state. Here, $X_{C}$ denotes the amplitude quadrature of the optical field, while $X_{M}$ and $Y_{M}$ represent the amplitude and phase quadratures of the magnon mode, respectively. Reprinted figures (a)-(c) with permission from~\cite{cat} Copyright 2021 by the American Physical Society.}
		\label{fig3}
	\end{figure}

	\subsubsection{Magnonic Fock states}

	Magnonic Fock or number states are a type of typical quantum states of magnetization in a ferromagnetically ordered system. Their realization is crucial for achieving single-quantum-level control of collective spin excitations. Reference~\cite{Fock} proposes a probabilistic heralding scheme to prepare a single-magnon state by means of single-photon detection. In this protocol, two WGMs $a_{1}$ and $a_{2}$ are simultaneously coupled to a magnon mode $m$ (figure \ref{fig2}(a)), and it contains two steps: the ``write" and ``read" process (figure \ref{fig2}(b)). In the former process, the $a_{2}$ mode is resonantly driven by a write pulse to weakly activate the optomagnonic Stokes scattering, where a single pump photon converts into a single Stokes photon entering the $a_{1}$ mode by emitting a magnon. 
	The single-magnon state $|1\rangle_{m}$ is thus prepared conditioned that a single Stokes photon is detected. Subsequently, a reading pulse is sent to pump the $a_{1}$ mode to activate the anti-Stokes scattering, where a pump photon converts into an anti-Stokes photon by absorbing a single magnon (the one created in the former process). The effective BS interaction (equation (\ref{bs})) in this process maps the magnonic state to the anti-Stokes field. Therefore, the single-magnon state can be verified by measuring the second-order correlation function $g^{(2)}_{\rm Read}(0)$ of the anti-Stokes field.
	Figure \ref{fig2}(c) shows $g^{(2)}_{\rm Read}(0)$ as a function of the time duration of the reading process and the mean thermal magnon number. The dashed curve corresponding to $g^{(2)}_{\rm Read}(0)=1$ marks the transition from antibunching ($g^{(2)}_{\rm Read}(0)<1$) to bunching ($g^{(2)}_{\rm Read}(0)>1$). 
	The parameter regime corresponding to $g^{(2)}_{\rm Read}(0)\ll 1$ indicates that a high-fidelity single-magnon state is prepared.

	\subsubsection{Magnonic cat states}

	Based on cavity optomagnonics, Ref.~\cite{cat} proposes a scheme to remotely generate and manipulate a magnonic cat state by performing non-Gaussian operations on optical photons exploiting strong optomagnonic entanglement and Einstein--Podolsky--Rosen (EPR) steering, as illustrated in figure \ref{fig3}(a). The YIG sphere is driven by a light beam to activate the optomagnonic Stokes scattering, which creates strong quantum entanglement and EPR steering between the magnon mode and the Stokes field. Then, by remotely performing appropriate single-photon operations and projective measurements on the optical output Stokes field, the magnon mode collapses to an even or odd cat state.
	
	Figures \ref{fig3}(b)-(c) show the Wigner functions of the projected magnon mode with an outcome of the optical projective measurement $X_{C}=0$ after a single
	photon subtraction performed on the optical output field. It can be clearly seen that two distinct peaks appear in the direction of $X_{M}$, and the Wigner function at the origin ($X_{M}=Y_{M}=0$) is negative, revealing that the magnon collapses to an odd cat state. Moreover, when a single-photon subtraction and a subsequent single-photon addition are performed on the optical output, an even cat state of the magnon mode with a positive Wigner function at the origin can be generated. A larger TMS parameter $r$ gives rise to a larger size of the magnonic cat state.

	\subsubsection{Single-MACS and -MATS}
	
	\begin{figure}[t]
		\centering
		\includegraphics[width=0.98\textwidth]{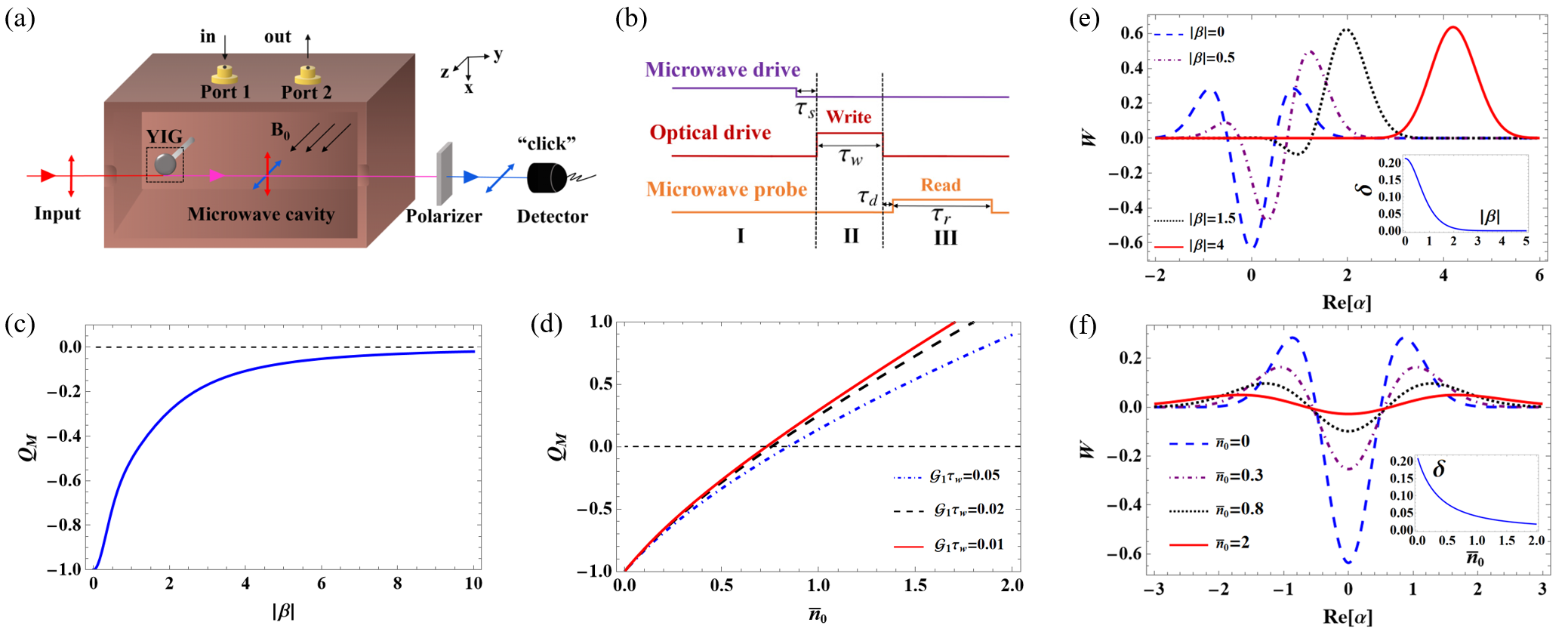}
		\caption{(a) The microwave-optomagnonic system used to generate single-MACS and -MATS, consisting of a YIG sphere coupled to a microwave cavity. (b) Time sequence of the optical and microwave pulses to prepare single-MACS and -MATS. (c) Mandel parameter $Q_{M}$ of the single-MACS versus the amplitude $|\beta|$ of the initial coherent state. (d) Mandel $Q_{M}$ parameter of the single-MATS versus initial thermal occupation $\bar{n}_{0}$ for different values of $\mathcal{G}_{1} \tau_{w}$. (e) Wigner function (the Im$(\alpha)=0$ plane) of the single-MACS for different coherent amplitude $|\beta|$. Inset shows the Wigner negativity $\delta$ versus $|\beta|$. (f) Wigner function (the Im$(\alpha)=0$ plane) of the single-MATS for different thermal occupation $\bar{n}_{0}$. Inset shows the Wigner negativity $\delta$ versus $\bar{n}_{0}$. Reprinted figures (a)-(f) from~\cite{Lu1} licensed under the \href{https://creativecommons.org/licenses/by/4.0/}{CC BY 4.0}.}
		\label{fig4}
	\end{figure}

	The addition of a single excitation onto Gaussian states, such as coherent or thermal states, provides a simple yet efficient approach to generating non-Gaussian states~\cite{GS91,GS92}. Reference~\cite{Lu1} applies this idea to cavity optomagnonics and proposes to prepare two types of magnonic non-Gaussian states, namely single-MACS and -MATS, by adding a single magnon in a heralded way onto coherent and thermal states, respectively. The setup is shown in figure~\ref{fig4}(a).	
	
	The protocol consists of three steps (figure \ref{fig4}(b)): $i$) State initialization. The whole system is placed at a low bath temperature to ensure that the thermal excitation of both the microwave cavity field and the magnon mode is negligibly small and their noise is essentially vacuum noise. To prepare a magnonic thermal state, one simply increases the bath temperature to have a nonzero thermal occupation, while to prepare a magnonic coherent state, a weak microwave coherent field is used to drive the cavity and the magnon mode is thereby displaced to a coherent state due to the cavity--magnon BS interaction (equation \eqref{Hma}).  $ii$) Adding a single magnon. Once the magnon mode is in the desired Gaussian state, the microwave drive is switched off. After a short period $\kappa_{a}^{-1}\ll\tau_{s}\ll\kappa_{m}^{-1}$, during which all microwave cavity photons decay, while the magnon state remains almost unchanged, an optical write pulse with duration $\tau_{w}$ is sent to resonantly drive the WGM $a_{2}$ to enable a weak TMS interaction (equation (\ref{tms})). After a single photon is detected in a short interval $\tau_{d}$, the magnon mode is successfully added a single magnon. $iii$) Readout of magnonic non-Gaussian states. The single-MACS and -MATS can be read out by sending a weak probe field with duration $\tau_{r}$ to the microwave cavity and measuring the cavity output field, which again uses the cavity--magnon BS interaction. 
	
	Figures~\ref{fig4}(c)-(d) show the Mandel $Q_{M}$ parameter of the single-MACS versus initial coherent amplitude $|\beta|$ and of the single-MATS versus initial thermal occupation $\bar{n}_{0}$ for different values of $\mathcal{G}_{1} \tau_{w}$ (where $\mathcal{G}_{1}=G_{1}^{2}/\kappa_{1}$), respectively. A negative $Q_{M}<0$ represents the sub-Poissonian statistics of the state and is a signature of nonclassicality, since classical states, such as coherent and thermal states, correspond to Poissonian ($Q_{M}=0$) or super-Poissonian ($Q_{M}>0$) statistics. Figure \ref{fig4}(c) clearly shows that the Mandel $Q_{M}$ of the single-MACS is negative in the full range of $|\beta|$ and a small amplitude $|\beta|$ is preferred for seeing a strong sub-Poissonian feature. Moreover, the single-MACS exhibits quadrature squeezing when $|\beta|>1$, while the single-MATS has no any squeezing~\cite{Lu1}. Figure \ref{fig4}(d) shows that for the single-MATS, a negative $Q_{M}<0$ exists only for a small thermal occupation $\bar{n}_{0}$. As $\bar{n}_{0}$ increases, the state exhibits a smooth transition from the sub-Poissonian ($Q_{M}<0$) to the super-Poissonian ($Q_{M}>0$) statistics, which can be interpreted as a quantum-to-classical transition.
	Both the single-MACS and -MATS exhibit a negative Wigner function (figures \ref{fig4}(e)-(f)), which is an indicator of both nonclassicality and non-Gaussianity. The Wigner negativity of the single-MACS (MATS), defined as the volume $\delta$ of negative Wigner distributions in phase space, reduces as $|\beta|$ ($\bar{n}_{0}$) increases, as shown in the inset of figure \ref{fig4}(e) (\ref{fig4}(f)).

	\begin{figure}[t]
		\centering
		\includegraphics[width=0.9\textwidth]{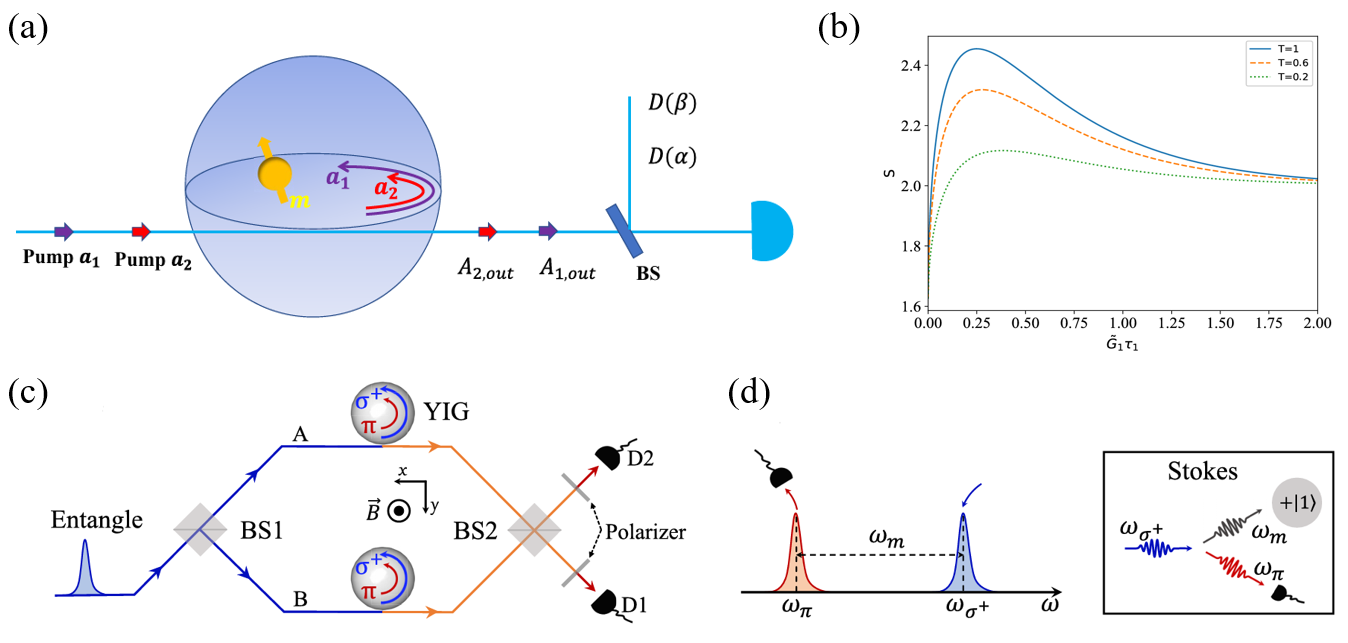}
		\caption{(a) Protocol of the optomagnonic Bell test. {The output fields of WGMs $a_{1}$ and $a_{2}$ are measured with an on-off photon detector preceded by a displacement, which can be realized by injecting a coherent field into a BS.} (b) Optimal values of $S$ over the measurement settings $\alpha_{1,2}$ and $\beta_{1,2}$ as a function of $\tilde{G}_{1}\tau_{1}$ for various magnon--photon state conversion efficiencies $T$. Reprinted figures (a)-(b) with permission from~\cite{Xie22} Copyright 2022 by the American Physical Society.
		(c) Schematic diagram of the optomagnonic DLCZ protocol. {A weak laser pulse with a certain polarization is sent into the interferometer and a subsequent single-photon detection (via $D_{1}$ or $D_{2}$) in the output of the interferometer projects the two magnon modes onto a path-entangled state.} (d) Mode frequencies associated with the optomagnonic Stokes scattering. A $\sigma^{+}$-polarized (TM-polarized) photon of frequency $\omega_{\sigma^{+}}$ is converted into a $\pi$-polarized (TE-polarized) Stokes photon of frequency $\omega_{\pi}$ by creating a magnon of frequency $\omega_{m}$. Reprinted figures (c)-(d) with permission from~\cite{WJA} Copyright 2021 by the American Physical Society.}
		\label{fig5}
	\end{figure}

	\subsection{Optomagnonic Bell test}\label{be}

	The Bell test~\cite{Bell} is a genuine test of nonclassicality without any quantum assumptions, which has been performed in various systems, such as photonic systems~\cite{Cla}, atomic ions~\cite{Win}, mesoscopic superconducting qubits~\cite{Martinis}, and macroscopic optomechanical systems~\cite{Bellop}. It provides foundational evidence against local hidden-variable theories, supporting the nonlocal nature of quantum mechanics, and serves as critical tools in quantum science and technology. The optomagnonic Bell test was first studied in Ref.~\cite{Xie22}, which proposes to violate the Clauser--Horne--Shimony--Holt (CHSH) inequality by utilizing entanglement between optical photons and magnons. As shown in figure \ref{fig5}(a), the first resonant pulse is used to drive the WGM $a_{2}$ to activate the optomagnonic Stokes scattering and generate the TMS interaction between the magnon mode and the WGM $a_{1}$ (equation (\ref{tms})). The magnonic state is subsequently mapped into the output anti-Stokes field of the WGM $a_{2}$ by sending the second pulse to resonantly driving the $a_{1}$ mode using the optomagnonic state-swap interaction (equation (\ref{bs})). Therefore, the quantum correlation shared between magnons and Stokes photons is mapped to that between Stokes photons and subsequent anti-Stokes photons in the outputs of the WGMs. The correlation of the photon--photon pairs is measured by an on-off photon detector, preceded by a displacement operation $D(\alpha)$ or $D(\beta)$ which can be implemented by injecting a coherent beam into a BS (figure \ref{fig5}(a)). This measurement process can be described by the positive-operator-valued measure with two orthogonal projection operators $P_{\alpha}=D(\alpha)|0\rangle\langle 0|D^{\dagger}(\alpha)=|\alpha\rangle\langle\alpha|$ with an outcome $+1$, and $Q_{\alpha}=\mathbb{I}-|\alpha\rangle\langle\alpha|$ with an outcome $-1$, so the observable of the system is described by $2P_{\alpha}-\mathbb{I}$.
	
	For the local hidden-variable model, the four correlation functions between pairs of measurements obey the CHSH	inequality
	\begin{equation}
		S=|E_{\alpha_{1}\beta_{1}}+E_{\alpha_{1}\beta_{2}}+E_{\alpha_{2}\beta_{1}}-E_{\alpha_{2}\beta_{2}}|\leqslant 2,
	\end{equation}
	where $E_{\alpha\beta}=\langle (2P_{\alpha}-\mathbb{I})\otimes(2P_{\beta}-\mathbb{I})\rangle$ is the correlation function of the two output fields of the WGMs $a_{1}$ and $a_{2}$. Violation of this inequality implies the nonlocal nature of the created magnon--photon pairs. Figure~\ref{fig5}(b) shows the optimal values of $S$ over the measurement settings $\alpha_{1,2}$ and $\beta_{1,2}$ as a function of $\tilde{G}_{1}\tau_{1}$ ($\tilde{G}_{1}=G_{1}^{2}/\kappa_{1}$ and $\tau_{1}$ is the first pulse duration) for various magnon-to-photon state conversion efficiencies $T$ ($T=1-e^{-2\tilde{G}_{2}\tau_{2}}$ with $\tilde{G}_{2}=G_{2}^{2}/\kappa_{2}$ and $\tau_{2}$ is the second pulse duration). Obviously, the violation of the CHSH inequality $S>2$ can be obtained with a proper choice of $\tilde{G}_{1}\tau_{1}$ and a high conversion efficiency $T$.

	\subsection{Optomagnonic DLCZ protocol}

	{The Duan--Lukin--Cirac--Zoller (DLCZ) protocol}~\cite{DLCZ} was first proposed to realize long-distance quantum communication with atomic ensembles based on linear optics operations, which finds broad applications in quantum teleportation, quantum cryptography, and the test of Bell inequalities. Until now it has been experimentally realized in various systems, such as atomic ensembles~\cite{Kimble05} and macroscopic mechanical resonators~\cite{Simon18}. The optomagnonic version of the DLCZ protocol was proposed in Ref.~\cite{WJA} using a system involving two massive YIG spheres placed in two arms of an optical interferometer formed by two $50/50$ BSs (figure \ref{fig5}(c)). First, a weak coherent pulse $|\alpha\rangle\simeq|0\rangle+\sqrt{p}|1\rangle$, where $p=|\alpha|^{2}\ll 1$ is the probability of the pulse being in the single-photon state, is sent into one input of the first BS, which with equal probability goes into one arm of the interferometer (the two arms are termed as paths $A$ and $B$) to couple to a certain $\sigma^{+}$-polarized (TM-polarized) WGM $a_{2}$ of the YIG sphere. Therefore, the two optomagnonic systems in the two paths are in the state
	\begin{equation}
		|\psi\rangle_{a_{2}}\simeq|00\rangle_{\sigma^{+}_{A}\sigma^{+}_{B}}+\sqrt{\dfrac{p}{2}}\left(|01\rangle_{\sigma^{+}_{A}\sigma^{+}_{B}}+|10\rangle_{\sigma^{+}_{A}\sigma^{+}_{B}}\right),
	\end{equation}
	where the subscript $\sigma^{+}_{j}$ ($j=A, B$) denotes the $\sigma^{+}$-polarized WGM $a_{2}$ in path $j$. Due to the optomagnonic interaction, the $\sigma^{+}$-polarized photon converts into a $\pi$-polarized (TE-polarized) Stokes photon entering the WGM $a_{1}$ by creating a magnon (figure \ref{fig5}(d)). After that, the optomagnonic systems  are in the state
	\begin{equation}\label{EPR}
		|\psi\rangle_{ma_{1}}\simeq|0000\rangle_{m_{A}m_{B}\pi_{A}\pi_{B}}+\sqrt{\dfrac{p}{2}}\left(|0101\rangle_{m_{A}m_{B}\pi_{A}\pi_{B}}+|1010\rangle_{m_{A}m_{B}\pi_{A}\pi_{B}}\right),
	\end{equation}
	where $|0101\rangle_{m_{A}m_{B}\pi_{A}\pi_{B}}$ denotes the coexistence of a magnon residing in the YIG sphere and a $\pi$-polarized Stokes photon in path $B$, and similarly for $|1010\rangle_{m_{A}m_{B}\pi_{A}\pi_{B}}$. The generated $\pi$-polarized Stokes photon, with equal probability in path $A$ or $B$, then enters the second $50/50$ BS. The polarizers in the outputs of the BS select the TE Stokes photon over the TM photons that failed to effectively activate the Stokes scattering. A single-photon detection ($D_{1}$ or $D_{2}$) in the output of the BS, which realizes the measurement $M_{\pm}=(|01\rangle_{\pi_{A}\pi_{B}}\pm|10\rangle_{\pi_{A}\pi_{B}})^{\dagger}/\sqrt{2}$, then projects the two magnon modes onto the state
	\begin{equation}
		|\psi'\rangle_{m}=\dfrac{1}{\sqrt{2}}\left(|01\rangle_{m_{A}m_{B}}\pm|10\rangle_{m_{A}m_{B}}\right).
	\end{equation}
	This is a path-entangled state of the two magnon modes sharing a single magnon excitation, and ``$\pm$" correspond to the detection of the TE photon in different outputs of the BS.

	\subsection{Optomagnonically induced entangled traveling fields}

\begin{figure}[t]
		\centering
		\includegraphics[width=0.95\textwidth]{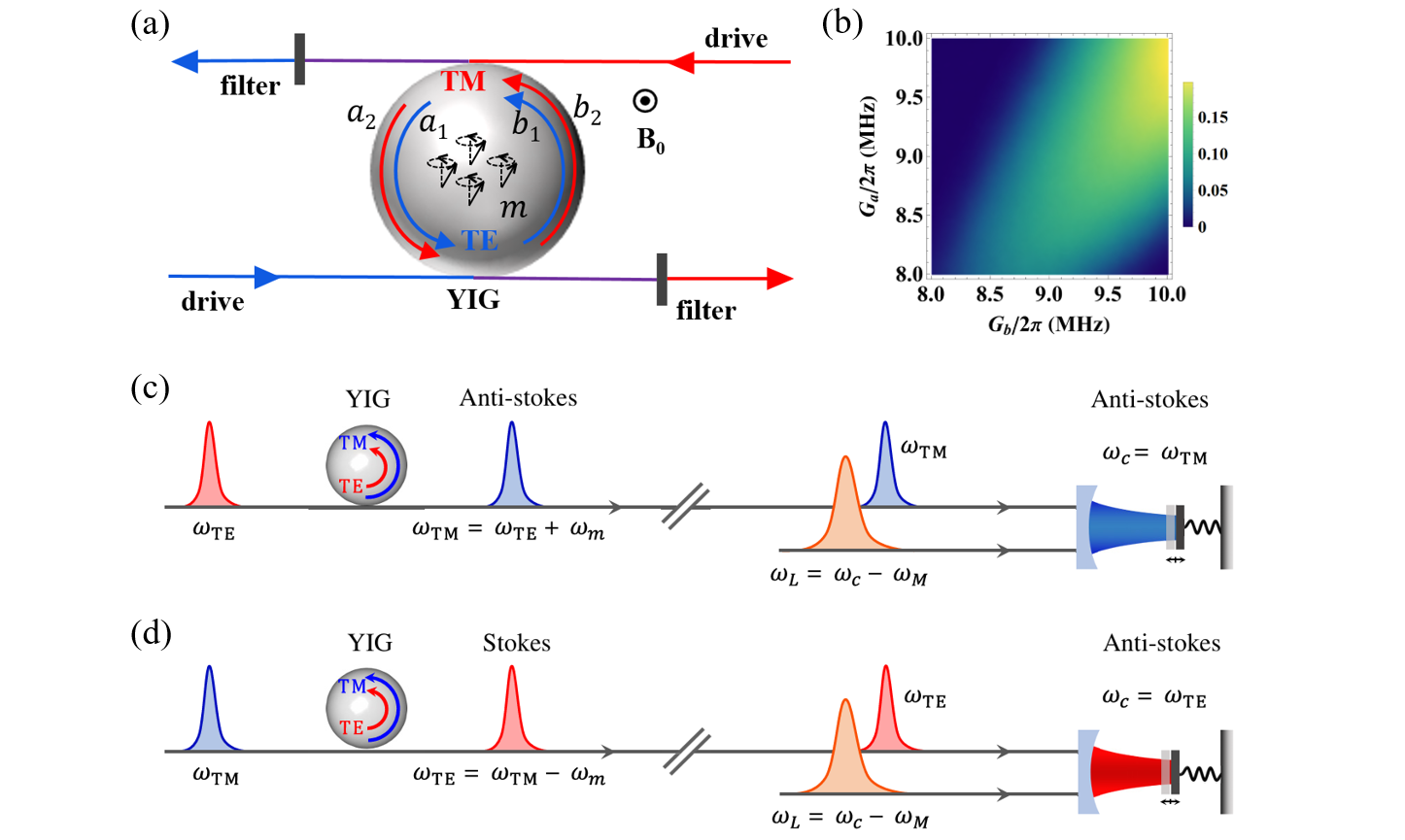}
		\caption{(a) Protocol of optomagnonically induced entanglement between two traveling fields. The system consists of a YIG micro disk, supporting a magnon mode and two pairs of WGMs, coupled to two fibers. The entangled Stokes and anti-Stokes output fields are extracted from pump fields by means of two filters. (b) Stationary traveling entanglement $E_{N}$ versus effective optomagnonic coupling rates $G_{a}$ and $G_{b}$. Reprinted figures (a)-(b) from~\cite{Lu3} licensed under the \href{https://creativecommons.org/licenses/by/4.0/}{CC BY 4.0}. (c)-(d) Quantum network protocol based on magnonic and mechanical nodes connected through a fiber. (c) The quantum state transfer protocol, exploiting the optomagnonic and optomechanical anti-Stokes scatterings, realizes a remote magnon-to-phonon state transfer. (d) The entanglement distribution protocol, exploiting the optomagnonic Stokes and optomechanical anti-Stokes scatterings, achieves nonlocal magnon--phonon entanglement. Reprinted figures (c)-(d) from~\cite{Jie21X} licensed under the \href{https://creativecommons.org/licenses/by/4.0/}{CC BY 4.0}.}
		\label{fig6}
	\end{figure}

	It is shown in section \ref{be} that the activation of optomagnonic Stokes and anti-Stokes scatterings in sequence can yield quantum-correlated Stokes and anti-Stokes output fields. Here, Ref.~\cite{Lu3} proposes an optomagnonic configuration that involves two pairs of WGMs, denoted as ($a_{1}$, $a_{2}$) and ($b_{1}$, $b_{2}$), and two pump fields (figure \ref{fig6}(a)). By simultaneously activating the two scattering processes involving a single magnon mode, this protocol can generate entangled Stokes and anti-Stokes output fields in the steady state and thus breaks the temporal constraint required in section \ref{be}. In one pair, a strong laser field is used to drive the TM-polarized WGM to activate the optomagnonic Stokes scattering, while in the other, the TE-polarized WGM is pumped and the anti-Stokes scattering is activated. The two laser fields are applied via two fibers coupled to a YIG micro sphere or disk. The effective Hamiltonian in the interaction picture is given by
	\begin{equation}\label{tra}
		H_{\rm eff}/\hbar=G_{a}\left(a_{2}m^{\dagger}+a_{2}^{\dagger}m \right)+G_{b} \left(b_{1}^{\dagger}m^{\dagger}+b_{1}m \right),
	\end{equation}
	with $G_{a}$ and $G_{b}$ being the effective optomagnonic coupling strengths. Given that the magnon mode simultaneously involves the above two scattering processes, the Stokes and anti-Stokes photons in the two WGMs $a_{2}$ and $b_{1}$ get entangled due to the mediation of the magnon mode. However, intracavity entanglement is inaccessible---what can be accessed is that of the output fields propagating in the fibers. Then two filters are used to define two output modes and extract the stationary entanglement shared between the Stokes and anti-Stokes output fields (figure \ref{fig6}(a)). To optimize traveling entanglement, the central frequencies of the filters should be resonant with the Stokes and anti-Stokes sidebands, and the filter bandwidth should be narrow~\cite{Lu3}.
	
	Figure \ref{fig6}(b) shows the stationary traveling entanglement $E_{N}$ versus the effective optomagnonic coupling rates $G_{a}$ and $G_{b}$. As explained above, the entanglement is a result of the simultaneous activation of both the Stokes and anti-Stokes scatterings. In the limit case of $G_{b}\gg G_{a}$, i.e., the TMS interaction being much stronger than the state-swap interaction, the former would generate strong entanglement between the modes $b_{1}$ and $m$. However, due to the slow state swapping rate between the modes $a_{2}$ and $m$, the created entanglement cannot be efficiently transferred to the two WGMs $b_{1}$ and $a_{2}$, leading to weak entanglement. In the opposite case of $G_{a}\gg G_{b}$, although the state swapping rate between $a_{2}$ and $m$ is fast, a weak TMS interaction creates weak entanglement between $m$ and $b_{1}$, which also gives weak entanglement. The trade-off between these two interactions yields an optimal situation of two close coupling rates $G_{a} \simeq G_{b}$ (figure \ref{fig6}(b)).
	
	{Another work~\cite{Xieoe} also proposes to achieve optomagnonically induced optical entanglement, where only one pair of WGMs is adopted. By simultaneously activating the Stokes and anti-Stokes scatterings, the TMS and BS interactions are realized, generating entanglement between the two WGMs in one pair. Although this protocol is simpler than the protocol~\cite{Lu3} employing two pairs of WGMs, it cannot entangle two optical fields with a tunable frequency difference, especially much larger than the magnon frequency.}

	\subsection{Quantum networks}

	Quantum networks~\cite{QN}, as a physical system in which the nodes function as interconnected components that interact with one another through quantum channels, require realizing distant quantum state transfer and nonlocal entanglement distribution among quantum nodes, and provide vast opportunities across a range of intellectual and technical frontiers, including quantum computation, communication and metrology. Until now there are several protocols~\cite{Jie21X,Zhao23,Tan25,WuY} for constructing quantum networks with magnons as quantum nodes to realize deterministic multipartite magnonic entanglement~\cite{Tan25} or asymmetric quantum state transfer~\cite{WuY}. Reference~\cite{Jie21X} proposes a quantum network with magnonic and mechanical resonators as quantum nodes and they are connected by light, which is an optimal carrier for transmitting quantum information over a long distance. 
	
	In the first part, the protocol realizes a distant magnon-to-phonon quantum state transfer, with the aim of solving the problem of the currently short-lived magnonic quantum states (figure \ref{fig6}(c)). Assume that the optomagnonic and mechanical systems are in an initial state
	\begin{equation}\label{net1}
		\rho_{m,2,b}(0)=\sum_{n,s=0}^{\infty}c_{n,s}|n\rangle\langle s|_{m}\otimes|0\rangle\langle 0|_{2}\otimes|0\rangle\langle 0|_{b}.
	\end{equation}
	Here, to be generic, the magnon mode is assumed in an arbitrary state, which can be expanded in the Fock-state basis with arbitrary real coefficients $c_{n,s}\geqslant 0$, i.e., $\rho_{m}(0)=\sum_{n,s=0}^{\infty}c_{n,s}|n\rangle\langle s|_{m}$, and the WGM $a_{2}$ and mechanical mode $b$ are in the vacuum state. By using a laser pulse to activate the anti-Stokes scatterings, i.e., the BS interaction (equation (\ref{bs})) between the magnon mode $m$ and WGM $a_{2}$, an arbitrary magnonic state can be transferred to the anti-Stokes output field of the WGM $a_{2}$, i.e., the system is in the state
	\begin{equation}
		\rho_{m,2,b}(\tau_{2})=\sum_{n,s=0}^{\infty}c_{n,s}(-i)^{n}i^{s}S^{(n+s)/2}|0\rangle\langle 0|_{m}\otimes|n\rangle\langle s|_{2}\otimes|0\rangle\langle 0|_{b},
	\end{equation}
	where $S=1-e^{-2\mathcal{G}_{2}\tau_{2}}$ (with $\mathcal{G}_{2}=G_{2}^{2}/\kappa_{2}$ and pulse duration $\tau_{2}$) is the optomagnonic state conversion efficiency ($0<S<1$). Then the anti-Stokes output field transmits, through a fiber, to a distant optomechanical system and acts as a resonant input field into the optomechanical cavity. Meanwhile, the cavity is driven by another red-detuned pulse, which activates the optomechanical state-swap interaction between the cavity field and the mechanical resonator~\cite{RMP14}, leading the system to the state
	\begin{equation}\label{net2}
		\rho_{m,2,b}=\sum_{n,s=0}^{\infty}c_{n,s}(SW)^{(n+s)/2}|0\rangle\langle 0|_{m}\otimes|0\rangle\langle 0|_{2}\otimes|n\rangle\langle s|_{b},
	\end{equation}
	with $W$ being the optomechanical state conversion efficiency ($0<W<1$). Therefore, after two state-swap operations (magnon-to-photon and photon-to-phonon), an arbitrary magnon state is successfully transferred to the mechanical mode that has a much longer lifetime and can act as a quantum memory for storing short-lived magnonic quantum states. Comparing equations (\ref{net1}) and (\ref{net2}), the fidelity in the magnon--photon--phonon state transfer depends on the product of the two conversion efficiencies $S$ and $W$ in the two anti-Stokes processes.
	
	In the second part, the protocol realizes nonlocal entanglement distribution between the magnonic and mechanical nodes (figure \ref{fig6}(d)). Now two laser pulses are sent to successively activate the optomagnonic Stokes scattering and the optomechanical anti-Stokes scattering. The former generates an entangled state of the magnons and the Stokes field, which transmits to the distant optomechanical cavity, and the latter maps the state of the Stokes field to the mechanical resonator. Therefore, through the traveling Stokes field, nonlocal entanglement between remote magnonic and mechanical nodes is established. The degree of the entanglement increases with the strength of the optomagnonic TMS interaction and the optomechanical state conversion efficiency.

	\subsection{Optomagnonic quantum teleportation}
	
	Quantum teleportation~\cite{Bennett93,Zeil,Kim,Simon21} describes the transfer of an unknown input state onto a remote quantum system without physical transfer of the carrier of quantum information, which is an essential component for quantum repeaters and distributed quantum computing. Recent proposals ~\cite{Fan23,Lu2,Yuan26} indicate the possibility of realizing quantum teleportation in magnonic systems through either discrete variables (DV) or continuous variables (CV). Below we review two proposals for realizing the DV and CV quantum teleportation in cavity optomagnonics.

	\subsubsection{DV quantum teleportation}

\begin{figure}[t]
		\centering
		\includegraphics[width=\textwidth]{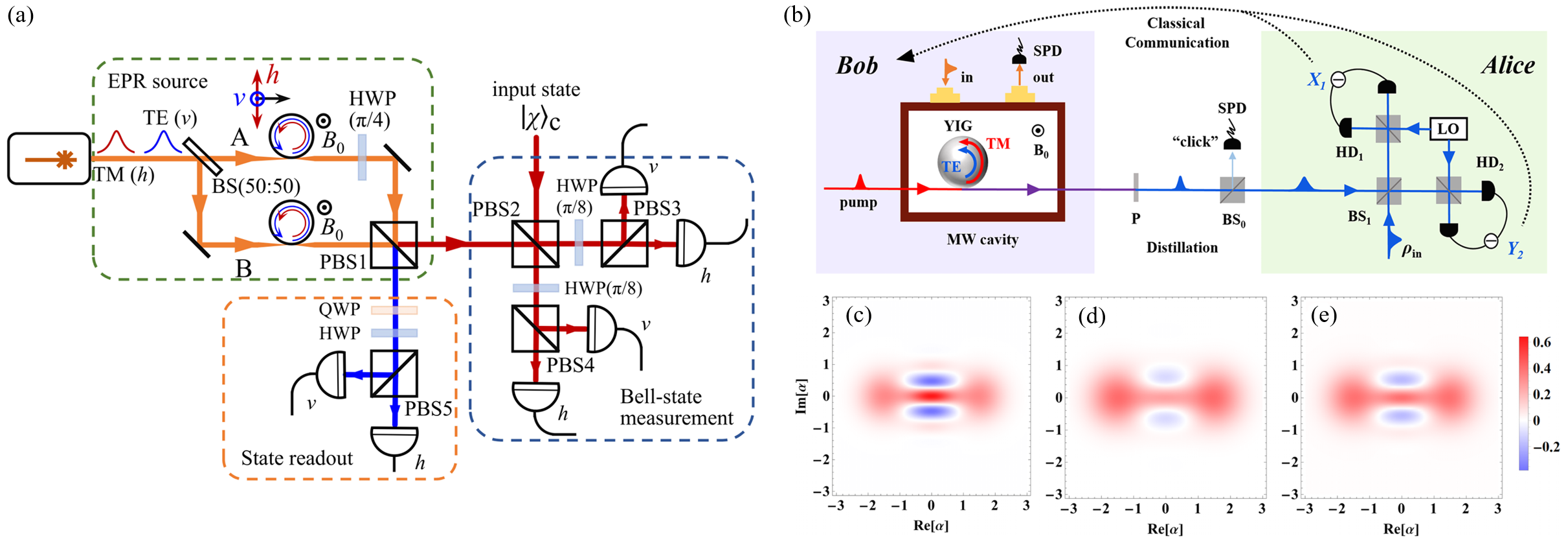}
		\caption{(a) Schematic diagram of the optomagnonic DV quantum teleportation based on an optical interferometer with a YIG sphere in each arm. The protocol contains three steps: generation of the optomagnonic EPR state, Bell-state measurement, and readout of the teleported magnonic state. PBS denotes polarizing beam splitter, HWP means half-wave plate, and QWP stands for quarter-wave plate. Note that in~\cite{Fan23}, the definition of TE- and TM-polarized modes is contrary to this review. Reprinted figure (a) from~\cite{Fan23} licensed under the \href{https://creativecommons.org/licenses/by/4.0/}{CC BY 4.0}. (b) Schematic diagram of the optomagnonic CV quantum teleportation. The protocol contains four steps as detailed in the text. (c)-(e) Wigner function of (c) the initial cat state $\rho_{\rm in}$, and of the teleported magnon state with (d) an undistilled entangled state or (e) a distilled non-Gaussian entangled state. Reprinted figures (b)-(e) from~\cite{Lu2} licensed under the \href{https://creativecommons.org/licenses/by/4.0/}{CC BY 4.0}.}
		\label{fig7}
	\end{figure}

	Reference~\cite{Fan23} proposes an optomagnonic DV quantum teleportation (figure \ref{fig7}(a)) that can transfer an arbitrary photonic qubit state onto a magnonic system consisting of two optomagnonic devices placed in two arms of an optical interferometer (the so-called dual-rail encoding~\cite{Jie20A}). A coherent laser pulse is sent into one input port of the first $50/50$ BS to activate the optomagnonic Stokes scattering. For simplicity, the protocol assumes at most single excitations in the optomagnonic devices. This is the case of using a weak coherent pulse, where the probability of creating higher-than-one-excitation states in the Stokes scattering is negligible (equation (\ref{EPR})). Since the vacuum component in equation (\ref{EPR}) will not trigger any coincidence in the Bell-state detection, leading to unsuccessful trials for the teleportation, by selecting trials with successful scattering events, the Stokes scattering prepares an optomagnonic Bell state
	\begin{equation}
		|\psi\rangle_{ma_{1}}=\dfrac{1}{\sqrt{2}}\left(|H\rangle_{a_{1}}|L\rangle_{m}+|V\rangle_{a_{1}}|U\rangle_{m}\right),
	\end{equation}
	where $|H\rangle_{a_{1}}$ ($|V\rangle_{a_{1}}$) denotes the generated Stokes photon in the horizontal (vertical) polarization in the lower path B (the upper path A), and $|L\rangle_{m}$ ($|U\rangle_{m}$) represents the generated single magnon in the corresponding path. Then an input photonic qubit state $|\chi\rangle_{c}=\alpha|H\rangle_{c}+\beta|V\rangle_{c}$, with complex coefficients $\alpha$ and $\beta$ satisfying $|\alpha|^{2}+|\beta|^{2}=1$, is injected into one input of the PBS2 {(PBS: polarizing beam splitter)}, and meanwhile, the output of the PBS1 enters the other input port (figure \ref{fig7}(a)). Thereby, the joint state before the Bell-state measurement is
	\begin{equation}
		|\psi\rangle_{ma_{1}}\otimes|\chi\rangle_{c}=\dfrac{1}{\sqrt{2}}\left(\alpha|H\rangle_{a_{1}}|H\rangle_{c}|L\rangle_{m}+\beta|H\rangle_{a_{1}}|V\rangle_{c}|L\rangle_{m}+\alpha|V\rangle_{a_{1}}|H\rangle_{c}|U\rangle_{m}+\beta|V\rangle_{a_{1}}|V\rangle_{c}|U\rangle_{m}\right).
	\end{equation}
	The Bell-state measurement, e.g., $M=(|H\rangle_{a_{1}}|H\rangle_{c}+|V\rangle_{a_{1}}|V\rangle_{c})^{\dagger}/\sqrt{2}$, which is performed onto the input photonic qubit state and the output Stokes photon from the interferometer, projects the two magnon modes onto the state
	\begin{equation}
		|\psi'\rangle_{m}=\alpha|L\rangle_{m}+\beta|U\rangle_{m},
	\end{equation}
	which indicates the successful teleportation of the input photonic qubit state $|\chi\rangle_{c}$ to a dual-rail encoding magnonic system. The teleported magnonic state $|\psi'\rangle_{m}$ can be read out by activating the optomagnonic BS interaction~\cite{Fan23}.

	\subsubsection{CV quantum teleportation}

	Compared with the DV teleportation, which is, by its nature probabilistic, the CV teleportation is deterministic. Moreover, the CV teleportation can transfer more information benefitting from the capability of the CV system to encode more information in an infinite-dimensional Hilbert space. Reference~\cite{Lu2} provides the first optomagnonic CV quantum teleportation protocol and it also shows how the protocol can be enhanced by non-Gaussian distillation operations (figure \ref{fig7}(b)). 
	
	The protocol contains four steps: $i$) Creating the photon--magnon entanglement. A strong optical pulse is used to pump the WGM $a_{2}$ to activate the optomagnonic Stokes scattering, yielding entanglement between the magnon mode and TE-polarized Stokes field, which reaches a low-reflectivity beam splitter ($\mathrm{BS}_{0}$) after passing through a polarizer (P). $ii$) Distilling the entanglement with non-Gaussian operations, explicitly the single-magnon and -photon subtraction. A single photon is subtracted from the Stokes field when the single-photon detector (SPD) clicks at the reflection port of $\mathrm{BS}_{0}$. Then the single-photon-subtracted Stokes field is mixed with an input optical state $\rho_{\rm in}$ at a $50/50$ beam splitter ($\mathrm{BS}_{1}$). On the other side, a single magnon is also subtracted from the magnon mode by sending a weak pulse into the microwave cavity and conditioned on the detection of a single photon in the output port.  $iii$) Performing the homodyne detection (HD) and magnon displacement operation. Alice subsequently performs the HD on the two output fields of the $\mathrm{BS}_{1}$ to measure a pair of quadratures $X_{1}$ and $Y_{2}$, and she then tells the results to Bob via classical communication. Based on the results $X_{1}$ and $Y_{2}$,  Bob implements a displacement operation onto the magnon mode by sending another resonant pulse into the microwave cavity, which completes the teleportation. $iv$) Reading out the final magnon state. The magnon state can be read out by sending a weak microwave pulse into the microwave cavity and the cavity--magnon state-swap interaction maps the magnon state to the cavity output field, from which the state can be reconstructed by performing microwave tomography.
	
	Figures~\ref{fig7}(c)-(e) show the Wigner functions of the initial optical cat state $\rho_{\rm in}$ and the teleported magnon state with the shared Gaussian entanglement or distilled non-Gaussian entanglement. Although both the magnonic cat states in figures~\ref{fig7}(d)-(e) show unambiguous negativity and interference fringes, the teleported state with non-Gaussian distillation is of higher fidelity to the initial cat state. Moreover, the protocol unveils a powerful capability for preparing diverse magnonic quantum states, such as single-magnon, squeezed, and cat states, by exploiting the photon-to-magnon CV quantum teleportation~\cite{Lu2}.

	\subsection{Microwave quantum illumination}

\begin{figure}[t]
		\centering
		\includegraphics[width=0.95\textwidth]{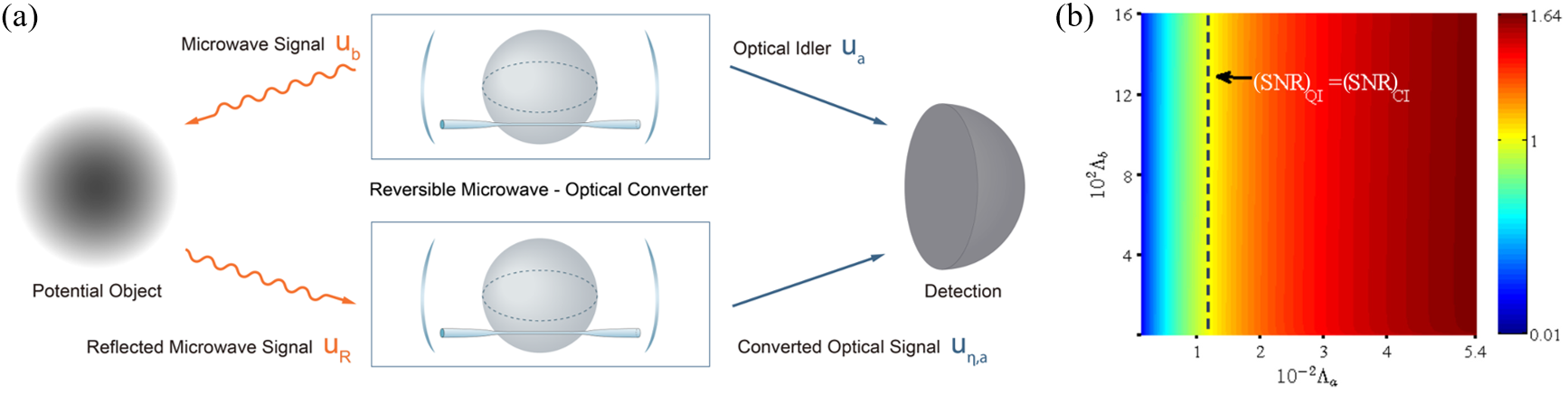}
		\caption{(a) Schematic diagram of microwave quantum illumination based on a microwave-optomagnonic system. A YIG sphere is coupled to a microwave cavity and an optical nanofiber, which serves as a magnonic reversible microwave-optical converter and generates entangled microwave-optical fields. The receiver collects the reflected microwave signal and upconverts it into the optical frequency. Then, the converted optical signal together with the retained optical idler is fed into the detector for measuring the presence or absence of the target. (b) The ratio $(\rm SNR)_{\rm QI}/(\rm SNR)_{\rm CI}$ versus optomagnonic cooperativity $\Lambda_{a}=G_{1}^{2}/(\kappa_{2}\kappa_{m})$ and cavity--magnon cooperativity $\Lambda_{b}=g_{\rm ma}^{2}/(\kappa_{a}\kappa_{m})$. Reprinted figures (a)-(b) with permission from~\cite{Zhou21} Copyright 2021 by the American Physical Society.}
		\label{fig8}
	\end{figure}

	Quantum illumination~\cite{QI} refers to a quantum sensing protocol which uses entangled signal--idler photon pairs to enhance the detection efficiency of low-reflectivity objects that are immersed in thermal noisy environments. Reference~\cite{Zhou21} proposes a scheme based on cavity optomagnonics to perform microwave quantum illumination (figure \ref{fig8}(a)). First, magnons as the intermediary serve to entangle optical and microwave fields by using the optomagnonic Stokes scattering and magnon--cavity BS interaction. The microwave field, as a signal, is sent to probe the target, and the entangled optical field is retained as an idler field. Then, the reflected microwave signal is collected by another reversible microwave-optical converter and
	upconverted into the optical domain. The optical signal is fed into the detector together with the idler field for target detection. The quantum-enhanced sensitivity can be estimated by the signal-to-noise ratio of the system  $(\rm SNR)_{\rm QI}$. In figure \ref{fig8}(b), the advantage of using optomagnonic system over the classical coherent-state microwave radar is reflected in the ratio $(\rm SNR)_{\rm QI}/(\rm SNR)_{\rm CI}$, where $(\rm SNR)_{\rm CI}$ is the signal-to-noise ratio in the classical illumination. It shows that the quantum-enhanced sensitivity increases with the optomagnonic cooperativity $\Lambda_{a}$.
	
	\subsection{Others}
	
	Besides the quantum-related protocols discussed above, cavity optomagnonics finds many other potential applications. These include nonreciprocal photon--magnon blockade~\cite{block1,block2,block3}, where the magnon-induced BLS between photons and magnons suppresses the excitation of a second quantum in a direction-dependent manner---blockade occurs in one propagation direction but not the reverse---thus enabling chiral quantum networks and nonreciprocal single-photon (magnon) devices; all-optical polarization-state engineering~\cite{Ying23}, where magnon-mediated polarization conversion with broken time-reversal symmetry allows the output polarization to be mapped onto the entire Poincar\'{e} sphere purely by tuning an external driving laser; and ultrahigh nonreciprocal optical transmission through rotational Sagnac effect~\cite{Zhong}.


	\section{Magnon--spin hybrid system}\label{mag-spin}
	{
	In hybrid quantum architectures, superconducting qubits offer fast operations and mature microwave control but face scalability challenges, while solid-state spin defects, such as nitrogen-vacancy (NV) centers, provide long coherence times and nanoscale addressability, but suffer from weak direct coupling due to their small magnetic dipole moments. Magnons in nanoscale ferromagnets, with high spin density, low dissipation, and subwavelength mode volumes, can concentrate microwave magnetic fields and act as quantum intermediaries that coherently connect superconducting qubits, NV centers, microwave and optical photons, and acoustic phonons (figure~\ref{hybrid}), thus complementing these platforms~\cite{Neuman2020,Hei2021,Xiong2022}.
	The coupling between magnons and solid-state spins is mediated by the magnetic-dipole coupling between the localized spin qubit and the magnetic stray field generated by the magnon mode of a nearby YIG nanosphere~\cite{Neuman2020,Hei2021,Xiong2022}.}
	This novel coupling mechanism opens up new possibilities for constructing hybrid quantum systems that combine the long coherence time of solid-state spins with the unique advantages of magnonic systems, such as high spin density and great tunability via external magnetic fields~\cite{Neuman2020}.

	Solid-state spin defects, particularly the NV center in diamond, have emerged as promising qubit candidates owing to their atom-like long coherence times, high-fidelity optical initialization and readout, and compatibility with nanostructured environments~\cite{zhao2025magnon,Kumar25}.  {As a prominent magnetic dipolar emitter, the NV center features a spin-1 ground state with a zero-field splitting of \(D_0 \approx 2\pi \times 2.87\)~GHz. By applying a static magnetic field \(B_z\) along the NV symmetry axis, the degeneracy of the \(m_s = \pm 1\) states is lifted via the Zeeman effect, allowing one to select the \(m_s = 0\) and \(m_s = -1\) levels to form an effective two-level spin qubit. Despite these advantages, constructing a multiphysics interface for individual spins at the single-quantum level remains an outstanding challenge, primarily because the intrinsic magnetic dipole moment of a single spin is too weak to achieve strong direct coupling to other quantum systems.} Magnons, however, provide a natural and compelling solution to this coupling barrier. Owing to the high spin density and low dissipation rate of nanoscale YIG structures, magnonic excitations can confine the microwave magnetic field deeply into subwavelength mode volumes, thereby dramatically enhancing the local magnetic environment experienced by a nearby spin~\cite{Neuman2020,Xiong2022}. This unique capability positions the magnon as an ideal quantum intermediary, enabling strong and coherent magnon--spin interactions at the single-quantum level through physically realizable manners. The resulting hybrid magnon--spin platform not only overcomes the long-standing weak-coupling bottleneck of individual solid-state spins, but also opens up an appealing pathway toward scalable quantum information processing based on magnonic interfaces.

	\begin{figure}[t]
		\centering
		\includegraphics[width=0.95\linewidth]{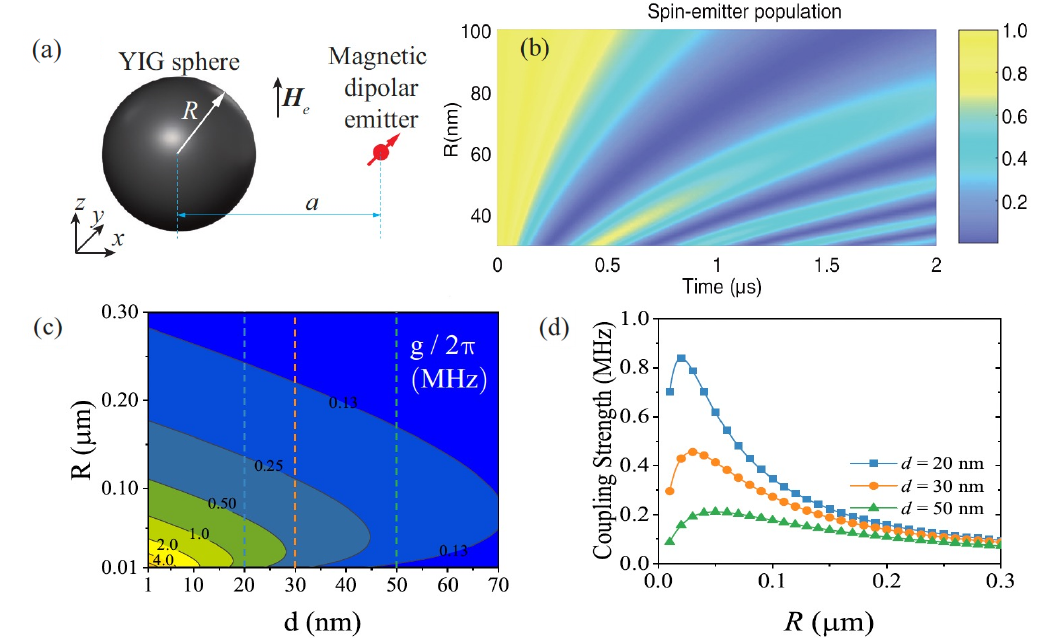}
		\caption{\label{fig:f1}{(a) Schematic of the interaction between magnons and magnetic dipolar emitters. (b) {Time evolution of the spin excited-state population \(\lvert c_e \rvert^2\) as a function of the sphere radius. The spin and magnon are in resonance.} Reprinted figure (b) with permission from~\cite{Neuman2020} Copyright 2020 by the American Physical Society. (c) and (d) Coupling strength $g$ between the individual NV center spin and the Kittel mode of the YIG sphere versus distance $d$ and radius $R$. Reprinted figures (c)-(d) with permission from~\cite{Hei2021} Copyright 2021 by the American Physical Society.}}
	\end{figure}

	\subsection{Theory of the magnon--spin coupling}

	The magnon--spin coupling is considered in a model in which a magnetic dipolar emitter is positioned at a distance $d=a-R$ from the surface of a uniformly magnetized YIG nanosphere of radius $R$ (figure \ref{fig:f1}(a))~\cite{Neuman2020}. Magnon modes in a magnetic sphere generate a stray magnetic field in the surrounding space. When subjected to this field, a magnetic dipolar emitter (with transition frequency  $\omega_0$) undergoes a spin flip---a process that is governed by the Zeeman effect. To clarify the mechanism, one can employ the Weisskopf-Wigner approach to obtain the full non-Markovian dynamics of the excited state. The excited-state coefficient ($c_e$) satisfies the following integro-differential equation:
	\begin{equation}
		\dot{\tilde{c}}_e = -\int_0^t \int_{-\infty}^{\infty} J(\omega) \, e^{i(\omega_0 - \omega)(t - t')}  d\omega  \tilde{c}_e(t')  dt' ,
	\end{equation}
	with $ \tilde{c}_e(t') e^{-i\omega_0 t'} = c_e(t') $ and
	\begin{eqnarray}
		J(\omega) = \frac{\mu_0 |\mu_B|^2}{\hbar \pi}  k_0^2 \left( \operatorname{Im}\{[G_m]_{xx} + [G_m]_{yy}\}  + \operatorname{Re}\{[G_m]_{xy} - [G_m]_{yx}\} \right).  \nonumber
	\end{eqnarray}
	Here, {$\mu_0$ is the vacuum permeability, $k_0$ is the free-space wavenumber}, and the spectral density $ J(\omega) $ depends on the magnetic Green's tensor \( G_m(\mathbf{r}, \mathbf{r}') \). This tensor characterizes the magnetic response field generated at a distant point \(\mathbf{r}\) by a magnetic point dipole \(\mathbf{m}_{r'}\) located at \(\mathbf{r}'\) and oscillating harmonically at frequency \(\omega\). Furthermore, the Green's tensor can be expanded into a superposition of magnon modes using a complete set of solid harmonic functions (Walker modes). Within a specific range of the bias magnetic field, the frequency of the Kittel mode is far-detuned from those of other modes, ensuring negligible spectral overlap. When the emitter and the Kittel mode are resonant and the geometric factor is set to be $a = 1.2R$, numerical results reveal that the temporal evolution of the emitter's excited-state population exhibits clear Rabi flopping~\cite{Neuman2020}. Such oscillatory dynamics constitute unambiguous evidence for the coherent magnon--spin coupling, and the Rabi oscillations gradually become less pronounced as the radius of the magnetic sphere increases (figure \ref{fig:f1}(b)).

	{Take NV centers as an example.}
	The YIG sphere supports a uniform precession mode, i.e., the Kittel mode, whose magnetization generates a magnetic field felt by the NV spin.
	Under the RWA, the magnon--spin interaction takes the JC form:
	\begin{equation}\label{eq:JC}
		{H}_{\text{int}} = -\hbar g \left({s}_K {\sigma}^+ + {s}_K^\dagger {\sigma}^- \right),
	\end{equation}
	where ${s}_K$ (${s}_K^\dagger$) is the annihilation (creation) operator of the Kittel mode and ${\sigma}^+ = |\!-\!1\rangle\langle 0|$ (${\sigma}^- = |0\rangle\langle -1|$) is the spin raising (lowering) operator. This Hamiltonian describes a coherent excitation exchange between the magnon and the spin. The coupling strength $g$ originates from the overlap integral of the magnetic field of the Kittel mode with the magnetic dipole moment of the NV center. Based on the canonical quantization of the magnetostatic field, an analytical expression of $g$ was derived~\cite{Hei2021}:
	\begin{equation}\label{eq:g_coupling}
		g = \sqrt{\frac{|\gamma| M_s}{24 \pi \hbar}} \frac{g_e \mu_0 \mu_B R^{3/2}}{(R+d)^3},
	\end{equation}
	where $\gamma$ is the gyromagnetic ratio, and $M_s$ is the saturation magnetization.
	This formula reveals the strong geometrical dependence of the coupling---it initially increases with the volume of the sphere ($\propto R^{3/2}$) but quickly reduces with the spin-surface separation ($\propto (R+d)^{-3}$).
	As shown in figures~\ref{fig:f1}(c)-(d), for a nanosphere with $R\sim 50$ nm and a small gap of $d\sim 10$ nm, $g/2\pi$ is estimated to reach the order of 1 MHz, which can exceed the typical dissipation rates of the magnon (in the YIG) and the NV spin, thus entering the strong-coupling regime.

	\subsection{Quantum protocols based on the magnon--spin coupling}

\begin{figure}[t]
		\centering
		\includegraphics[width=0.95\linewidth]{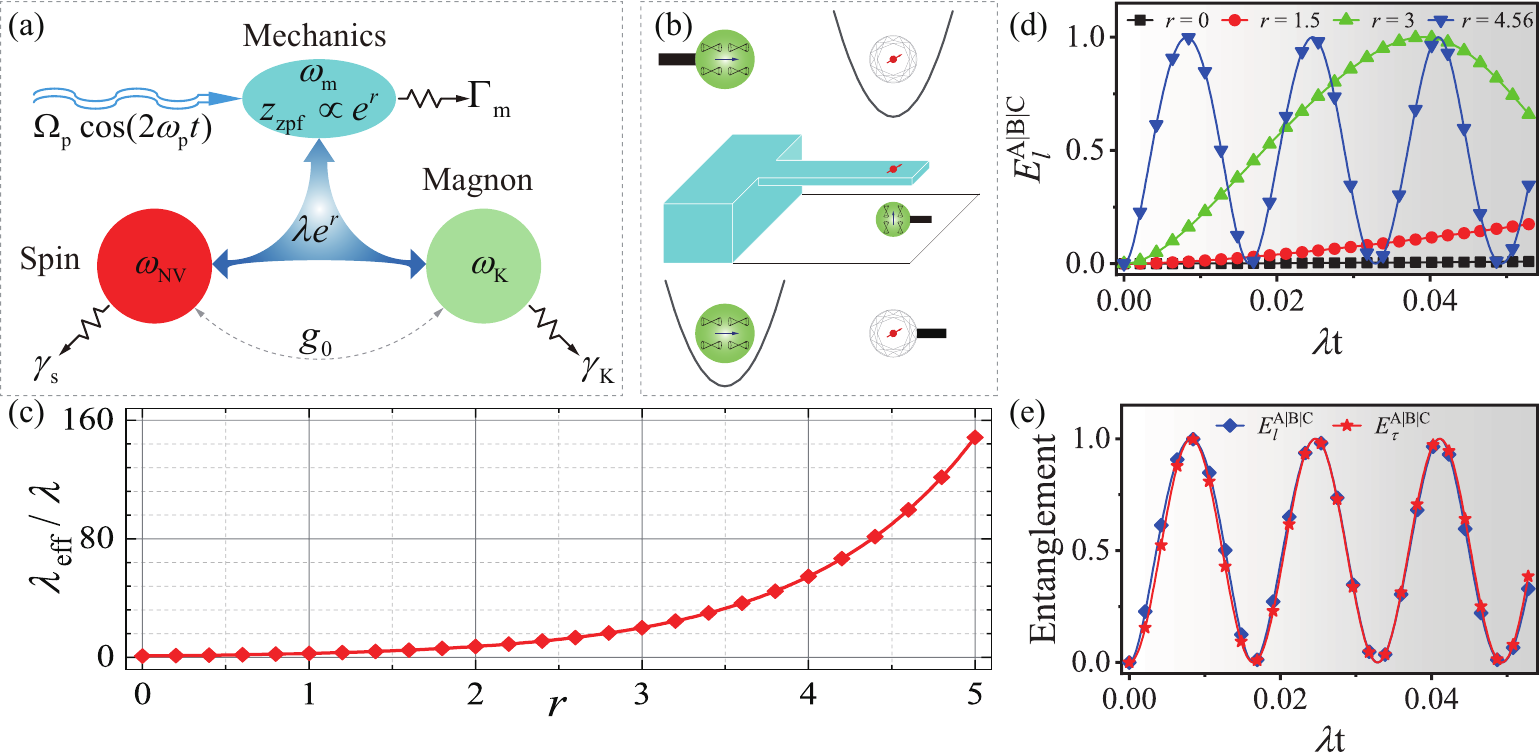}
		\caption{\label{fig:f2}{(a) Spin--magnon--phonon tripartite coupling. The spin qubit, the magnon mode, and the phonon mode are simultaneously coupled, with an enhanced tripartite coupling  via a two-phonon driving. (b) Three possible configurations realizing the proposal: a diamond particle with a single NV spin in an electrical trap (top); an NV center embedded in a cantilever (middle);  and a YIG microsphere levitated in a magnetic trap (bottom). (c) Enhanced effective tripartite coupling $\lambda_{\mathrm{eff}}/\lambda$ versus squeezing parameter $r$ associated with the parametric drive. (d) Genuine tripartite entanglement (minimum residual contangle) versus time for different values of $r$.  (e) Two measures of tripartite entanglement, minimum residual contangle and three-tangle, versus time. Reprinted figures (a)-(e) with permission from~\cite{Hei2023} Copyright 2023 by the American Physical Society. } }
	\end{figure}

	We now introduce a series of quantum protocols with such a magnon--spin coupling as a core component.  Reference~\cite{Hei2023} predicts a direct tripartite coupling in a hybrid system consisting of an NV spin, a magnon mode ($m$), and a mechanical mode ($b$), taking the form $\lambda ({b} + {b}^\dagger)({m}^\dagger {\sigma}^- + {m} {\sigma}^+)$, with   the tripartite coupling strength $\lambda$ (figure~\ref{fig:f2}(a)). The protocol is general in the sense that it applies to different coupling structures, as schematically shown in figure~\ref{fig:f2}(b).  When applying a parametric drive, the tripartite coupling strength can be exponentially enhanced (figure~\ref{fig:f2}(c)), which facilitates the generation of genuine tripartite entanglement among the spin, magnon, and phonon degrees of freedom (figures~\ref{fig:f2}(d)-(e)).
	In terms of scalable quantum architectures, Ref.~\cite{Hei2024b} demonstrates that an array of YIG nanospheres, coupled via a common microwave cavity and subjected to periodic magnetic field modulations, can simulate the Su--Schrieffer--Heeger (SSH) model within the magnonic domain.
	When NV spins are coupled to this topological magnon chain, the magnon-mediated spin--spin interactions become chiral and tunable, leading to the formation of unidirectional bound states and nonreciprocal entanglement between distant spins.
	Including the mechanical degree of freedom, e.g., the {center-of-mass motion} of a levitated micromagnet, Ref.~\cite{Pan2023} proposes a scheme to significantly enhance the spin-mechanical coupling strength between an NV spin and a levitated micromagnet. A driving electrical current is used to modulate the mechanical motion of the micromagnet, which induces a two-phonon drive and can exponentially enhance the spin--phonon and phonon-mediated spin--spin coupling strengths, enabling the preparation of mechanical Schr\"{o}dinger cat states and high-fidelity geometric quantum gates.
	

	An important application of the spin--magnon hybrid is the realization of strong coupling between a single spin and microwave photons via the magnon mode~\cite{Hei2021}. In the dispersive regime, the magnon mode can be adiabatically eliminated, yielding a single-quantum-level effective spin--photon interaction, of which the strength can exceed the direct spin--photon coupling strength by orders of magnitude and overcome both the spin decoherence and cavity dissipation~\cite{Hei2021}. This reveals the role of magnons as an efficient quantum transducer for realizing coherent energy exchange between hybrid systems.
	More importantly, the magnon interface can be used to realize a long-range strong coupling between two distant spin qubits. The mechanism is exemplified in the nanomagnonic cavity scheme~\cite{Neuman2020}: two NV spins that are positioned on opposite sides of a YIG nanosphere and detuned from the Kittel mode interact via a virtual-magnon transition, thus mitigating the intrinsic losses of the magnon. The magnon-mediated effective coupling can be about three orders of magnitude stronger than the direct dipole-dipole coupling and can exceed the spin decoherence rate, leading to high-fidelity periodic exchange of their quantum states. This concept was extended later in Ref.~\cite{Xiong2022} by exploiting the magnon Kerr effect, where the amplified coupling allows the inter-spin separation to reach the micrometer scale while maintaining a strong effective spin--spin coupling up to tens of kHz. This enables quantum state transfer between the two spins and the realization of the two-qubit iSWAP gate.
	The controllable spin--spin interaction mediated by magnons provides a powerful means for generating long-range entanglement of spin qubits. Along this line, Ref.~\cite{Fukami21} adopts a YIG waveguide structure to establish magnon‑mediated long-range entanglement between two NV centers, and predicts a strong spin--spin coupling over micrometer distances with cooperativities exceeding unity. It also systematically compares on-resonant transduction and off-resonant virtual-magnon exchange protocols under cryogenic conditions ($T < 150$ mK), and outlines a practical roadmap for realizing solid-state spin entanglement via the magnonic platform. Besides, in the topological magnon chain protocol~\cite{Hei2024b}, both the driving parameter $\eta$ in the effective SSH model and the spin--magnon coupling sites modulate the strength and chirality of the emergent spin--spin coupling.
	
	As a conceptual extension, Ref.~\cite{Hei2024a} incorporates the skyrmion qubit into magnon-based hybrid quantum systems as a distinct qubit platform built on quantized skyrmion helicity. The proposal is conceptually analogous to the spin--magnon coupling, as both rely on a magnon mode interacting coherently with a quantized degree of freedom (i.e., the spin or qubit).  The strong coupling between the magnon mode of the micromagnet and the skyrmion qubit follows the same physical principle as discussed above. Consequently, the hybrid system allows magnon-mediated nonreciprocal interactions between distant skyrmion qubits, or between skyrmion qubits and superconducting qubits.
	
	{
	In addition, recent theories have exploited equilibrium magnon squeezing to study magnon--spin hybrid systems. Reference~\cite{Skogvoll2021} shows that a spin qubit exchange-coupled to an anisotropic ferromagnet can realize a tunable anisotropic quantum Rabi model that can reach the deep-strong coupling regime, where the composite squeezed magnon can enable joint excitation of three qubits for the GHZ states. Reference~\cite{Romling2023} demonstrates that the dispersive coupling between a qubit and a noneigenmode magnon enables the detection of the quantum superposition of magnonic number states, the control of the equilibrium magnon squeezing, and the generation of squeezed even Fock states. Further, Ref.~\cite{Romling2025} shows that spin-nonconserving and beyond-RWA interactions in antiferromagnets yield strong squeezing of the magnon pseudospin, which is controllable by the qubit state. These works highlight equilibrium magnon squeezing as an important resource for quantum control and sensing.}

	\section{Coupling to the COM motion of levitated YIG spheres}\label{mag-COM}

	The advances in cavity magnomechanics~\cite{Zuo24} have opened up a promising avenue for exploring macroscopic quantum phenomena. Hybrid systems that combine microwave photons, magnons, and mechanical degrees of freedom have demonstrated extraordinary capabilities, ranging from ground-state cooling to the generation of nonclassical states~\cite{Li18,Kani22,Jing21,Asjad23,Isart20}. The mechanical motion can be the geometric deformation (acoustic modes) induced by magnetostriction (see review article~\cite{Zuo24}), or the {center-of-mass (COM)} motion of a levitated ferromagnet~\cite{Kani22,Isart21}.  In particular, levitated YIG spheres have emerged as a versatile platform, because their COM motion can serve as an ultrahigh-$Q$ mechanical oscillator that interacts with both cavity fields and magnon modes~\cite{Pan2023,Isart20prb,FuwaM23}.

	\subsection{Theory of the magnon--COM coupling}

\begin{figure}[t]
		\centering
		\includegraphics[width=\linewidth]{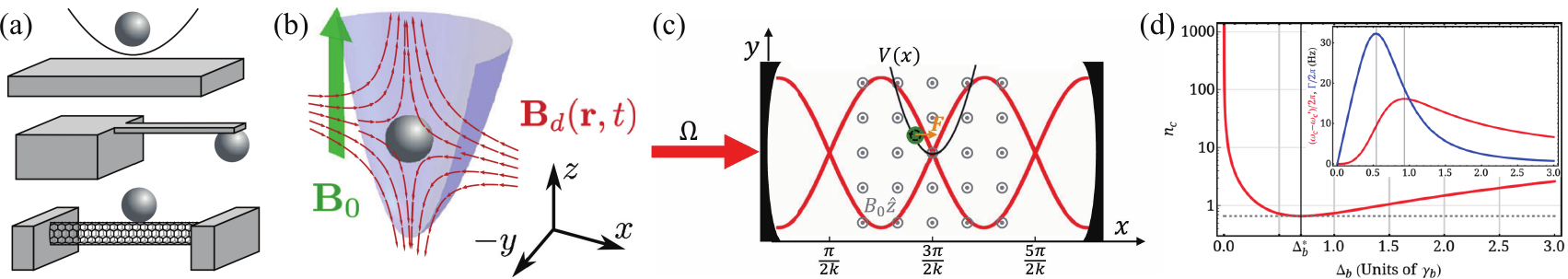}
		\caption{\label{fig:f3}{(a)-(b) Schematic diagram of the coupling between magnons and the COM motion of a magnetic sphere. Reprinted figures (a)-(b) with permission from~\cite{Isart20} Copyright 2020 by the American Physical Society.  (c) A YIG sphere is trapped in a harmonic potential $V(x)$ at a node of the microwave cavity magnetic field $\mathbf{B}_{\text{cav}} = {y}B \cos(kx)$. A uniform bias magnetic field $B_0$ is applied in the $z$ direction. The cavity field gradient couples the magnon mode and the COM motion. (d) Mean phonon occupancy of the COM motion as a function of magnon-drive detuning. Inset shows the induced mechanical frequency shift (red), and the effective cooling rate (blue). Reprinted figures (c)-(d) from~\cite{Kani22} licensed under the \href{https://creativecommons.org/licenses/by/4.0/}{CC BY 4.0}. } }
	\end{figure}

	The magnon--COM coupling arises from the magnetic field gradient experienced by a vibrating micromagnet~\cite{Isart20} (figure~\ref{fig:f3}(a)), providing an efficient interface between the magnonic and mechanical degrees of freedom.  When a magnetic sphere is levitated in an inhomogeneous magnetic field (figure~\ref{fig:f3}(b)), its COM oscillation modulates the local magnetic field experienced by the magnon modes confined in the sphere, thereby inducing a change in its magnetization state. Conversely, the alteration in magnetization exerts a reciprocal back-action on the COM motion via the inhomogeneous magnetic field.
	
	A remarkable advantage of the levitated magnetic sphere is that the coupling among the COM motion, the microwave cavity mode, and the magnon mode is independent of the size of the sphere.  This property originates from the collective nature of the magnon excitation: the photon--magnon coupling strength $g_{\rm ab}\propto\sqrt{N}$ ($N$ is the total number of spins) exactly compensates the reduction of the zero-point motion $x_{\mathrm{ZPF}}\propto 1/\sqrt{m}$ for large masses, yielding a tripartite coupling $g_{\rm abc}=g_{\rm ab} k x_{\mathrm{ZPF}}$ ($k$ is the microwave field wave number) remaining constant for spheres much smaller than the microwave wavelength~\cite{Kani22}.
	Such a size-independent coupling is a crucial component in seeing macroscopic quantum effects in the system, such as ground-state cooling of the COM motion.
	
	The physical model of a trapped YIG sphere inside a driven microwave cavity (figure~\ref{fig:f3}(c)) is described by the Hamiltonian, which in the frame rotating at the drive frequency $\omega_l$ reads~\cite{Kani22}
	\begin{equation}
		\begin{aligned}
			{{H}}/\hbar =& \Delta_a {a}^\dagger{a} + \Delta_m {m}^\dagger{m} + \omega_c {c}^\dagger{c} + g_{\rm ab}\cos \left(kx_0 \right) \left({a}{m}^\dagger + {a}^\dagger{m} \right) \\
			&- g_{\rm ab}k\sqrt{\frac{\hbar}{2\rho_m V\omega_c}}\,\sin(kx_0) \left({a}{m}^\dagger + {a}^\dagger{m} \right) \left({c}+{c}^\dagger \right) + \Omega \left({a}+{a}^\dagger \right), 
		\end{aligned}
	\end{equation}
	where ${a}$, ${m}$, and ${c}$ are the annihilation operators of the cavity, magnon, and COM modes, respectively, $\Delta_{a(m)}=\omega_{a(m)}-\omega_l$ is the cavity (magnon)-drive detuning with $\omega_j$, $j=a,m,c$, being the corresponding mode frequencies, $\rho_m$ ($V$) is the mass density (volume) of the YIG sphere, and $\Omega$ denotes the drive amplitude of the cavity mode.  The tripartite coupling stems from the magnetic-dipole interaction between the cavity field ${\bf B}_{\mathrm{cav}}={y}B\cos(kx)$ (the COM motion along the $x$ direction, figure~\ref{fig:f3}(c)) and the magnon mode of the sphere. Expanding $\cos(kx)$ up to first order in particle position around the position of the trap minimum $x_0$ gives rise to a trilinear coupling among the photon, magnon, and COM motion.  When the sphere is trapped at a node of the cavity magnetic field ($kx_0=n\pi + \frac{\pi}{2}$), giving $\cos(kx_0)=0$, the direct photon--magnon coupling vanishes, and the interaction reduces to an effective two-mode magnomechanical interaction after linearization under a strong driving field.  If instead the trap deviates from the node of the magnetic field, $\cos(kx_0)$ remains finite, and one then obtains a tunable three-mode interaction described by the linearized Hamiltonian ${H}_{\mathrm{int}}/\hbar =  \left[(G_m\delta{m} + G_a\delta{a}) + \mathrm{H.c.}\right](\delta{c}+\delta{c}^\dagger)$, where $G_m$ and $G_a$ are the effective optomechanical-like couplings and can be tuned by the driving field and the position of the sphere.

	\subsection{Quantum protocols based on the magnon--COM coupling}

	The magnon--COM coupling achieved in levitated magnetic systems provides a new and efficient means to generate and control quantum states of mechanical motion. In particular, Ref.~\cite{Kani22} predicts that ground-state cooling of the COM motion of a levitated YIG sphere is possible under the resolved-sideband condition. The mean phonon occupancy can be reduced to below unity (figure~\ref{fig:f3}(d)) even for a millimeter-sized YIG sphere, benefitting from the fact that the cooling rate becomes an intensive property, i.e., size-independent.   
	Theory also indicates that, by exploiting the magnon Kerr effect,  simultaneous cooling of the COM motion and an internal acoustic mode is possible even in the unresolved sideband regime~\cite{Zhong25b}.  Furthermore, a single cavity can be employed to simultaneously cool the COM motions of two levitated YIG spheres~\cite{Zhong25}.  These cooling results provide a prerequisite for observing macroscopic quantum coherence of levitated magnetic spheres.  
	It is worth noting that the magnon--COM coupling does not necessarily require the presence of a microwave cavity.  For instance, a static but inhomogeneous magnetic field $\mathbf{B}=(B_0+\xi x){z}$ ($\xi$ is the gradient) results in a single-magnon magnomechanical coupling strength $\propto\xi\,x_{\mathrm{ZPF}}$, which, together with magnon Kerr effect, can be used to achieve ground-state cooling of the COM motion~\cite{XiongH25}.
	In a different configuration, the COM motion of a levitated micromagnet can strongly couple to an acoustic mode via the magnetoelastic interaction, where magnons act as a passive mediator. By tuning the external magnetic field, such that the magnon frequency is nearly resonant with the acoustic mode, and by further driving the magnon with an inhomogeneous microwave field, an effective acoustomechanical coupling is engineered~\cite{Isart20,Isart20prb}.
	These strategies expand the toolbox of levitodynamics and enable hybrid systems that combine magnetic, COM, and acoustic degrees of freedom.

	Besides mechanical cooling, entangled states can also be generated in the system.  By properly choosing the detunings of the cavity and magnon modes, one can activate the BS and TMS interactions that lead to bipartite and tripartite entanglement among the photons, magnons, and COM phonons~\cite{NieW25,Yang2025}.
	The entanglement can be further enhanced by driving the magnon with the parametric amplification~\cite{Bayati24}, or exploiting magnon dissipation as a resource in a reservoir-engineering approach~\cite{Huo25}.  Interestingly, in the strong and even ultrastrong coupling regimes, the system can exhibit genuine tripartite entanglement without any pairwise entanglement, which emerges near the critical point of a quantum phase transition~\cite{Twamley25}.  Moreover, the COM motions of two YIG spheres can be entangled via a common cavity field, establishing a long-range quantum correlation between massive objects~\cite{Rao25}.
	These results demonstrate that levitated magnomechanics offers a versatile platform for exploring macroscopic quantum correlations.

	In addition, the strong magnon--COM interaction enables sophisticated control of microwave propagation and nonreciprocal behavior.
	Magnomechanically induced transparency, similar to that of the acoustic mode~\cite{ZhangX16}, was investigated in a levitated-sphere system, where the width of the transparency window can be tuned by the position of the sphere and the driving parameters, and switching between slow and fast microwave propagation can be realized~\cite{Bayati2024}.
	Furthermore, by applying independent Coulomb forces to two levitated YIG spheres placed in coupled microwave cavities, the time-reversal symmetry of the system can be broken, yielding an extremely high isolation ratio in the nonreciprocal microwave transmission~\cite{Nie25}. These findings hold promise for developing nonreciprocal transmission devices.

	The levitated magnomechanical platform has transitioned swiftly from a theoretical concept to a fertile ground for both experimental and theoretical studies. {Its coupling strength being largely size-independent in certain regimes, excellent isolation from the environment, and compatibility with microwave quantum optics make it a promising platform for testing the foundations of quantum mechanics at macroscopic scales and for developing ultra-sensitive sensors and novel quantum technologies. Despite these advantages, several experimental challenges must be overcome to realize strong magnon--COM coupling. Reducing the size of the magnetic sphere to enhance the coupling tends to exacerbate the surface-induced decoherence and damping, while on-chip integration remains difficult because stable trapping and precise positioning typically require bulky optical or magnetic setups that are challenging to integrate with planar photonic and superconducting circuits~\cite{Hensen26}. Efficient coupling to high-finesse cavities and preserving long coherence time in the submicron regime pose further obstacles. Besides, scaling to larger masses and multiple particles adds additional complexity~\cite{FuwaM23, Tamegai23}. All of these make the ground-state cooling and mechanics-based quantum precision measurement challenging tasks~\cite{Isart21}, and addressing them becomes crucial for exploiting the levitated-sphere platforms~\cite{Gieseler20,Huillery20} to achieve the envisioned magnonic quantum control.}

	\section{Conclusion and outlook}\label{conc}

	To conclude, we have introduced the coupling of magnons, particularly in YIG spheres, to various quantum systems, and reviewed both quantum-related experiments and theoretical proposals, focusing on the generation of magnonic quantum states and their applications in many related fields. The experimental realization of strong coupling is the starting point for enabling those quantum protocols and applications introduced in each section, and a larger cooperativity gives rise to an enhanced performance. Up to now, strong coupling has been achieved in the cavity--magnon~\cite{Tabuchi14,Zhang14} and magnon--qubit~\cite{tabuchi2015} systems, and the latter is the only platform that has generated and characterized magnonic quantum states in a macroscopic magnonic system, thanks to the strong nonlinearity introduced from the superconducting qubit. Note that recently the cavity magnomechanical system has also entered the strong-coupling regime~\cite{Shen25}. In their experiment, however, the strong-coupling regime was reached by significantly reducing the effective decay rate of the polariton mode, rather than greatly enhancing the magnomechanical coupling strength, which limits its quantum applications.  Similarly to the enhancement of the optomagnonic coupling (section~\ref{CEnh}), the guideline is to reduce the mode volume and increase the mode overlap between the magnon and optical/mechanical modes, which means miniaturization. This is towards the practicalization of many quantum applications, but meanwhile this reduces the size of the macroscopic quantum states that could be created, so there is a trade-off between fundamental studies, e.g., macroscopic quantum states, and practical applications. {On the one hand, a larger size of the YIG crystal gives rise to a larger number of spins, yielding a stronger coupling to the microwave cavity or the superconducting qubit. This can enhance the coupling with the axion field, and thereby improve the detection sensitivity of dark-matter axions using, e.g., YIG spheres~\cite{Crescini20sq}. On the other hand, as mentioned above, miniaturization is required to enhance both the optomagnonic  and magnomechanical coupling strength, and to integrate with superconducting microwave circuits~\cite{Yi19L}. Nevertheless, miniaturization, such as YIG films, reduces the coupling strength to the microwave resonator and increases the magnon dissipation rate. An example is the on-chip magnomechanical platform, which promises many applications, including microwave phonon manipulation, magnon-phonon transducer, and novel phononic devices~\cite{Hatanaka26}. However, the on-chip devices give rise to surface-acoustic-wave phonons with a much higher frequency in the gigahertz range, in which the dominant magnon-phonon coupling becomes a linear form, making it difficult to prepare magnonic quantum states which typically requires nonlinearity~\cite{Zuo24}. }

	Apart from strong coupling, a more challenging task is the manipulation of magnons at the single-magnon level.  Although, so far, this has only been realized on the {cavity--magnon--qubit (CMQ)} platform, there is a strong motivation to achieve this in the optomagnonic system, as this will enable numerous quantum protocols (section~\ref{optomag}), such as the {Duan--Lukin--Cirac--Zoller} and the {discrete-variable} quantum teleportation, and the preparation of magnonic non-Gaussian states, including Fock states, {single-magnon-added coherent and thermal states}.  A crucial step in accelerating this progress is the significant improvement of the optomagnonic coupling strength, e.g., using a YIG micro disk.   
	Nevertheless, even without requiring the demanding single-magnon manipulations, theories indicate that magnonic quantum states, e.g., cat, squeezed, and entangled states, can be generated using currently available parameters from the CMQ experiments (section~\ref{mag-qubit}).  A direction for future research in this system could be the generation of multi-component or multi-mode magnonic cat states, which can improve the detection sensitivity of dark-matter axions~\cite{Crescini20sq,Zheng26}. The current magnon squeezing experiment and protocols based on the CMQ system achieved only small or moderate squeezing. Therefore, properly designing schemes to obtain strong magnon squeezing is of great importance for quantum sensing. Besides, preparing multipartite macroscopic entanglement involving multiple YIG spheres via manipulating a single superconducting qubit would be fascinating for the study of macroscopic quantum phenomena.

	Beyond that, theoretical results confirm that strong magnon--spin coupling at the single-quantum level is highly feasible (section~\ref{mag-spin}). The magnon--spin hybrid system bridges spins with other quantum systems that are inherently weakly coupled, enabling indirect strong coupling between microwave photons and single spins, and provides a route to single-magnon quantum control for applications such as magnon blockade and magnon qubits. However, most existing theories are based on submicron-sized magnetic spheres, and the main experimental obstacles lie in the high-precision micro/nano-fabrication of magnetic spheres and the preservation of magnon coherence at reduced sizes, because magnon dissipation strongly depends on crystal defects and surface structures of the materials. In addition, strong magnon--{center-of-mass} coupling also requires smaller sizes of the magnetic sphere (section~\ref{mag-COM}). The levitated magnetic sphere platform is already relatively mature in experiments~\cite{Gieseler20,Huillery20}. This coupling connects magnetic and mechanical degrees of freedom with potential applications, including ground-state cooling of the mechanical motion and mechanics-based quantum precision measurement~\cite{Isart21}, while the well-developed manipulation of levitated systems also offers a new pathway for realizing magnonic quantum control.   


	\section*{Acknowledgments}
	
	We thank Dr. Da Xu for the careful reading and useful feedback on the qubit section. X.L.H. was supported by the Young Scientists Fund of the National Natural Science Foundation of China (No. 12505029), the China Postdoctoral Science Foundation (No. 2025M773347), and the Fundamental Research Funds for the Central Universities of Ministry of Education of China (No. xzy012025077).  Z.X.L. was supported by the National Natural Science Foundation of China (No. 12105047). Q.G. was supported by the National Natural Science Foundation of China (No. 12274274). P.B.L. was supported by the National Natural Science Foundation of China (No. W2411002 and No. 12375018). J.L. was supported by the National Natural Science Foundation of China (No. 12474365 and No. 92265202), the National Key Research and Development Program of China (No. 2024YFA1408900 and No. 2022YFA1405200), and the Zhejiang Provincial Natural Science Foundation of China (No. LR25A050001).



\begin{thebibliography}{99}
		
		
		\bibitem{Naka19}
		Lachance-Quirion D, Tabuchi Y, Gloppe A, Usami K and Nakamura Y 2019 Hybrid quantum systems based on magnonics \textit{Appl. Phys. Express} \href{https://doi.org/10.7567/1882-0786/ab248d}{{\bf 12} 070101}
		
		\bibitem{Clerk20}
		Clerk A A, Lehnert K W, Bertet P, Petta J R and Nakamura Y 2020 Hybrid quantum systems with circuit quantum electrodynamics \textit{Nat. Phys.} \href{https://doi.org/10.1038/s41567-020-0797-9}{{\bf 16} 257}
		
		\bibitem{Li20}
		Li Y, Zhang W, Tyberkevych V, Kwok W-K, Hoffmann A and Novosad V 2020 Hybrid magnonics: physics, circuits and applications for coherent information processing \textit{J. Appl. Phys.} \href{https://doi.org/10.1063/5.0020277}{{\bf 128} 130902}
		
		\bibitem{Aws21}
		Awschalom D D et al 2021 Quantum engineering with hybrid magnonic systems and materials  \textit{IEEE Trans. Quantum Eng.} \href{https://doi.org/10.1109/TQE.2021.3057799}{{\bf 2} 1}
		
		\bibitem{Yuan22}
		Yuan H Y, Cao Y, Kamra A, Duine R A and Yan P 2022 Quantum magnonics: when magnon spintronics meets quantum information science \textit{Phys. Rep.} \href{https://doi.org/10.1016/j.physrep.2022.03.002}{{\bf 965} 1}
		
		\bibitem{Bauer22}
		Rameshti B Z, Kusminskiy S V, Haigh J A, Usami K, Lachance-Quirion D, Nakamura Y, Hu C-M, Tang H X, Bauer G E W and Blanter Y M 2022 Cavity magnonics {\it Phys. Rep.} \href{https://doi.org/10.1016/j.physrep.2022.06.001}{{\bf 979} 1}
		
		\bibitem{Zuo24}
		Zuo X, Fan Z-Y, Qian H, Ding M-S, Tan H, Xiong H and Li J 2024 Cavity magnomechanics: from classical to quantum {\it New J. Phys.} \href{https://doi.org/10.1088/1367-2630/ad327c}{{\bf 26} 031201}
		
		\bibitem{Huebl13}
		Huebl H, Zollitsch C W, Lotze J, Hocke F, Greifenstein M, Marx A, Gross R and Goennenwein S T B 2013 High cooperativity in coupled microwave resonator ferrimagnetic insulator hybrids {\it Phys. Rev. Lett.} \href{https://doi.org/10.1103/PhysRevLett.111.127003}{{\bf 111} 127003}
		
		\bibitem{Tabuchi14}
		Tabuchi Y, Ishino S, Ishikawa T, Yamazaki R, Usami K and Nakamura Y 2014 Hybridizing ferromagnetic magnons and microwave photons in the quantum limit {\it Phys. Rev. Lett.} \href{https://doi.org/10.1103/PhysRevLett.113.083603}{{\bf 113} 083603}
		
		\bibitem{Zhang14}
		Zhang X, Zou C-L, Jiang L and Tang H X 2014 Strongly coupled magnons and cavity microwave photons {\it Phys. Rev. Lett.} \href{https://doi.org/10.1103/PhysRevLett.113.156401}{{\bf 113} 156401}
		
		\bibitem{Chumak26}
		Serha R O, Dubs C and Chumak A V 2026 Magnetic materials for quantum magnonics {\it APL Mater.} \href{https://doi.org/10.1063/5.0306423}{{\bf 14} 030901}
		
		\bibitem{tabuchi2015}
		Tabuchi Y, Ishino S, Noguchi A, Ishikawa T, Yamazaki R, Usami K and Nakamura Y 2015 Coherent coupling between a ferromagnetic magnon and a superconducting qubit \textit{Science} \href{https://doi.org/10.1126/science.aaa3693}{\textbf{349} 405}
		
		\bibitem{Shen25}
		Shen R-C, Li J, Sun Y-M, Wu W-J, Zuo X, Wang Y-P, Zhu S-Y and You J Q 2025 Cavity-magnon polaritons strongly coupled to phonons {\it Nat. Commun.} \href{https://doi.org/10.1038/s41467-025-60799-x}{{\bf 16} 5652}
		
		\bibitem{Nakamura16}
		Osada A, Hisatomi R, Noguchi A, Tabuchi Y, Yamazaki R, Usami K, Sadgrove M, Yalla R, Nomura M and Nakamura Y 2016 Cavity optomagnonics with spin-orbit coupled photons \textit{Phys. Rev. Lett.} \href{https://doi.org/10.1103/PhysRevLett.116.223601}{{\bf 116} 223601}		
		\bibitem{Zhang16}
		Zhang X, Zhu N, Zou C-L and Tang H X 2016 Optomagnonic whispering gallery microresonators \textit{Phys. Rev. Lett.} \href{https://doi.org/10.1103/PhysRevLett.117.123605}{{\bf 117} 123605}
		\bibitem{Haigh16}
		Haigh J A, Nunnenkamp A, Ramsay A J and Ferguson A J 2016 Triple-resonant brillouin light scattering in magneto-optical cavities \textit{Phys. Rev. Lett.} \href{https://doi.org/10.1103/PhysRevLett.117.133602}{{\bf 117} 133602}
		
		\bibitem{Disk26}
		Demirchyan S S, Krichevsky D M and Belotelov V I 2026 Magnon–photon coupling in the YIG-based disk and ring microcavities \textit{J. Appl. Phys.} \href{https://doi.org/10.1063/5.0315725}{{\bf 139} 113903}
		
		
		\bibitem{lachance-quirion2020}
		Lachance-Quirion D, Wolski S P, Tabuchi Y, Kono S, Usami K and Nakamura Y 2020 Entanglement-based single-shot detection of a single magnon with a superconducting qubit \textit{Science} \href{https://doi.org/10.1126/science.aaz9236}{\textbf{367} 425}
		
		\bibitem{xu2023}
		Xu D, Gu X-K, Li H-K, Weng Y-C, Wang Y-P, Li J, Wang H, Zhu S-Y and You J Q 2023 Quantum control of a single magnon in a macroscopic spin system \textit{Phys. Rev. Lett.}\href{https://doi.org/10.1103/PhysRevLett.130.193603}{\textbf{130} 193603}
		
		\bibitem{weng2026}
		Weng Y-C, Xu D, Chen Z, Tan L-Z, Gu X-K, Li J, Yu H-F, Zhu S-Y, Hu X, Nori F and You J Q 2026 Magnon squeezing in the quantum regime \textit{Nat. Commun.} \href{https://doi.org/10.1038/s41467-026-69312-4}{\textbf{17} 2679}
		
		\bibitem{Li18}
		Li J, Zhu S-Y and Agarwal G S 2018 Magnon-photon-phonon entanglement in cavity magnomechanics {\it Phys. Rev. Lett.} \href{https://doi.org/10.1103/PhysRevLett.121.203601}{{\bf 121} 203601}
		
		\bibitem{LJ20}
		Yu M, Shen H, and Li J 2020 Magnetostrictively Induced Stationary Entanglement between Two Microwave Fields {\it Phys. Rev. Lett.} \href{https://doi.org/10.1103/PhysRevLett.124.213604}{{\bf 124} 213604} 
		
		\bibitem{Hei2023}
		Hei X-L, Li P-B, Pan X-F and Nori F 2023 Enhanced tripartite interactions in spin-magnon-mechanical hybrid systems {\it Phys. Rev. Lett.} \href{https://doi.org/10.1103/PhysRevLett.130.073602}{{\bf 130} 073602}
		
		\bibitem{Kani22}
		Kani A, Sarma B and Twamley J 2022 Intensive cavity-magnomechanical cooling of a levitated macromagnet {\it Phys. Rev. Lett.} \href{https://doi.org/10.1103/PhysRevLett.128.013602}{{\bf 128} 013602}
		
		
		
		
		\bibitem{Stancil09}
		Stancil D D and Prabhakar A 2009 Spin Waves: Theory and Applications  {\it Springer}
		
		\bibitem{Kittel48}
		Kittel C 1948 On the theory of ferromagnetic resonance absorption {\it Phys. Rev.} \href{https://doi.org/10.1103/PhysRev.73.155}{{\bf 73} 155}
		
		
		
		
		\bibitem{Leggett02}
		Leggett A J 2002 Testing the limits of quantum mechanics: motivation, state of play, prospects {\it J. Phys.: Condens. Matter} \href{https://doi.org/10.1088/0953-8984/14/15/201}{{\bf 14} R415}
		
		
		\bibitem{Bassi13}
		Bassi A, Lochan K, Satin S, Singh T P and Ulbricht H 2013 Models of wave-function collapse, underlying theories, and experimental tests {\it Rev. Mod. Phys.} \href{https://doi.org/10.1103/RevModPhys.85.471}{{\bf 85} 471}
		
		\bibitem{Weaver18}
		Weaver M J, Newsom D, Luna F, L\"offler W, and Bouwmeester D 2018 Phonon interferometry for measuring quantum decoherence {\it Phys. Rev. A} \href{https://doi.org/10.1103/PhysRevA.97.063832}{{\bf 97} 063832}
		
		\bibitem{Flower19sq}
		Flower G, Bourhill J, Goryachev M and Tobar M E 2019 Broadening frequency range of a ferromagnetic axion haloscope with strongly coupled cavity--magnon polaritons {\it Phys. Dark Universe} \href{https://doi.org/10.1016/j.dark.2019.100306}{{\bf 25} 100306}
		
		\bibitem{Crescini20sq}
		Crescini N, Alesini D, Braggio C, Carugno G, Agostino D D, Gioacchino D D, Falferi P, Gambardella U, Gatti C, Iannone G, Ligi C, Lombardi A, Ortolan A, Pengo R, Ruoso G and Taffarello L 2020 Axion search with a quantum-limited ferromagnetic haloscope {\it Phys. Rev. Lett.} \href{https://doi.org/10.1103/PhysRevLett.124.171801}{{\bf 124} 171801}
		
		
		
		
		\bibitem{Vitali07}
		Vitali D, Gigan S, Ferreira A, Böhm H R, Tombesi P, Guerreiro A, Vedral V, Zeilinger A and Aspelmeyer M 2007 Optomechanical entanglement between a movable mirror and a cavity field {\it Phys. Rev. Lett.} \href{https://doi.org/10.1103/PhysRevLett.98.030405}{{\bf 98} 030405}
		
		\bibitem{Vidal02}
		Vidal G and Werner R F 2002 Computable measure of entanglement {\it Phys. Rev. A} \href{https://doi.org/10.1103/PhysRevA.65.032314}{{\bf 65} 032314}
		
		\bibitem{Plenio05}
		Plenio M B 2005 Logarithmic negativity: A full entanglement monotone that is not convex {\it Phys. Rev. Lett.} \href{https://doi.org/10.1103/PhysRevLett.95.090503}{{\bf 95} 090503}
		
		\bibitem{Yuan20Bell} 
		Yuan H Y, Yan P, Zheng S, He Q Y, Xia K and Yung M-H 2020 Steady Bell state generation via magnon-photon coupling {\it Phys. Rev. Lett.} \href{https://doi.org/10.1103/PhysRevLett.124.053602}{{\bf 124} 053602}
		
		
		
		\bibitem{Li19MME} 
		Li J and Zhu S-Y 2019 Entangling two magnon modes via magnetostrictive interaction {\it New J. Phys.} \href{https://doi.org/10.1088/1367-2630/ab3508}{{\bf 21} 085001}
		
		\bibitem{Zhang19}
		Zhang Z, Scully M O and Agarwal G S 2019 Quantum entanglement between two magnon modes via Kerr nonlinearity driven far from equilibrium {\it Phys. Rev. Res.} \href{https://doi.org/10.1103/PhysRevResearch.1.023021}{{\bf 1} 023021}
		
		\bibitem{Wang16}
		Wang Y-P, Zhang G-Q, Zhang D, Luo X-Q, Xiong W, Wang S-P, Li T-F, Hu C-M and You J Q 2016 Magnon Kerr effect in a strongly coupled cavity--magnon system {\it Phys. Rev. B} \href{https://doi.org/10.1103/PhysRevB.94.224410}{{\bf 94} 224410}
		
		\bibitem{Shen22}
		Shen R-C, Li J, Fan Z-Y, Wang Y-P, and You J Q  2022 Mechanical bistability in Kerr-modified cavity magnomechanics {\it Phys. Rev. Lett.} \href{https://doi.org/10.1103/PhysRevLett.129.123601}{{\bf 129} 123601}
		
		\bibitem{Wu22}
		Wu W-J, Xu D, Qian J, Li J, Wang Y-P and You J-Q 2022 Observation of nonlinearity and heating-induced frequency shifts in cavity magnonics {\it Chinese Phys. B} \href{https://doi.org/10.1088/1674-1056/ac9b02}{{\bf 31} 127503}
		
		\bibitem{Nair20}
		Nair J M P and Agarwal G S 2020 Deterministic quantum entanglement between macroscopic ferrite samples {\it Appl. Phys. Lett.} \href{https://doi.org/10.1063/5.0015195}{{\bf 117} 084001} 
		
		\bibitem{Yu20}
		Yu M, Zhu S-Y and Li J 2020 Macroscopic entanglement of two magnon modes via quantum correlated microwave fields {\it J. Phys. B: At. Mol. Opt. Phys.} \href{https://doi.org/10.1088/1361-6455/ab68b5}{{\bf 53} 065402}
		
		\bibitem{Xie23}
		Xie J, Yuan H, Ma S, Gao S, Li F and Duine R A 2023 Stationary quantum entanglement and steering between two distant macromagnets {\it Quantum Sci. Technol.} \href{https://doi.org/10.1088/2058-9565/acd576}{{\bf 8} 035022}
		
		\bibitem{Hu24}
		Hu N and Tan H 2024 Steady-state magnon entanglement and backaction-evading of a weak magnetic signal via two-tone modulated cavity electromagnonics {\it Opt. Express} \href{https://doi.org/10.1364/OE.536838}{{\bf 32} 35419}
		
		\bibitem{Kamra19sq}
		Kamra A, Thingstad E, Rastelli G, Duine R A, Brataas A, Belzig W and Sudb{\o} A 2019 Antiferromagnetic magnons as highly squeezed Fock states underlying quantum correlations {\it Phys. Rev. B} \href{https://doi.org/10.1103/PhysRevB.100.174407}{{\bf 100} 174407}
		
		\bibitem{Yuan20}
		Yuan H Y, Zheng S, Ficek Z, He Q Y and Yung M-H 2020 Enhancement of magnon-magnon entanglement inside a cavity {\it Phys. Rev. B} \href{https://doi.org/10.1103/PhysRevB.101.014419}{{\bf 101} 014419}
		
		\bibitem{Mousolou21}
		Mousolou V A, Liu Y, Bergman A, Delin A, Eriksson O, Pereiro M, Thonig D and Sj\"{o}qvist E 2021 Magnon-magnon entanglement and its quantification via a microwave cavity {\it Phys. Rev. B} \href{https://doi.org/10.1103/PhysRevB.104.224302}{{\bf 104} 224302}
		
		
		
		
		\bibitem{Li19b}
		Li J, Zhu S-Y and Agarwal G S 2019 Squeezed states of magnons and phonons in cavity magnomechanics {\it Phys. Rev. A} \href{https://doi.org/10.1103/PhysRevA.99.021801}{{\bf 99} 021801}
		
		\bibitem{Kamra20sq}
		Kamra A, Belzig W and Brataas A 2020 Magnon-squeezing as a niche of quantum magnonics {\it Appl. Phys. Lett.} \href{https://doi.org/10.1063/5.0021099}{{\bf 117} 090501}
		
		\bibitem{Li23sq}
		Li J, Wang Y-P, You J-Q and Zhu S-Y 2023 Squeezing microwaves by magnetostriction {\it Natl. Sci. Rev.} \href{https://doi.org/10.1093/nsr/nwac247}{{\bf 10} nwac247}
		
		\bibitem{Fabre94sq}
		Fabre C, Pinard M, Bourzeix S, Heidmann A, Giacobino E and Reynaud S 1994 Quantum-noise reduction using a cavity with a movable mirror {\it Phys. Rev. A} \href{https://doi.org/10.1103/PhysRevA.49.1337}{{\bf 49} 1337}
		
		\bibitem{Mancini94sq}
		Mancini S and Tombesi P 1994 Quantum noise reduction by radiation pressure {\it Phys. Rev. A} \href{https://doi.org/10.1103/PhysRevA.49.4055}{{\bf 49} 4055}
		
		\bibitem{CostaFilho00sq}
		Costa Filho R N, Cottam M G and Farias G A 2000 Microscopic theory of dipole-exchange spin waves in ferromagnetic films: Linear and nonlinear processes {\it Phys. Rev. B} \href{https://doi.org/10.1103/PhysRevB.62.6545}{{\bf 62} 6545}
		
		\bibitem{Kamra16sq} 
		Kamra A and Belzig W 2016 Super-poissonian shot noise of squeezed-magnon mediated spin transport {\it Phys. Rev. Lett.} \href{https://doi.org/10.1103/PhysRevLett.116.146601}{{\bf 116} 146601}
		
		\bibitem{Kamra17sq}
		Kamra A, Agrawal U and Belzig W 2017 Noninteger-spin magnonic excitations in untextured magnets {\it Phys. Rev. B} \href{https://doi.org/10.1103/PhysRevB.96.020411}{{\bf 96} 020411(R)}
		
		\bibitem{Hioki26sq}
		Hioki T, Tojo K, Elyasi M, Horibe S, Shimizu H, Hoshi K, Makiuchi T, Bauer G E W and Saitoh E 2026 Single- and two-mode magnon thermal squeezing \textit{Nat. Phys.} \href{https://doi.org/10.1038/s41567-026-03294-4}{{\bf 22} 1107}
		
		\bibitem{Schrodinger}
		Schr\"odinger E 1935 Die gegenw\"{a}rtige Situation in der Quantenmechanik {\it Naturwissenschaften} \href{https://doi.org/10.1007/BF01491891}{{\bf 23} 807}
		
		\bibitem{Vlastakis13}
		Vlastakis B, Kirchmair G, Leghtas Z, Nigg S E, Frunzio L, Girvin S M, Mirrahimi M, Devoret M H and Schoelkopf R J 2013 Deterministically encoding quantum information using 100-photon Schr\"odinger cat states {\it Science} \href{https://doi.org/10.1126/science.1243289}{{\bf 342} 607}
		
		\bibitem{Tatsuta19}
		Tatsuta M, Matsuzaki Y and Shimizu A 2019 Quantum metrology with generalized cat states {\it Phys. Rev. A} \href{https://doi.org/10.1103/PhysRevA.100.032318}{{\bf 100} 032318}
		
		\bibitem{Zheng26}
		Zheng P, Cai Y, Xu B, Wen S, Zhang L, Ni Z, Mai J, Zeng Y, Lin L, Hu L, Deng X, Liu S, Shu J, Xu Y and Yu D 2026 Quantum-enhanced dark matter search using cat states {\it Phys. Rev. Lett.} \href{https://doi.org/10.1103/wbhn-v1sw}{{\bf 136} 171002}
		
		\bibitem{Sharma21Cat} 
		Sharma S, Bittencourt V A S V, Karenowska A D and Kusminskiy S V 2021 Spin cat states in ferromagnetic insulators {\it Phys. Rev. B} \href{https://doi.org/10.1103/PhysRevB.103.L100403}{{\bf 103} L100403}
		
		\bibitem{Shomroni20} 
		Shomroni I, Qiu L and Kippenberg T J 2020 Optomechanical generation of a mechanical catlike state by phonon subtraction {\it Phys. Rev. A} \href{https://doi.org/10.1103/PhysRevA.101.033812}{{\bf 101} 033812}
		
		\bibitem{Zhan20}
		Zhan H, Li G and Tan H 2020 Preparing macroscopic mechanical quantum superpositions via photon detection {\it Phys. Rev. A} \href{https://doi.org/10.1103/PhysRevA.101.063834}{{\bf 101} 063834}
		
		\bibitem{Li26}
		Li H-T, Wang H-B, Lu Z-X and Li J 2026 Generation of mechanical cat-like states via optomagnomechanics {\it Quantum Rev. Lett.} \href{https://doi.org/10.1016/j.qrl.2026.03.001}{{\bf 2} 35}
		
		
		
		
		\bibitem{blais2004}
		Blais A, Huang R-S, Wallraff A, Girvin S M and Schoelkopf R J 2004 Cavity quantum electrodynamics for superconducting electrical circuits: An architecture for quantum computation \textit{Phys. Rev. A} \href{https://doi.org/10.1103/PhysRevA.69.062320}{ \textbf{69} 062320}
		
		\bibitem{blais2021}
		Blais A, Grimsmo A L, Girvin S M and Wallraff A 2021 Circuit quantum electrodynamics \textit{Rev. Mod. Phys.} \href{https://doi.org/10.1103/RevModPhys.93.025005}{\textbf{93} 025005}
		
		\bibitem{lachance-quirion2017}
		Lachance-Quirion D, Tabuchi Y, Ishino S, Noguchi A, Ishikawa T, Yamazaki R and Nakamura Y 2017 Resolving quanta of collective spin excitations in a millimeter-sized ferromagnet \textit{Sci. Adv.} \href{https://doi.org/10.1126/sciadv.1603150}{\textbf{3} e1603150}
		
		
		\bibitem{schuster2007}
		Schuster D I, Houck A A, Schreier J A, Wallraff A, Gambetta J M, Blais A, Frunzio L, Majer J, Johnson B, Devoret M H, Girvin S M and Schoelkopf R J 2007 Resolving photon number states in a superconducting circuit \textit{Nature} \href{https://doi.org/10.1038/nature05461}{ \textbf{445} 515}
		
		\bibitem{wolski2020}
		Wolski S P, Lachance-Quirion D, Tabuchi Y, Kono S, Noguchi A, Usami K and Nakamura Y 2020 Dissipation-based quantum sensing of magnons with a superconducting qubit \textit{Phys. Rev. Lett.} \href{https://doi.org/10.1103/PhysRevLett.125.117701}{ \textbf{125} 117701}
		
		
		\bibitem{autler1955}
		Autler S H and Townes C H 1955 Stark effect in rapidly varying fields \textit{Phys. Rev.} \href{https://doi.org/10.1103/PhysRev.100.703}{\textbf{100} 703}
		
		
		
		\bibitem{kerr_squeezing}
		Kitagawa M and Yamamoto Y 1986 Number-phase minimum-uncertainty state with reduced number uncertainty in a Kerr nonlinear interferometer \textit{Phys. Rev. A} \href{https://doi.org/10.1103/PhysRevA.34.3974}{\textbf{34} 3974}
		
		\bibitem{kerr_state}
		Wilson-Gordon A D, Buek V and Knight P L 1991 Statistical and phase properties of displaced Kerr states \textit{Phys. Rev. A} \href{https://doi.org/10.1103/PhysRevA.44.7647}{ \textbf{44} 7647}
		
		\bibitem{kerrwigner}
		Stobi\'{n}ska M, Milburn G J and W\'{o}dkiewicz K 2008 Wigner function evolution of quantum states in the presence of self-Kerr interaction \textit{Phys. Rev. A} \href{https://doi.org/10.1103/PhysRevA.78.013810}{\textbf{78} 013810}
		
		
		\bibitem{he2023cat}
		He S, Xin X, Zhang F-Y and Li C 2023 Generation of a Schr\"{o}dinger cat state in a hybrid ferromagnet-superconductor system
		\textit{Phys. Rev. A} \href{https://doi.org/10.1103/PhysRevA.107.023709}{\textbf{107} 023709}
		
		\bibitem{he2024scat}
		He S, Xin X, Wang Z, Zhang F-Y and Li C 2024 Generation of a squeezed Schr\"{o}dinger cat state in an anisotropic ferromagnet-superconductor coupled system \textit{Phys. Rev. A} \href{https://doi.org/10.1103/PhysRevA.110.053710}{\textbf{110} 053710}
		
		\bibitem{he20264cat}
		He S, Yang Z-L, Jin S, Zhang F-Y and Li C 2026 Generation of four-component magnonic Schr\"{o}dinger cat states via Floquet engineering \textit{Phys. Rev. A} \href{https://doi.org/10.1103/sw7f-syvg}{\textbf{113} 013739}
		
		\bibitem{li2026arxiv}
		Li G, Liu G, Yang R-C and Li J 2026 Preparing two-mode magnonic Schr\"{o}dinger cat states in a cavity--magnon--qubit system (arXiv:\href{https://arxiv.org/abs/2606.25511}{2606.25511})
		
		\bibitem{hou2024cat}
		Hou Y-B, Hei X-L, Pan X-F, Xie J-K, Ren Y-L, Ma S-L, Li F-L and Li P-B 2024 Robust generation of a magnonic cat state via a superconducting flux qubit \textit{Phys. Rev. A} \href{https://doi.org/10.1103/PhysRevA.110.013711}{\textbf{110} 013711}
		
		\bibitem{liu2025cat}
		Liu G, Li G, Tan H and Li J 2025 Magnon cat states in a cavity--magnon--qubit system via two-magnon driving and dissipation \textit{Phys. Rev. A} \href{https://doi.org/10.1103/zhgm-p3ss}{\textbf{112} 023709}
		
		\bibitem{kounalakis2022cat}
		Kounalakis M, Bauer G E W and Blanter Y M 2022 Analog quantum control of magnonic cat states on a chip by a superconducting qubit  \textit{Phys. Rev. Lett.} \href{https://doi.org/10.1103/PhysRevLett.129.037205}{\textbf{129} 037205}
		
		\bibitem{first_Exp_blockade}
		Imamo\u{g}lu A, Schmidt H, Woods G and Deutsch M 1997 Strongly interacting photons in a nonlinear cavity \textit{Phys. Rev. Lett.} \href{https://doi.org/10.1103/PhysRevLett.79.1467}{\textbf{79} 1467}
		
		\bibitem{liu2026quantum}
		Liu Z-X, Zuo X-J, Peng J-X and Xiong H 2026 Quantum sensing with cavity optomechanics \textit{Appl. Phys. Rev.} \href{https://doi.org/10.1063/5.0237048}{\textbf{13} 011307}
		
		\bibitem{liu2019magnon}
		Liu Z-X, Xiong H and Wu Y 2019 Magnon blockade in a hybrid ferromagnet-superconductor quantum system \textit{Phys. Rev. B} \href{https://doi.org/10.1103/PhysRevB.100.134421}{\textbf{100} 134421}
		
		\bibitem{xie2020quantum}
		Xie J-K, Ma S-L and Li F-L 2020 Quantum-interference-enhanced magnon blockade in an yttrium-iron-garnet sphere coupled to superconducting circuits \textit{Phys. Rev. A} \href{https://doi.org/10.1103/PhysRevA.101.042331}{\textbf{101} 042331}
		
		\bibitem{liu2025dispersive}
		Liu Z-X, Wu Y-H and Sun J-H 2025 Dispersive-induced magnon blockade with a superconducting qubit \textit{Front. Phys.} \href{https://journal.hep.com.cn/fop/EN/10.15302/frontphys.2025.063200}{\textbf{20} 063200}
		
		\bibitem{jin2023magnon}
		Jin Z-Y and Jing J 2023 Magnon blockade in magnon--qubit systems \textit{Phys. Rev. A} \href{https://doi.org/10.1103/PhysRevA.108.053702}{\textbf{108} 053702}
		
		\bibitem{zhao2025magnon}
		Zhao S, Ren Y-L, Hei X-L, Pan X-F and Li P-B 2025 Magnon blockade in spin-magnon systems with frequency detuning \textit{Phys. Rev. A} \href{https://journals.aps.org/pra/abstract/10.1103/8jy6-fp5x}{\textbf{112} 013712}
		
		\bibitem{li2021tunable}
		Li X, Wang X, Wu Z, Yang W-X and Chen A 2021 Tunable magnon antibunching in a hybrid ferromagnet-superconductor system with two qubits \textit{Phys. Rev. B} \href{https://doi.org/10.1103/PhysRevB.104.224434}{\textbf{104} 224434}
		
		\bibitem{xie2025unconventional}
		Xie H, He L-W, Lin X, Shi Z-G and Lin X-M 2025 Unconventional magnon blockade in an anisotropic ferromagnetic system \textit{Phys. Rev. B} \href{https://journals.aps.org/prb/abstract/10.1103/77sf-w3k4}{\textbf{112} 024406}
		
		\bibitem{xu2021conventional}
		Xu Y-J, Yang T-L, Lin L and Song J 2021 Conventional and unconventional magnon blockades in a qubit--magnon hybrid quantum system  \textit{J. Opt. Soc. Am. B} \href{https://opg.optica.org/josab/abstract.cfm?uri=josab-38-3-876}{\textbf{38} 876}
		
		\bibitem{wu2021phase}
		Wu K, Zhong W-X, Cheng G-L and Chen A-X 2021 Phase-controlled multimagnon blockade and magnon-induced tunneling in a hybrid superconducting system \textit{Phys. Rev. A} \href{https://doi.org/10.1103/PhysRevA.103.052411}{\textbf{103} 052411}
		
		\bibitem{fan2023nonclassical}
		Fan Y, Li J and Wu Y 2023 Nonclassical magnon pair generation and Cauchy--Schwarz inequality violation \textit{Phys. Rev. A} \href{https://doi.org/10.1103/PhysRevA.108.053715}{\textbf{108} 053715}
		
		\bibitem{wang2022dissipation}
		Wang Y, Xiong W, Xu Z, Zhang G-Q and You J-Q 2022 Dissipation-induced nonreciprocal magnon blockade in a magnon-based hybrid system \textit{Sci. China Phys. Mech. Astron.} \href{https://link.springer.com/article/10.1007/s11433-021-1880-7}{\textbf{65} 260314}
		
		\bibitem{huang2024nonreciprocal}
		Huang K-W, Wang X, Qiu Q-Y and Xiong H 2024 Nonreciprocal magnon blockade via the Barnett effect \textit{Opt. Lett.} \href{https://opg.optica.org/ol/abstract.cfm?uri=ol-49-3-758}{\textbf{49} 758}
		
		\bibitem{zhang2025nonreciprocal}
		Zhang W, Liu S, Zhang S and Wang H-F 2025 Nonreciprocal unconventional magnon blockade induced by Barnett effect and parametric amplification \textit{Opt. Express} \href{https://opg.optica.org/oe/fulltext.cfm?uri=oe-33-2-3339}{\textbf{33} 3339}
		
		\bibitem{zhang2024nonreciprocal}
		Zhang H-Q, Chu S-S, Zhang J-S, Zhong W-X and Cheng G-L 2024 Nonreciprocal magnon blockade based on nonlinear effects \textit{Opt. Lett.} \href{https://opg.optica.org/ol/abstract.cfm?URI=ol-49-8-2009}{\textbf{49} 2009}
		
		\bibitem{feng2026nonreciprocal}
		Feng S, Du Y, Cheng G, Zhong W and Chen A 2026 Nonreciprocal magnon blockade via combining Barnett and Sagnac effects in a nonlinear system \textit{Opt. Express} \href{https://doi.org/10.1364/oe.585451}{\textbf{34} 5079}
		
		
		\bibitem{guo2023pra}
		Guo Q, Cheng J, Tan H and Li J 2023 Magnon squeezing by two-tone driving of a qubit in cavity--magnon--qubit systems \textit{Phys. Rev. A} \href{https://doi.org/10.1103/PhysRevA.108.063703}{\textbf{108} 063703}
		
		\bibitem{liu2026pra}
		Liu G, Li G, Yang R-C, Xiong W and Li J 2026 Magnon squeezing near a quantum critical point in a cavity--magnon--qubit system \textit{Phys. Rev. A} \href{https://doi.org/10.1103/d2st-rr91}{\textbf{113} 033707}
		
		\bibitem{luo2021nonlocal}
		Luo D-W, Qian X-F and Yu T 2021 Nonlocal magnon entanglement generation in coupled hybrid cavity systems \textit{Opt. Lett.} \href{https://opg.optica.org/ol/abstract.cfm?uri=ol-46-5-1073}{\textbf{46} 1073}
		
		\bibitem{ren2022prb}
		Ren Y-L, Xie J-K, Li X-K, Ma S-L and Li F-L 2022 Long-range generation of a magnon-magnon entangled state \textit{Phys. Rev. B} \href{https://doi.org/10.1103/PhysRevB.105.094422}{\textbf{105} 094422}
		
		\bibitem{yang2026arxiv}
		Yang R-C, Liu G, Li G and Li J 2026 Macroscopic entanglement between two magnon modes via two-tone driving of a superconducting qubit \textit{Phys. Rev. A} \href{https://doi.org/10.1103/jtyx-xsgc}{\textbf{114} 043707}
		
		\bibitem{qi2022pra}
		Qi S-F and Jing J 2022 Generation of Bell and Greenberger-Horne-Zeilinger states from a hybrid qubit-photon-magnon system \textit{Phys. Rev. A} \href{https://doi.org/10.1103/PhysRevA.105.022624}{\textbf{105} 022624}
		
		\bibitem{qi2023pra}
		Qi S-F and Jing J 2023 Floquet generation of a magnonic NOON state \textit{Phys. Rev. A} \href{https://doi.org/10.1103/PhysRevA.107.013702}{\textbf{107} 013702}
		
		\bibitem{kong2021prb}
		Kong D, Hu X, Hu L and Xu J 2021 Magnon-atom interaction via dispersive cavities: Magnon entanglement \textit{Phys. Rev. B} \href{https://doi.org/10.1103/PhysRevB.103.224416}{\textbf{103} 224416}
		
		\bibitem{kong2022prr}
		Kong D, Xu J, Tian Y, Wang F and Hu X 2022 Remote asymmetric Einstein-Podolsky-Rosen steering of magnons via a single pathway of Bogoliubov dissipation \textit{Phys. Rev. Res.} \href{https://doi.org/10.1103/PhysRevResearch.4.013084}{\textbf{4} 013084}
		
		\bibitem{yan2024generating}
		Yan J-S and Jing J 2024 Generating magnon Bell states via parity measurement \textit{APL Quantum} \href{https://doi.org/10.1063/5.0201228}{\textbf{1} 026103}
		
		
		
		
		\bibitem{Nakamura18}
		Osada A, Gloppe A, Hisatomi R, Noguchi A, Yamazaki R, Nomura M, Nakamura Y and Usami K 2018 Brillouin light scattering by magnetic quasivortices in cavity optomagnonics \textit{Phys. Rev. Lett.} \href{https://doi.org/10.1103/PhysRevLett.120.133602}{{\bf 120} 133602} 
		
		\bibitem{Haigh21}
		Haigh J A, Nunnenkamp A and Ramsay J A 2021 Polarization dependent scattering in cavity optomagnonics \textit{Phys. Rev. Lett.} \href{https://doi.org/10.1103/PhysRevLett.127.143601}{{\bf 127} 143601}
		
		\bibitem{Sharma17}
		Sharma S, Blanter Y M and Bauer G E W 2017 Light scattering by magnons in whispering gallery mode cavities \textit{Phys. Rev. B} \href{https://doi.org/10.1103/PhysRevB.96.094412}{{\bf 96} 094412}
		
		\bibitem{PAP17}
		Pantazopoulos P A, Stefanou N, Almpanis E and Papanikolaou N 2017 Photomagnonic nanocavities for strong light-spin-wave interaction \textit{Phys. Rev. B} \href{https://doi.org/10.1103/PhysRevB.96.104425}{{\bf 96} 104425}
		
		\bibitem{Nakamuranjp}
		Osada A, Gloppe A, Nakamura Y and Usami K 2018 Orbital angular momentum conservation in brillouin light scattering within a ferromagnetic sphere \textit{New J. Phys.} \href{https://doi.org/10.1088/1367-2630/aae4b1}{{\bf 20} 103018}
		
		\bibitem{Haigh18}
		Haigh J A, Lambert N J, Sharma S, Blanter Y M, Bauer G E M and Ramsay A J 2018 Selection rules for cavity-enhanced brillouin light scattering from magnetostatic modes \textit{Phys. Rev. B} \href{https://doi.org/10.1103/PhysRevB.97.214423}{{\bf 97} 214423}
		
		\bibitem{Alm18}
		Almpanis E 2018 Dielectric magnetic microparticles as photomagnonic cavities: Enhancing the modulation of near-infrared light by spin waves \textit{Phys. Rev. B} \href{https://doi.org/10.1103/PhysRevB.97.184406}{{\bf 97} 184406}
		
		\bibitem{SV18}
		Graf J, Pfeifer H, Marquardt F and Kusminskiy S V 2018 Cavity optomagnonics with magnetic textures: Coupling a magnetic vortex to light \textit{Phys. Rev. B} \href{https://doi.org/10.1103/PhysRevB.98.241406}{{\bf 98} 241406 (R)}
		
		\bibitem{SV21}
		Graf J, Sharma S, Hueb H and Kusminskiy S V 2021 Design of an optomagnonic crystal: Towards optimal magnon-photon mode matching at the microscale \textit{Phys. Rev. Res.} \href{https://doi.org/10.1103/PhysRevResearch.3.013277}{{\bf 3} 013277}
		
		\bibitem{SV22}
		Bittencourt V A S V, Liberal I and Kusminskiy S V 2022 Optomagnonics in dispersive media: Magnon-photon coupling enhancement at the epsilon-near-zero frequency \textit{Phys. Rev. Lett.} \href{https://doi.org/10.1103/PhysRevLett.128.183603}{{\bf 128} 183603}
		
		\bibitem{Vahala}
		Eichenfield M, Chan J, Camacho R M, Vahala K J and Painter O 2009 Optomechanical crystals \textit{Nature} \href{https://doi.org/10.1038/nature08524}{{\bf 462} 78}
		
		
		\bibitem{NakamuraPRB}
		Hisatomi R, Osada A, Tabuchi Y, Ishikawa T, Noguchi A, Yamazaki R, Usami K and Nakamura Y 2016 Bidirectional conversion between microwave and light via ferromagnetic magnons \textit{Phys. Rev. B} \href{https://doi.org/10.1103/PhysRevB.93.174427}{{\bf 93} 174427}
		
		\bibitem{Op20}
		Zhu N, Zhang X, Han X, Zou C-L, Zhong C, Wang C-H, Jiang L and Tang H X 2020 Waveguide cavity optomagnonics for microwave-to-optics conversion \textit{Optica} \href{https://doi.org/10.1364/OPTICA.397967}{{\bf 7} 1291}
		
		\bibitem{GS22}
		Mukhopadhyay D, Nair J M P and Agarwal G S 2022 Anti-PT symmetry enhanced interconversion between microwave and optical fields \textit{Phys. Rev. B} \href{https://doi.org/10.1103/PhysRevB.105.064405}{{\bf 105} 064405}
		
		\bibitem{LiF23}
		Xie J, Ma S, Ren Y, Li X, Gao S and Li F 2023 Nonreciprocal single-photon state conversion between microwave and optical modes in a hybrid magnonic system \textit{New J. Phys.} \href{https://doi.org/10.1088/1367-2630/ace3eb}{{\bf 25} 073009}
		
		\bibitem{LPR24}
		Wu W-J, Wang Y-P, Li J, Li G and You J Q 2024 Microwave-to-optics conversion using magnetostatic modes and a tunable optical cavity \textit{Laser Photon. Rev.} \href{https://doi.org/10.1002/lpor.202400648}{{\bf 19} 2400648}
		
		\bibitem{Libb}
		Liu J-F, Hu Z-G, Zhang Y-X and Li B-B 2025 Integrated and tunable microwave-to-optical transduction via magno-optomechanics \textit{Phys. Rev. A} \href{https://doi.org/10.1103/fwpw-f81y}{{\bf 112} 033512}
		
		\bibitem{NG}
		Walschaers M 2021 Non-Gaussian quantum states and where to find them \textit{PRX Quantum} \href{https://doi.org/10.1103/PRXQuantum.2.030204}{{\bf 2} 030204}
		
		\bibitem{Fock}
		Bittencourt V A S V, Feulner V and Kusminskiy S V 2019 Magnon heralding in cavity optomagnonics \textit{Phys. Rev. A} \href{https://doi.org/10.1103/PhysRevA.100.013810}{{\bf 100} 013810}
		
		
		\bibitem{cat}
		Sun F-X, Zheng S-S, Xiao Y, Gong Q, He Q and Xia K 2021 Remote generation of magnon Schrödinger cat state via magnon-photon entanglement \textit{Phys. Rev. Lett.} \href{https://doi.org/10.1103/PhysRevLett.127.087203}{{\bf 127} 087203} 
		
		\bibitem{GS91}
		Agarwal G S and Tara K 1991 Nonclassical properties of states generated by the excitations on a coherent state \textit{Phys. Rev. A} \href{https://doi.org/10.1103/PhysRevA.43.492}{{\bf 43} 492}
		
		\bibitem{GS92}
		Agarwal G S and Tara K 1992 Nonclassical character of states exhibiting no squeezing or sub-Poissonian statistics \textit{Phys. Rev. A} \href{https://doi.org/10.1103/PhysRevA.46.485}{{\bf 46} 485}
		
		\bibitem{Lu1}
		Lu Z-X, Zhu H-B, Zuo X and Li J 2025 Preparing magnonic non-Gaussian states by adding a single magnon onto Gaussian states \textit{Phys. Rev. Res.} \href{https://doi.org/10.1103/jr6g-dcnp}{{\bf 7} 023242}
		
		\bibitem{Bell}
		Bell J S 1964 On the Einstein Podolsky Rosen paradox \textit{Physics Physique Fizika} \href{https://doi.org/10.1103/PhysicsPhysiqueFizika.1.195}{{\bf 1} 195}
		
		\bibitem{Cla}
		Freedman S J and Clauser J F 1972 Experimental test of local hidden-variable theories \textit{Phys. Rev. Lett.} \href{https://doi.org/10.1103/PhysRevLett.28.938}{{\bf 28} 938}
		
		\bibitem{Win}
		Rowe M A, Kielpinski D, Meyer V, Sackett C A, Itano M W, Monroe C and Wineland D J 2001 Experimental violation of a Bell's inequality with efficient detection \textit{Nature} \href{https://doi.org/10.1038/35057215}{{\bf 409} 791}
		
		\bibitem{Martinis}
		Ansmann M, Wang H, Bialczak R C, Hofheinz M, Lucero E, Neeley M, O'Connell A D, Sank D, Weides M, Wenner J, Cleland A N and Martinis J M 2009 Violation of Bell's inequality in Josephson phase qubits \textit{Nature} \href{https://doi.org/10.1038/nature08363}{{\bf 461} 504}
		
		
		\bibitem{Bellop}
		Marinkovi\'{c} I, Wallucks A, Riedinger R, Hong S, Aspelmeyer M and Gr\"{o}blacher S 2018 Optomechanical Bell test \textit{Phys. Rev. Lett.} \href{https://doi.org/10.1103/PhysRevLett.121.220404}{{\bf 121} 220404}
		
		\bibitem{Xie22}
		Xie H, Shi Z-G, He L-W, Chen X, Liao C-G and Lin X-M 2022 Proposal for a Bell test in cavity optomagnonics \textit{Phys. Rev. A} \href{https://doi.org/10.1103/PhysRevA.105.023701}{{\bf 105} 023701}
		
		
		\bibitem{DLCZ}
		Duan L-M, Lukin M D, Cirac J I and Zoller P 2001 Long-distance quantum communication with atomic ensembles and linear optics \textit{Nature} \href{https://doi.org/10.1038/35106500}{{\bf 414} 413}
		
		\bibitem{Kimble05}
		Chou C W, Riedmatten H D, Felinto D, Polyakov S V, Enk S J V and Kimble H J 2005 Measurement-induced entanglement for excitation stored in remote atomic ensembles \textit{Nature} \href{https://doi.org/10.1038/nature04353}{{\bf 438} 828}
		
		\bibitem{Simon18}
		Riedinger R, Wallucks A, Marinkovi\'{c} I, L\"{o}schnauer C, Aspelmeyer M, Hong S and Gr\"{o}blacher S 2018 Remote quantum entanglement between two micromechanical oscillators \textit{Nature} \href{https://doi.org/10.1038/s41586-018-0036-z}{{\bf 556} 473}
		
		\bibitem{WJA}
		Wu W-J, Wang Y-P, Wu J-Z, Li J and You J Q 2021 Remote magnon entanglement between two massive ferrimagnetic spheres via cavity optomagnonics \textit{Phys. Rev. A} \href{https://doi.org/10.1103/PhysRevA.104.023711}{{\bf 104} 023711}
		
		\bibitem{Lu3}
		Lu Z-X, Zhu H-B, Zuo X and Li J 2026 Optomagnonic generation of entangled traveling fields with different polarizations \textit{Quantum Sci. Technol.} \href{https://doi.org/10.1088/2058-9565/aeaf7c}{\textbf{11} 045070}
		
		\bibitem{Xieoe}
		Xie H, He L-W, Liao C-G, Chen Z-H and Lin X-M 2023 Generation of robust optical entanglement in cavity optomagnonics \textit{Opt. Express} \href{https://doi.org/10.1364/OE.478963}{{\bf 31} 7994}
		
		\bibitem{QN}
		Kimble H J 2008 The quantum internet \textit{Nature} \href{https://doi.org/10.1038/nature07127}{{\bf 453} 1023}
		
		\bibitem{Jie21X}
		Li J, Wang Y-P, Wu W-J, Zhu S-Y and You J Q 2021 Quantum network with magnonic and mechanical nodes \textit{PRX Quantum} \href{https://doi.org/10.1103/PRXQuantum.2.040344}{{\bf 2} 040344}
		
		\bibitem{Zhao23}
		Zhao C, Yang Z, Wang D, Yan Y, Li C, Wang Z and Zhou L 2023 Quantum networks assisted by dark modes in optomagnonic systems \textit{Phys. Rev. A} \href{https://doi.org/10.1103/PhysRevA.108.043703}{{\bf 108} 043703}
		
		\bibitem{Tan25}
		Hu N and Tan H 2025 Deterministic multipartite magnonic entanglement in a cavity optomagnonic network \textit{Phys. Rev. A} \href{https://doi.org/10.1103/xpdp-zwjy}{{\bf 112} 053708}
		
		\bibitem{WuY}
		Li Z-H, Huang W, Gan S-Q, Wu Y and L\"{u} X-Y 2026 Chiral dark mode for nonreciprocal transmission and asymmetric quantum state transfer in a magneto-optical hybrid system \textit{Phys. Rev. A} \href{https://doi.org/10.1103/yc3j-ffyj}{{\bf 113} 063718}
		
		\bibitem{RMP14}
		Aspelmeyer M, Kippenberg T J and Marquardt F 2014 Cavity optomechanics \textit{Rev. Mod. Phys.} \href{https://doi.org/10.1103/RevModPhys.86.1391}{{\bf 86} 1391}
		
		\bibitem{Bennett93}
		Bennett C H, Brassard G, Cr\'{e}peau C, Jozsa R, Peres A and Wootters W K 1993 Teleporting an unknown quantum state via dual classical and Einstein-Podolsky-Rosen channels \textit{Phys. Rev. Lett.} \href{https://doi.org/10.1103/PhysRevLett.70.1895}{{\bf 70} 1895}
		
		\bibitem{Zeil}
		Bouwmeester D, Pan J-W, Mattle K, Eibl M, Weinfurter H and Zeilinger A 1997 Experimental quantum teleportation \textit{Nature} \href{https://doi.org/10.1038/37539}{{\bf 390} 575}
		
		\bibitem{Kim}
		Braunstein S L and Kimble H J 1998 Teleportation of continuous quantum variables \textit{Phys. Rev. Lett.} \href{https://doi.org/10.1103/PhysRevLett.80.869}{{\bf 80} 869}
		
		\bibitem{Simon21}
		Fiaschi N, Hensen B, Wallucks A, Benevides R, Li J, Alegre T P M and Gr\"{o}blacher S 2021 Optomechanical quantum teleportation \textit{Nat. Photon.} \href{https://doi.org/10.1038/s41566-021-00866-z}{{\bf 15} 817}
		
		\bibitem{Fan23}
		Fan Z-Y, Zuo X, Qian H and Li J 2023 Proposal for optomagnonic teleportation and entanglement swapping \textit{Photonics} \href{https://doi.org/10.3390/photonics10070739}{{\bf 10} 739}
		
		\bibitem{Lu2}
		Lu Z-X, Zuo X, Fan Z-Y and Li J 2025 Optomagnonic continuous-variable quantum teleportation enhanced by non-Gaussian distillation \textit{Phys. Rev. Res.} \href{https://doi.org/10.1103/7215-zhbv}{{\bf 7} 043148}
		
		\bibitem{Yuan26}
		Xia Z, Ma R, Shu C, Yuan H and Xiao J 2026 Quantum entanglement and teleportation of magnons in coupled spin chains (arXiv:\href{https://doi.org/10.48550/arXiv.2601.13242}{2601.13242})
		
		\bibitem{Jie20A}
		Li J, Wallucks A, Benevides R, Fiaschi N, Hensen B, Alegre T P M and Gr\"{o}blacher S 2020 Proposal for optomechanical quantum teleportation \textit{Phys. Rev. A} \href{https://doi.org/10.1103/PhysRevA.102.032402}{{\bf 102} 032402}
		
		\bibitem{QI}
		Lloyd S 2008 Enhanced sensitivity of photodetection via quantum illumination \textit{Science} \href{https://doi.org/10.1126/science.1160627}{{\bf 321} 1463}
		
		
		\bibitem{Zhou21}
		Cai Q, Liao J, Shen B, Guo G and Zhou Q 2021 Microwave quantum illumination via cavity magnonics \textit{Phys. Rev. A} \href{https://doi.org/10.1103/PhysRevA.103.052419}{{\bf 103} 052419}
		
		\bibitem{block1}
		Xie H, He L-W, Shang X, Lin G-W and Lin X-M 2022 Nonreciprocal photon blockade in cavity optomagnonics \textit{Phys. Rev. A} \href{https://doi.org/10.1103/PhysRevA.106.053707}{{\bf 106} 053707}
		
		\bibitem{block2}
		Yuan Z-H, Chen Y-J, Han J-X, Wu J-L, Li W-Q, Xia Y, Jiang Y-Y and Song J 2023 Periodic photon-magnon blockade in an optomagnonic system with chiral exceptional points \textit{Phys. Rev. B} \href{https://doi.org/10.1103/PhysRevB.108.134409}{{\bf 108} 134409}
		
		\bibitem{block3}
		Deng X, Zhang K-K, Shui T and Yang W-X 2024 Nonreciprocal unconventional magnon blockade via the Sagnac-Fizeau shift in an optomagnonic system \textit{Phys. Rev. A} \href{https://doi.org/10.1103/PhysRevA.110.063711}{{\bf 110} 063711}
		
		\bibitem{Ying23}
		Liang Z, Li J and Wu Y 2023 All-optical polarization-state engineering in quantum cavity optomagnonics \textit{Phys. Rev. A} \href{https://doi.org/10.1103/PhysRevA.107.033701}{{\bf 107} 033701}
		
		\bibitem{Zhong}
		Yang Z-Q, Lin S-Q and Zhong Z-R 2025 Nonreciprocal transmission in a cavity--magnon system by rotational Sagnac effect \textit{Phys. Rev. Res.} \href{https://doi.org/10.1103/27cz-jqd4}{{\bf 7} 043256}
		
		
		
		
		\bibitem{Neuman2020}
		Neuman T, Wang D S and Narang P 2020 Nanomagnonic cavities for strong spin-magnon coupling and magnon-mediated spin-spin interactions {\it Phys. Rev. Lett.} \href{https://doi.org/10.1103/PhysRevLett.125.247702}{{\bf 125} 247702}
		
		\bibitem{Hei2021}
		Hei X-L, Dong X-L, Chen J-Q, Shen C-P, Qiao Y-F and Li P-B 2021 Enhancing spin-photon coupling with a micromagnet {\it Phys. Rev. A} \href{https://doi.org/10.1103/PhysRevA.103.043706}{{\bf 103} 043706}
		
		\bibitem{Xiong2022}
		Xiong W, Tian M, Zhang G Q and You J Q 2022 Strong long-range spin-spin coupling via a Kerr magnon interface {\it Phys. Rev. B} \href{https://doi.org/10.1103/PhysRevB.105.245310}{{\bf 105} 245310}
		
		
		\bibitem{Kumar25}
		Ribeiro S and Kumar P 2025 Dual-domain coherent emission: Photon-magnon laser in a hybrid quantum system {\it Phys. Rev. A} \href{https://doi.org/10.1103/2vkc-y6ww}{{\bf 112} 043711}
		
		
		\bibitem{Hei2024b}
		Hei X-L, Dong X-L, Chen J-Q, Qiao Y-F, Pan X-F, Yao X-Y, Zheng J-C, Ren Y-M, Huo X-W and Li P-B 2024 Topological simulation and chiral spin-spin interaction in driven cavity magnonics {\it Phys. Rev. Applied} \href{https://doi.org/10.1103/PhysRevApplied.22.044025}{{\bf 22} 044025}
		
		
		\bibitem{Pan2023}
		Pan X-F, Hei X-L, Dong X-L, Chen J-Q, Shen C-P, Ali H and Li P-B 2023 Enhanced spin-mechanical interaction with levitated micromagnets {\it Phys. Rev. A} \href{https://doi.org/10.1103/PhysRevA.107.023722}{{\bf 107} 023722}
		
		
		\bibitem{Fukami21}
		Fukami M, Candido D R, Awschalom D D and Flatt\'{e} M E 2021 Opportunities for long-range magnon-mediated entanglement of spin qubits via on- and off-resonant coupling {\it PRX Quantum} \href{https://doi.org/10.1103/PRXQuantum.2.040314}{{\bf 2} 040314} 
		
		
		\bibitem{Hei2024a}
		Pan X-F, Li P-B, Hei X-L, Zhang X, Mochizuki M, Li F-L and Nori F 2024 Magnon-Skyrmion hybrid quantum systems {\it Phys. Rev. Lett.} \href{https://doi.org/10.1103/PhysRevLett.132.193601}{{\bf 132} 193601}
		
		\bibitem{Skogvoll2021}
		Skogvoll I C, Lidal J, Danon J and Kamra A 2021 Tunable anisotropic quantum Rabi model via a magnon-spin-qubit ensemble \textit{Phys. Rev. Appl.} \href{https://doi.org/10.1103/PhysRevApplied.16.064008}{\textbf{16} 064008}
		
		\bibitem{Romling2023}
		R{\"o}mling A-L E, Vivas-Via{\~n}a A, S{\'a}nchez Mu{\~n}oz C and Kamra A 2023 Resolving nonclassical magnon composition of a magnetic ground state via a qubit \textit{Phys. Rev. Lett.} \href{https://doi.org/10.1103/PhysRevLett.131.143602}{\textbf{131} 143602}
		
		\bibitem{Romling2025}
		R{\"o}mling A-L E, Feist J, Garc{\'i}a-Vidal F J and Kamra A 2025 Squeezing and quantum control of the antiferromagnetic magnon pseudospin \textit{Phys. Rev. B} \href{https://doi.org/10.1103/qjm2-d19g}{\textbf{112} 214444}
		
		
		\bibitem{Jing21}
		Lu T-X, Zhang H, Zhang Q and Jing H 2021 Exceptional-point-engineered cavity magnomechanics {\it Phys. Rev. A} \href{https://doi.org/10.1103/PhysRevA.103.063708}{{\bf 103} 063708}
		
		\bibitem{Asjad23}
		Asjad M, Li J, Zhu S-Y and You J Q 2023 Magnon squeezing enhanced ground-state cooling in cavity magnomechanics {\it Fundam. Res.} \href{https://doi.org/10.1016/j.fmre.2022.07.006}{{\bf 3} 3}
		
		\bibitem{Isart20}
		Gonzalez-Ballestero C, Gieseler J and Romero-Isart O 2020 Quantum acoustomechanics with a micromagnet {\it Phys. Rev. Lett.} \href{https://doi.org/10.1103/PhysRevLett.124.093602}{{\bf 124} 093602}
		
		
		\bibitem{Isart21}
		Gonzalez-Ballestero C, Aspelmeyer M, Novotny L, Quidant R and Romero-Isart O 2021 Levitodynamics: Levitation and control of microscopic objects in vacuum {\it Science} \href{https://doi.org/10.1126/science.abg3027}{{\bf 374} eabg3027}
		
		\bibitem{Isart20prb}
		Gonzalez-Ballestero C, H\"ummer D, Gieseler J and Romero-Isart O 2020 Theory of quantum acoustomagnonics and acoustomechanics with a micromagnet {\it Phys. Rev. B} \href{https://doi.org/10.1103/PhysRevB.101.125404}{{\bf 101} 125404}
		
		\bibitem{FuwaM23}
		Fuwa M 2023 Stable magnetic levitation of soft ferromagnets for macroscopic quantum mechanics {\it Phys. Rev. A} \href{https://doi.org/10.1103/PhysRevA.108.023523}{{\bf 108} 023523}
		
		\bibitem{Zhong25b}
		Chen L, Liu Y, Bin L, Ye S-Y and Zhong Z-R 2025 Simultaneous cooling of the internal and external degrees of freedom of a levitated micromagnet in a cavity magnomechanical system {\it Phys. Rev. Res.} \href{https://doi.org/10.1103/8pzl-6c5l}{{\bf 7} 033157}
		
		\bibitem{Zhong25}
		Chen L, Liu Y, Bin L, Ye S-Y, Luo R-X and Zhong Z-R 2025 Simultaneous cooling of two levitated macromagnets in cavity magnomechanical system {\it Quantum Inf. Process.} \href{https://doi.org/10.1007/s11128-025-04690-0}{{\bf 24} 76}
		
		\bibitem{XiongH25}
		Xiong H 2025 Center-of-mass magnomechanics beyond magnetostrictive limits {\it Sci. China Phys. Mech. Astron.} \href{https://doi.org/10.1007/s11433-024-2606-4}{{\bf 68} 250313}
		
		\bibitem{NieW25}
		Zhan H, Chen A and Nie W 2025 Entanglement characteristics of levitated magnomechanical systems {\it Phys. Rev. A} \href{https://doi.org/10.1103/PhysRevA.111.023522}{{\bf 111} 023522}
		
		\bibitem{Yang2025}
		Yang Z, Xiong B, Zhao C and Zhou L 2025 Generation of multipartite entanglement in a cavity-magnomechanical system {\it Opt. Express} \href{https://doi.org/10.1364/OE.550848}{{\bf 33} 5123}
		
		\bibitem{Bayati24}
		Bayati S, Mahdifar A and Bagheri Harouni M 2024 Entanglement formation between magnon and center-of-mass motion of a levitated particle in a magnomechanical system {\it Phys. Rev. A} \href{https://doi.org/10.1103/PhysRevA.110.013710}{{\bf 110} 013710}
		
		\bibitem{Huo25}
		Huo X-W, Yang S-N, Hei X-L, Chen J-Q, Qiao Y-F, Pan X-F, Yao X-Y, Zheng J-C, Ren Y-M, Shu G-Y and Li P-B 2025 Robust dynamical and steady-state entanglement of phonons in the magnetic microsphere {\it Phys. Rev. A} \href{https://doi.org/10.1103/b6kk-rd74}{{\bf 112} 042613}
		
		\bibitem{Twamley25}
		Kumar Chauhan A, Kani A and Twamley J 2025 Enhancing macroscopic multimode entanglement through many-body interactions in cavity magnomechanics {\it Phys. Rev. A} \href{https://doi.org/10.1103/PhysRevA.111.033505}{{\bf 111} 033505}
		
		\bibitem{Rao25}
		Liu W and Rao S 2025 Entanglement generation between different center-of-mass motions in double levitated micromagnet system {\it Sci. Rep.} \href{https://doi.org/10.1038/s41598-025-28242-9}{{\bf 15} 44693}
		
		
		\bibitem{ZhangX16}
		Zhang X, Zou C-L, Jiang L and Tang H X 2016 Cavity magnomechanics {\it Sci. Adv} \href{https://doi.org/10.1126/sciadv.1501286}{{\bf 2} e1501286}
		
		\bibitem{Bayati2024}
		Bayati S, Harouni M B and Mahdifar A 2024 Magnomechanically induced transparency and tunable slow-fast light via a levitated micromagnet {\it Opt. Express} \href{https://doi.org/10.1364/OE.515093}{{\bf 32} 14914}
		
		\bibitem{Nie25}
		Nie W, Zhan H, Shang X, Zhang H and Chen A 2025 Modulating nonreciprocal transmission in levitated magnomechanical systems {\it Opt. Commun.} \href{https://doi.org/10.1016/j.optcom.2024.131212}{{\bf 574} 131212}
		
		
		\bibitem{Hensen26}
		Janse M, Mattana M, van Doorn J, van der Bent E, Wagner R, Smit R and Hensen B 2026 On-chip levitation of ferromagnetic microparticles (arXiv:\href{https://doi.org/10.48550/arXiv.2605.00090}{2605.00090}) 
		
		\bibitem{Tamegai23}
		Fuwa M, Sakagami R and Tamegai T 2023 Ferromagnetic levitation and harmonic trapping of a milligram-scale yttrium iron garnet sphere {\it Phys. Rev. A} \href{https://doi.org/10.1103/PhysRevA.108.063511}{{\bf 108} 063511}
		
		
		
		
		\bibitem{Gieseler20}
		Gieseler J, Kabcenell A, Rosenfeld E, Schaefer J D, Safira A, Schuetz M J A, Gonzalez-Ballestero C, Rusconi C C, Romero-Isart O and Lukin M D 2020 Single-spin magnetomechanics with levitated micromagnets {\it Phys. Rev. Lett.} \href{https://doi.org/10.1103/PhysRevLett.124.163604}{{\bf 124} 163604}
		
		\bibitem{Huillery20}
		Huillery P, Delord T, Nicolas L, Van Den Bossche M, Perdriat M and H\'etet G 2020 Spin-mechanics with levitating ferromagnetic particles {\it Phys. Rev. B} \href{https://doi.org/10.1103/PhysRevB.101.134415}{{\bf 101} 134415}
		
		\bibitem{Yi19L}
		Li Y, Polakovic T, Wang Y-L, Xu J, Lendinez S, Zhang Z, Ding J, Khaire T, Saglam H, Divan R, Pearson J, Kwok W-K, Xiao Z, Novosad V, Hoffmann A and Zhang W 2019 Strong coupling between magnons and microwave photons in on-chip
		ferromagnet-superconductor thin-film devices {\it Phys. Rev. Lett.} \href{https://doi.org/10.1103/PhysRevLett.123.107701}{{\bf 123} 107701}
		
		\bibitem{Hatanaka26}
		Hatanaka D, Asano M and Yamaguchi H 2026 Recent advances in on-chip magnomechanics for microwave phonon manipulation \textit{Appl. Phys. Express} \href{https://doi.org/10.35848/1882-0786/aea39c}{\textbf{19} 100101}
		
		
	\end{thebibliography}
\end{document}